# Krisenmanagement in kerntechnischen Notfällen

## Eine Evaluierung am Beispiel der Nuklearkatastrophe von Fukushima

Elias Koschier

8B

Betreuerin: Manuela Kainer

Evangelisches Realgymnasium Donaustadt

Maculangasse 2, 1220 Wien

30.01.2022

# Abstract

Ziel dieser vorwissenschaftlichen Arbeit ist es, die Krisenbewältigung während der Nuklearkatastrophe von Fukushima im Jahr 2011 zu analysieren und daraus Lektionen für einen sichereren Reaktorbetrieb und ein resilienteres Krisenmanagement abzuleiten. Die eingesetzten Methoden waren dabei neben offiziellen Berichten und einschlägiger Fachliteratur, Interviews mit einem Safeguard der IAEA und einer Politikwissenschaftlerin. Sie brachten zutage, dass der Unfall Defizite in der Unabhängigkeit der Atomaufsichtsbehörden aufzeigte und eine Fehlkoordination der vorhandenen technischen Ressourcen zusammen mit einer mangelhaften Kommunikationsstrategie zu vermeidbaren Evakuierungen führten.

The objective of this pre-scientific paper is to analyze the crisis management during the Fukushima nuclear accident in the year of 2011 and to derive lessons for a safer operation of reactors along with a more resilient crisis management framework. In addition to official reports and subject literature, the conducted methods included interviews with an IAEA safeguard and a political scientist. The central findings were that the accident pointed out deficiencies regarding the independence of the regulatory body and that mismanagement of the available technical resources in conjunction with insufficient communication led to evitable evacuations.

# Vorwort

Diese Arbeit ist das Ergebnis eines nun eineinhalbjährigen Arbeitsprozesses und ich bin stolz, Ihnen liebe Leserin / lieber Leser, die Ergebnisse meiner Recherchen nun präsentieren zu dürfen. Hinter jeder Seite verbirgt sich nächte- und tagelange Arbeit im Vorbereiten, Recherchieren, Exzerpieren, Schreiben und schließlich Korrigieren und es ist ein überwältigendes Gefühl, dieses Projekt am Ende eines langen Weges als Ganzes verwirklicht zu sehen. Schon seit ich Sprechen kann war mir das Atomkraftwerk Zwentendorf, welches sich in Gehweite von meinem Elternhaus befindet, ein Begriff und seither hält mich die unsichtbare und unbändige Natur dieser umstrittenen Form der Energiegewinnung in ihrem Bann. Im Rahmen dieser Arbeit durfte ich meinen brennenden Fragen im Umgang mit der Atomkraft während der Nuklearkatastrophe von Fukushima nachgehen.

Ich wünsche viel Freude und viele Erkenntnisse beim Lesen und möchte bei dieser Gelegenheit den größten Dank meiner Betreuerin Manuela Kainer aussprechen. Ob in den Sommerferien, oder am Wochenende, sie war jederzeit offen und erreichbar und begleitete mich unterstützend von der ersten Recherche bis zur Einreichung. Herzlicher Dank gebührt auch meinen Interviewpartnern Oliver Ali und Daniela Ingruber, die sich die Zeit nahmen, gut vorbereitet und detailliert all meine Fragen zu beantworten.

# Inhaltsverzeichnis

# 1. Einleitung

In der Zeit der Energiewende, in der die Stromerzeugung mittels Kernspaltungsreaktoren eine Renaissance erlebt, steigt neben ihrem Anteil an der globalen Stromerzeugung auch ihre Präsenz in unserem Alltag. Obwohl die Technologie, ungeachtet ihres Rufes in der Öffentlichkeit, insgesamt als eine der sichersten Methoden der Stromerzeugung gilt (vgl. Rashad/Hammad 2000: 223), kann auch in den Reaktoren mit den modernsten Sicherheits- und Redundanzsystemen keine absolute Sicherheit gewährleistet werden, weshalb das unwahrscheinliche Ereignis eines Unfalls jederzeit eintreten kann. Die dabei potentiell eintretenden Folgen sind, wie die Vergangenheit bereits mehrfach und zuletzt in der japanischen Präfektur Fukushima bewies, verheerend und verändern sich im Gegensatz zu anderen Katastrophenereignissen laufend. Es ist daher diese Eigendynamik, die die Krisenbewältigung am Ort des Unfalls und den Strahlenschutz der betroffenen Bevölkerung zu einer gewaltigen Herausforderung machen. Ziel dieser Arbeit ist deshalb, anhand der Nuklearkatastrophe von Fukushima 2011 mittels offizieller Berichte herauszufinden, wie die Krisenreaktion sowohl vor Ort, im Atomkraftwerk Fukushima Daiichi, als auch auf Ebene der Präfektur ablief und diese mithilfe von einschlägiger Fachliteratur und Analysen von IAEA-Safeguard Oliver Ali und der Politikwissenschaftlerin Daniela Ingruber zu analysieren. Aus den Taten und Fehlern der Stakeholder sollen in weiterer Folge Lektionen für mögliche künftige Unfallszenarien abgeleitet werden.

# 2. Nuklearkatastrophe von Fukushima – Ein Überblick

## 2.1. Auslöseereignisse

Ausgelöst wurde die Unglücksserie im Kernkraftwerk Fukushima Daiichi durch ein Erdbeben der Stärke 9.0, dessen Epizentrum sich weniger als 100 Kilometer von der japanischen Küste befand (vgl. IAEA 2015: 19).

Durch die intensive Bewegung der Kontinentalplatten wurden folglich enorme Mengen an Meerwasser in Bewegung gesetzt, was zur Entstehung von mehreren Tsunamiwellen führte. Die erste dieser Wellen mit einer Anlaufhöhe von 4-5 m traf dabei 40 Minuten nach dem Erdbeben auf das Kraftwerksgelände (vgl. IAEA 2015: 30), war aber, wie für Tsunamis üblich nicht die höchste (vgl. IAEA 2015: 49), sodass die 5,5 m hohen Schutzmauern am Ufer ihr standhielten (vgl. IAEA 2015: 31). Die zweite Welle, die jedoch eine Anlaufhöhe von 14-15 m hatte, überflutete das Gelände und somit auch die Mehrheit der Reaktorgebäude, die durch Aushebungen beim Bau unter dem entsprechenden Küstenniveau, nämlich auf 10 m über dem Meeresspiegel, konstruiert wurden (vgl. IAEA 2015: 31f).

## 2.2. Schäden der Überflutung und Stromverlust

Das eintretende Wasser verursachte unmittelbar, dass entweder die Dieselnotstromaggregate selbst oder die für ihre Kühlung zuständigen Pumpen beschädigt wurden, sodass 5 der 6 Reaktoren ihre Wechselstromversorgung verloren. Es kam daher zu einer Umschaltung der Notstromversorgung auf Gleichstrom von Batterien in den Reaktorgebäuden, die darauf ausgelegt waren, die Reaktoren für acht Stunden mit Strom zu versorgen und somit grundlegende Überwachungs- und Steuerungsfunktionen hinsichtlich Temperatur- und Druckentwicklung zu gewährleisten.

Da jedoch auch die Batterien oder essentielle Verbindungen dieser vom eintretenden Meerwasser betroffen waren, ging in den Reaktoreinheiten 1,2 und 4 binnen 15 Minuten die Gleichstromversorgung verloren (vgl. IAEA 2015: 31f).

Nachdem der völlige Stromverlust bedeutete, dass die für die Reaktorsteuerung zuständigen Personen keinerlei Auskunft über die Funktionalität der Sicherheitssysteme der betreffenden Reaktoren erhielten, wurde angenommen, dass die Wasserkühlung des Reaktorkerns verloren war. Auf Basis einer gesetzlichen Vorgabe wurden daher in der Folge sowohl die Tokyo Electric Power Company (TEPCO) als auch die japanischen Behörden über die Notlage in den Reaktorblöcken 1 und 2 informiert (vgl. IAEA 2015: 33).

Nach ungefähr zweieinhalb Stunden fand man heraus, dass sich die Isolationsventile des Reaktorkerns des ersten Reaktorblocks nach dem Ausfall des Wechselstroms geschlossen hatten (vgl. IAEA 2015: 34), was bedeutete, dass der Dampf des Reaktorkerns nicht mehr durch den Leerlaufkondensator zirkulieren konnte und somit jegliche Form der Reaktorkühlung verloren war (vgl. IAEA 2015: 26). Die Temperaturen stiegen seit dem Stromausfall also kontinuierlich an, was eine alternative Form der Kühlung dringend erforderlich machte. Da der Druck im Reaktor aber beinahe um den Faktor 10 zu hoch war, um eine externe Kühlung mit Wasser durchzuführen, wurde um 19:03 Uhr, über eine Ansprache des Premierministers (vgl. National Research Council 2014: 207), von der japanischen Regierung der nukleare Notfall ausgerufen (vgl. IAEA 2015: 34f).

Ein ähnlicher Zustand wurde in weiterer Folge auch für den zweiten Reaktorblock angenommen, bei dem allerdings unsicher war, ob das „isolation cooling system" noch funktionierte. Im Falle einer Inaktivität prognostizierte der eingerichtete Krisenstab („on-site emergency response centre"), dass es gegen 21:40 (vgl. IAEA 2015: 35) zu einer Freilegung des Reaktorkerns (vgl. Hatamura et al. 2012: 148; vgl. IAEA 2015: 35), also einer Leckentstehung im inneren Druckbehälter (vgl. IAEA 2015: 19) kommen würde. Dementsprechend wurde nach Erhalt der Meldung des Premierministers um 21:23 Uhr ein Befehl zur Evakuierung im Umkreis von 3 km um das Kraftwerksgelände ausgerufen und alle Personen im Umkreis von 10 km aufgefordert, in Innenräumen zu bleiben. Damit wurde der ursprüngliche Evakuierungsbefehl, der um 20:50 Uhr für einen Radius von 2 km herausgegeben worden war, erweitert (vgl. IAEA 2015: 35).

## 2.3. Freilegung des Reaktorkerns und Venting in Block 1

Nur kurze Zeit später kam es bereits zur Freilegung eines Reaktorkerns, nämlich in Reaktor 1 (vgl. Hatamura et al. 2012: 66), wie man um 21:51 Uhr durch eine enorm erhöhte Strahlenbelastung (0.8 mSv) im zugehörigen Gebäude feststellte. Unmittelbar nachdem der Reaktorkern im umschließenden Druckbehälter freigelegt worden war, stieg in diesem der Druck an, sodass die höchste vorgesehene Druckgrenze mit den am frühen Morgen erreichten Maximalwerten deutlich überschritten wurde und der Leiter des Krisenstabes entschied, Vorbereitungsmaßnahmen für ein Venting zu treffen (vgl. IAEA 2015: 35). Auch wenn hierfür einerseits mit dem Dampf aus dem Reaktorinneren radioaktive Isotope in die Umwelt freigesetzt werden müssen, so ermöglicht die Senkung des Drucks im Reaktorinneren andererseits, das Risiko für das Auftreten von Schäden zu senken (vgl. IAEA 2015: 38) und zudem den Reaktor bei niedrigem Druck provisorisch zu kühlen (vgl. IAEA 2015: 34).

Um sicherzustellen, dass durch die Durchführung des Ventings keine unbeteiligten Personen unmittelbar in Gefahr gebracht würden, sprach sich der Krisenstab mit den lokalen Einsatzkräften ab und begann erst mit der erforderlichen Vorarbeit, als bestätigt wurde, dass auch die Evakuierung in der wenige Kilometer südlich gelegenen Stadt Okuma abgeschlossen war (vgl. IAEA 2015: 36; vgl. TEPCO 2012: 84). Bis das Venting erfolgreich war und ein

Druckabfall im Reaktor gemessen wurde, vergingen allerdings über fünf Stunden, in denen die nötigen Ventile manipulativ geöffnet wurden (vgl. IAEA 2015: 36f). Unmittelbar danach begann man über eine Einlassöffnung (vgl. IAEA 2015: 35), an welche man die Pumpe eines Feuerwehrfahrzeugs anschloss, den Reaktor mit Frischwasser zu kühlen (vgl. IAEA 2015: 36). Da diese Form der Kühlung jedoch eine kontinuierliche Auffüllung des Löschfahrzeugs erforderte, traf man nach elf Stunden die Entscheidung, das vom Tsunami angestaute Meerwasser in den Reaktor einzuspeisen (vgl. IAEA 2015: 37).

## 2.4. Konflikt über die Meerwassernotkühlung in Block 1

Laut einem Paper des ehemaligen Chefredakteurs der Zeitung Asahi Shimbun und einem Untersuchungskommissionsmitglied zum Kraftwerksunfall, welches vom „Bulletin of the Atomic Scientists", einer Non-Profit-Organisation, die sich seit den Atombombenabwürfen auf Hiroshima und Nagasaki für einen verantwortungsvollen Umgang mit kerntechnischen Erfindungen einsetzt (vgl. Bulletin of the Atomic Scientists 2021), herausgegeben wurde, habe die Möglichkeit des Einspeisens von Meerwasser in den ersten Reaktorblock zu großen Spannungen zwischen der Unternehmensführung von TEPCO und dem Kraftwerksdirektor Masao Yoshida geführt. Der Repräsentant von TEPCO im Büro des damaligen Premierministers, Ichiro Takekuro soll Yoshida aufgefordert haben die nur kurz zuvor initiierte Noteinspeisung von Meerwasser in den Reaktorblock 1 zu unterbinden, bis eine ausdrückliche Weisung der Behörden hierzu käme. Da der Kraftwerksdirektor jedoch fest darauf beharrt habe, dass die Kühlung fortgesetzt würde, soll er in einer Videokonferenz in Anwesenheit der Unternehmensleitung durch Flüstern die Kraftwerksmitarbeiter:innen aufgefordert haben, den Befehl, den er in weiterer Folge äußern würde, zu ignorieren. Obwohl er daraufhin laut erklärte, die Reaktornotkühlung solle unterbrochen werden, wurde sie daher fortgesetzt (vgl. Funabashi/Kitazawa 2012: 17).

> *During a teleconference, Yoshida called the employee in charge of the seawater injections to his side and whispered in his ear – so the microphone for the teleconference with the head office would not pick up his voice – that though he would now order a halt to the seawater injections, the employee should disregard the order and continue. Thereupon, Yoshida loudly declared to all teleconference participants that water injections would be interrupted (Funabashi/Kitazawa 2012: 17).*

Der exakte Wortlaut war hierbei: „Shortly there will be an order to terminate seawater injection; however, whatever you do, do not stop injecting seawater" (Kitazawa et al. 2013: 188). Obwohl die IAEA diesen internen Konflikt bestätigt, wird im Report in kürzerer Form lediglich erwähnt, dass

> *[...] ein leitender Angestellter von TEPCO, der das Unternehmen im Büro des Premierministers vertrat, den Kraftwerksdirektor telefonisch aufforderte, die Meerwassereinspeisung in Block 1 zu stoppen. Diese Anweisung wurde nicht befolgt und die Meerwassereinspritzung wurde nicht unterbrochen (IAEA 2015: 39, engl. Version im Anhang).*

## 2.5. Wasserstoffexplosion in Block 1

Die zuvor fehlenden Möglichkeiten, den Reaktor zu kühlen bewirkten allerdings nicht nur, dass das Wasser im Druckbehälter siedete und zugleich der Druck anstieg, sondern durch den abfallenden Wasserstand und die daraus resultierende Freilegung der Brennstäbe im Kern (vgl. TEPCO 2013: 15) wurden mehrere gefährliche Entwicklungen in Gang gesetzt:

Zum einen bewirkte die Tatsache, dass die Brennstäbe nicht mehr von flüssigem Wasser umgeben waren, dass sich der Brennstoff weiter erhitzte und sich schließlich verflüssigte, was in einer Kernschmelze (vgl. Suppes/Storvick 2007: 341) resultierte. Das dabei entstehende (lavaartige) Corium aus den geschmolzenen Bestandteilen des Reaktorkerns (vgl. IAEA 2020) sowie der hohe Druck führten daher zur Leckentstehung sowohl im umschließenden Druckbehälter als auch im Reaktorsicherheitsbehälter, sodass Radioisotope über das Reaktorgebäude in die Umwelt geleitet wurden (vgl. TEPCO 2013: 15f).

Zum anderen kam es durch die freigelegten Kernbrennstoffe dazu, dass das auf der Oberfläche der Brennstäbe befindliche Zirkonium durch den Wasserdampf zu Zirkoniumdioxid oxidiert wurde, sodass in gewaltigen Mengen Wasserstoff entstand. Dieser entwich über Lecks aus den Druck- und Reaktorsicherheitsbehältern in das Reaktorgebäude und staute sich an der Decke. Am 12. März um 15:36 Uhr entzündete sich der angestaute Wasserstoff und es kam zu einer Explosion, die dem Reaktorgebäude schwere Schäden zufügte (vgl. TEPCO 2013: 16) und die Arbeit an den anderen Reaktorblöcken durch die im Umfeld verteilten Trümmer erschwerte (vgl. TEPCO 2013: 13). Tatsächlich war die Druckwelle der Explosion von solcher Intensität, dass sie Fenster in Gebäuden zerschlug, die drei Kilometer vom Kraftwerksgelände entfernt waren (vgl. Cleveland et al. 2021: 128)

## 2.6. Kernschmelze und Wasserstoffexplosion in Block 3

Im Unterschied zum ersten Reaktorblock, der nach dem Auftreffen des Tsunami jede Möglichkeit der Reaktorkühlung verlor, lagen die Batterien für die Gleichstromversorgung in Block 3 höher (vgl. TEPCO 2013: 21), sodass die Hochdruckpumpe für das „high-pressure coolant injection system" (HPCI) (TEPCO 2013: 5) für eineinhalb Tage am Laufen gehalten und der Reaktor gekühlt werden konnte (vgl. TEPCO 2013: 21).

Am Morgen des 13. März 2011 kamen allerdings Sorgen über die Zuverlässigkeit jener Turbine auf, die die Einspeisungspumpe betrieb, da eine mögliche Beschädigung dieser dazu führen könnte, dass Lecks im Reaktorsicherheitsbehälter entstünden. Als sich in weiterer Folge offenbarte, dass sich die Turbine nach Unterschreitung eines festgelegten Mindestdrucks nicht ausschaltete, entschied man daher angesichts des niedrigen Reaktordrucks, wie bereits bei Block 1 auf eine Niedrigdruckkühlung durch eine externe Pumpe umzusteigen. Das Öffnen der hierfür nötigen Druckventile scheiterte jedoch mehrmals, sodass der Reaktordruck in diesem Zeitraum den Punkt überschritt, ab dem eine externe Kühlung nicht mehr möglich war (IAEA 2015: 39). Darüber hinaus fiel auch der Wasserstand drastisch ab, sodass es zu einer Offenlegung des Kerns kam und dieser zusammen mit den Druck- und Reaktorsicherheitsbehältern sukzessive Schaden nahm (vgl. TEPCO 2013: 23f).

Um wie bei Block 1 die Niedrigdruckkühlung des Reaktors über die für den Brandfall vorgesehene Einspeisöffnung (IAEA 2015: 36) zu ermöglichen, wurde mit der Vorbereitung für ein Venting begonnen. Drei Stunden später waren die erforderlichen Ventile geöffnet, doch es zeichnete sich kein Druckabfall ab, was darauf hindeutete, dass die sogenannte „rupture disk“ (IAEA 2015: 40), ein Bauteil, der voreiliges Venting verhindern sollte (IAEA 2015: 38), noch nicht zersprungen war. Dies führte letztlich dazu, dass mit der Einspeisung von Meerwasser in Reaktor 3 erst am frühen Nachmittag begonnen werden konnte. Da man jedoch um 14:15 Uhr eine äußerst hohe Strahlenbelastung von 1 mSv/h am Rande des Kraftwerksgeländes und kurz darauf über 300mSv/h am Eingang des dritten Reaktorgebäudes maß, hielt die Notkühlung des Reaktors nicht lange an, zumal die zuständigen Mitarbeiter:innen in das Gelände des vierten Reaktors evakuiert wurden (vgl. IAEA 2015: 41).

Obwohl der Evakuierungsbefehl um 17:00 Uhr wieder aufgehoben wurde und man die Meerwassereinspeisung fortsetzte (vgl. IAEA 2015: 41), verschlechterte sich der Zustand des Reaktors in den folgenden Stunden maßgeblich. Nicht nur fiel in der Nacht auf den 14. März im Zusammenhang mit einem Anstieg des Drucks der Wasserstand im Reaktor drastisch, sodass der Kern offengelegt wurde und die Druck- und Reaktorsicherheitsbehälter Schaden nahmen, sondern es kam wie bereits in Block 1 zur Bildung von Wasserstoff, der über Risse aus dem Reaktorsicherheitsbehälter entwich und sich im Reaktorgebäude anstaute (vgl. TEPCO 2013: 23f; vgl. IAEA 2015: 42). Da man am 14. März um 6:20 Uhr feststellte, dass der Wasserstand den Messbereich unterschritt, befahl Kraftwerksdirektor Yoshida, alle Mitarbeiter:innen im Umfeld von Block 3 aufgrund der Gefahr einer Wasserstoffexplosion erneut zu evakuieren (vgl. IAEA 2015: 42).

Nur wenige Stunden später, um 11:01 Uhr, kam es tatsächlich zu einer enormen Explosion im oberen Teil des dritten Reaktorgebäudes (vgl. TEPCO 2013: 23; vgl. IAEA 2015: 42). Trotz der Evakuierung wurden Mitarbeiter:innen verletzt und es stellte sich später heraus, dass sich jene Ventile, die man vorsorglich für den Fall eines Ventings in Block 2 geöffnet hatte, als Folge der Explosion geschlossen hatten und sich nicht mehr öffnen ließen (vgl. IAEA 2015: 42).

## 2.7. Venting und Kernschmelze in Block 2

Entgegen den Erwartungen des Einsatzstabes im Kraftwerk, der prognostiziert hatte, dass der Kern von Block 2 als erster freigelegt würde (vgl. IAEA 2015: 35), blieb der Reaktor bis zum 14. März in einem stabilen Zustand, da das von der Stromversorgung unabhängige Kühlsystem („reactor core isolation cooling system“), anders als angenommen noch funktionstüchtig war. Erst um am 14. März um 13:00 Uhr vernahm man, dass der Wasserspiegel im Reaktor abfiel und zugleich der Druck anstieg, was darauf hindeutete, dass man die Kühlung verloren hatte. Unmittelbar danach, um 13:05 Uhr, versuchte man wie bereits in Blöcken 1 und 3, eine Notkühlung über die Einspritzöffnung mittels Feuerwehrfahrzeug einzurichten, stellte aber fest, dass der Druck im Reaktor zu hoch war. Tatsächlich waren die Druckwerte so hoch, dass man annahm, dass es unmittelbar zur Offenlegung des Kernbrennstoffes kommen könnte, weshalb man entschied, ein zeitintensives Venting zu verhindern und stattdessen den Reaktordruck über Ventile („relief valves“) direkt am inneren Druckbehälter zu senken (vgl. IAEA 2015: 42f). Im Unterschied zum Venting, bei dem der Dampf aus dem Reaktorinneren zu einem großen Teil durch ein Wasserbecken („supression pool“) geleitet wird und somit beim Austritt in die Atmosphäre weniger Radioisotope enthält (vgl. IAEA 2015: 38), wurde Druck unmittelbar aus dem inneren Druckbehälter in den umliegenden Reaktorsicherheitsbehälter abgelassen, sodass die Isolierungsfunktion des Reaktors erheblich eingeschränkt wurde. Als Konsequenz stiegen im umliegenden Reaktorsicherheitsbehälter daher sowohl die Strahlenwerte als auch der Druck, der gegen 23 Uhr den vorgesehenen Bereich überschritt, sodass ein Venting ausgehend vom Reaktorsicherheitsbehälter erforderlich wurde. Da die Versuche, ein Venting über das Wasserbecken zu starten, aber erneut erfolglos waren, versuchten die zugewiesenen TEPCO-Mitarbeiter:innen durch ein Manipulieren der Ventile das Becken zu umgehen und somit den stark kontaminierten Dampf aus dem Reaktorsicherheitsbehälter auf direktem Wege freizusetzen. Als jedoch auch dies scheiterte, stieg der Reaktordruck bis zum frühen Morgen weiterhin an, bis es am 15. März um 6:14 Uhr zu einer Explosion, gefolgt von einem enormen Druckabfall in Block 2 bis auf Atmosphärendruck, kam. Man schloss daraus, dass es zu einem völligen Verlust der Isolierungsfunktion des Reaktorsicherheitsbehälters gekommen war, was eine unkontrollierte Freisetzung von Radioisotopen bedeuten würde. Aus diesem Grund wurden alle rund 700 Mitarbeiter:innen in das Gebäude des Einsatzstabes evakuiert, bevor mit 650 Personen die überwiegende Mehrheit von ihnen in das nahegelegene Atomkraftwerk Fukushima-Daini gebracht wurde (vgl. IAEA 2015: 43f, vgl. Interview I).

Aus beiden Berichten, sowohl von der IAEA als auch von TEPCO geht hervor, dass ausgehend von Reaktorblock 2 die meisten radioaktiven Substanzen in die Umwelt freigesetzt wurden (vgl. TEPCO 2013: 17), was dadurch belegt wird, dass ungefähr 2 Stunden nach dem Druckabfall der Austritt von „[...] weißem Rauch [...]“ beobachtet wurde und zudem mit beinahe 12 mSv/h am Haupteingang zum Kraftwerksgelände „[...] der höchste Messwert [der Dosisleistung] seit Beginn des Unfalls“ (IAEA 2015: 44, engl. Version im Anhang) gemessen wurde. Dies veranlasste die Behörden dazu, alle Bewohner:innen im Bereich zwischen 20 und 30 km um das Kraftwerk aufzufordern, sich in Innenräumen aufzuhalten (vgl. IAEA 2015: 44). Doch obwohl der Kern offengelegt war und andererseits stark kontaminierter Dampf unmittelbar aus dem Reaktorsicherheitsbehälter in die Umwelt entwich, kam es im Gegensatz zu den Blöcken 1 und 3 zu keiner Wasserstoffexplosion, da durch die Explosion in Block 1 in der aus Paneelen zusammengefügten Fassade des Reaktorgebäudes eine Öffnung entstanden war, durch die das Gas entweichen konnte (vgl. TEPCO 2013: 17).

## 2.8. Wasserstoffexplosion in Block 4

Im Unterschied zu den Reaktoreinheiten 1, 2 und 3 entgingen Blöcke 4, 5 und 6 einer Kernschmelze. Während die Abfolge der Ereignisse in Block 4 stark jener der ersten 3 Reaktoreinheiten ähnelte, konnte eine Eskalation der Lage in den Blöcken 5 und 6 bereits dadurch verhindert werden, dass sie ungefähr 3 Meter höher über dem Meeresspiegel (vgl. IAEA 2015: 32) errichtet worden waren. Dies bewirkte, dass die gesamte Gleichstromversorgung durch Batterien uneingeschränkt genutzt werden konnte und in Block 6 sogar das Dieselnotstromaggregat funktionsfähig war (vgl. TEPCO 2013: 44; vgl. IAEA 2015: 33). Durch die Nutzbarkeit der Kühlsysteme konnte sowohl die Temperatur des Kühlwassers erfolgreich unter den Siedepunkt gebracht und der Reaktordruck schließlich auf Atmosphärenniveau gesenkt werden, sodass ein sogenannter „cold shutdown“ (vgl. TEPCO: 5) erreicht wurde (vgl. TEPCO 2012: 34; vgl. USNRC 2021)

Zusätzlich zu den Reaktoren 5 und 6 war aber insbesondere der Reaktor 4 in einem Zustand, von welchem man keine Gefahr erwartete, zumal dieser aufgrund von Wartungsarbeiten keine Brennstäbe enthielt. Dennoch kam es beinahe zeitgleich mit dem Druckabfall im Reaktorblock 2, am frühen Morgen des 15. März, zu einer enormen Wasserstoffexplosion im Reaktorgebäude von

Block 4, von welcher man anfangs vermutete, dass sie von einer Freilegung der vom Reaktor entnommenen Brennstäbe im Kühlbecken ausging. Wie sich jedoch herausstellte, war die Ursache für die Explosion, die den gesamten oberen Gebäudeteil zerstörte, die Tatsache, dass jenes wasserstoffhaltige Gas, welches beim Venting von Reaktorblock 3 freigesetzt wurde, nicht nur über Rohre in die Atmosphäre, sondern auch in das benachbarte Reaktorgebäude von Reaktor 4 entwich, wo es sich schließlich entzündete.

## 2.9. Entschärfung der Reaktorzustände

Der 15. März stellte einen Wendepunkt im Verlauf der Krise dar, da man danach schrittweise wieder Kontrolle über die Situation gewann (vgl. Funabashi/Kitazawa 2012: 13). Am 20. März gelang es, Reaktor 2 extern wieder mit Strom zu versorgen, sodass auch in Block 2 die Stromversorgung über Verbindungskabel wiederhergestellt werden konnte und die Armaturen im Kontrollraum nach mehr als einer Woche wieder genutzt werden konnten. Obwohl die Unglücksserie seither stagnierte, war dies der Anfang einer Phase der Aufräumarbeiten und Dekontamination, welche auch heute noch vor keinem Ende steht (vgl. Kasperski 2021). Gemäß Kriterien der IAEA ist das Ende der Unfallphase jedoch erst auf den 16. Dezember 2011 zu datieren, da sich erst an diesem Datum sowohl alle Reaktoren in einem „cold shutdown“ befanden als auch ausreichend niedrige Strahlungsmesswerte erreicht wurden.

# 3. Organisationsstruktur und Aufgabenverteilung der beteiligten Institutionen

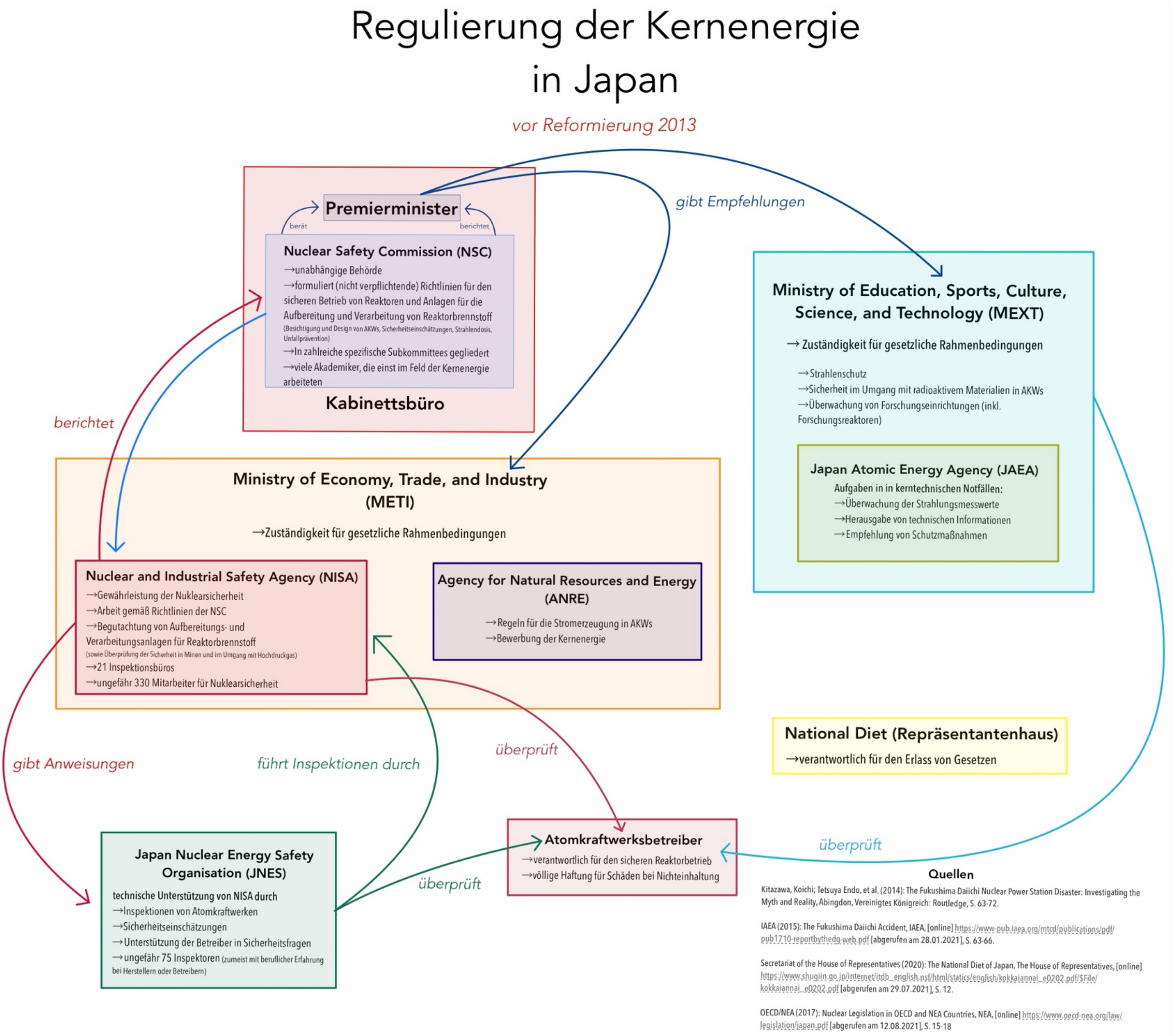

*Abbildung 1 – Übersichtsgrafik des Regulationssystems der Kernenergie*

Mit der Festlegung der Rahmenbedingungen und der Überwachung der japanischen Kernenergieindustrie wurden, wie in der Grafik ersichtlich, zahlreiche verschiedene Behörden beauftragt, wobei auch zwei unterschiedliche Ministerien (MEXT & METI) regulatorische Aufgaben übernahmen (vgl. Funabashi/Kitazawa 2012: 17; vgl. Hatamura et al. 2012:

72; vgl. IAEA 2015: 64). Im Kern lag die Rolle der Gewährleistung der Nuklearsicherheit jedoch im Verantwortungsbereich von NSC und NISA. Während erstere Organisation als Teil des Kabinettsbüros über den Premierminister (vgl. Hatamura et al. 2012: xxix; vgl. IAEA 2015: 64) Richtlinien zum Betrieb von Reaktoren und zugehöriger Infrastruktur wie Produktionsstätten und Brennstoffaufbereitungsanlagen herausgab (vgl. Hatamura et al. 2012: 66), arbeitete die NISA gemäß den Richtlinien der NSC als Kontrollorgan für Nuklearsicherheit (vgl. Hatamura et al. 2012: 66; vgl. IAEA 2015: 64) und musste der NSC, dem Gesetz entsprechend, zu Zwecken der Überwachung auf Anfrage Berichte übermitteln (vgl. IAEA 2015: 64). Obwohl die von der NSC festgelegten Richtlinien theoretisch betrachtet, nicht verpflichtend waren, zumal es sich lediglich um Empfehlungen handelte (vgl. Hatamura et al. 2012: 72), war deren Einhaltung für Betreiber:innen dennoch gewissermaßen verbindlich, da sie über die Instanz des Premierministers herausgegeben wurden (vgl. IAEA 2015: 64). Die Unabhängigkeit von NISA und NSC zeigte sich jedoch auch 2002, in jenem Jahr als ein Whistleblower ans Licht brachte, dass das Energieversorgungsunternehmen TEPCO zusammen mit mehreren anderen regionalen Energieversorgern in den 80er- und 90er-Jahren 29 Sicherheitsberichte gefälscht hatte, sodass die NISA ein einjähriges Anhalten aller TEPCO-Reaktoren erzwang (vgl. Hatamura et al. 2012: 81) und sie nach diesem Zeitraum erst wieder hochfuhr, wenn der Zustand aller Komponenten, den Vorgaben entsprechend, neuwertig war (vgl. Hatamura et al. 2012: 70). Obwohl die betreffenden Mängel, wie leichte Risse im Reaktorsicherheitsbehälter, die in Berichten bei internen Sicherheitsprüfungen verschwiegen wurden, keine unmittelbare Gefahr darstellten und laut europäischen oder US-amerikanischen Richtlinien nur Reparaturen oder gar keine Maßnahmen erfordert hätten (vgl. Hatamura et al. 2012: 81), wird durch die Fälschungscausa ersichtlich, dass die NISA unabhängig agieren konnte und Betreiber:innen bereits bei leichten

Verstößen gegen Richtlinien welche keinen Gesetzescharakter hatten, zur Rechenschaft ziehen konnte. Weiters zeigt sich dadurch aber auch, dass das Regulationssystem nur wenig Flexibilität bot, was dadurch bekräftigt wird, dass bereits kleinste Abweichungen von regulären Abläufen wie „ausgeleertes Wasser“ gemeldet hätten werden sollen und zudem viel Kritik von Kraftwerksmitarbeiter:innen und Industriellenvertreter:innen an der Bürokratie der Überprüfungen geübt wurde, welche „massive Abhakenübungen“ seien (vgl. Hatamura et al. 2012: 70, engl. Version im Anhang).

Auch wenn NISA, wie sowohl die IAEA als auch die OECD bestätigt (vgl. IAEA 2015: 64; vgl. OECD/NEA 2017: 15ff), in ihrer Arbeit unabhängig war, wurde die Souveränität der Behörde vielfach angezweifelt, was zumeist darauf zurückgeführt wird, dass das Ministerium für Wirtschaft, Handel und Industrie (METI) auch die Organisation ANRE (Agency for Natural Resources and Energy) beinhaltete, welche zuständig für die Bewerbung der Kernenergie war (vgl. IAEA 2015: 64). Dies führte jedoch zu der Schlussfolgerung, dass das Ministerium METI, dessen Teile die Organisationen ANRE und NISA waren, undifferenziert sowohl für die Bewerbung von Atomkraft als auch für die Gewährleistung der Nuklearsicherheit zuständig wäre Die IAEA empfahl daher, dass die Unabhängigkeit der Organisation NISA von METI gesetzlich eindeutiger festgelegt werden sollte (vgl. IAEA 2015: 64).

Obwohl die NISA die Aufgabe hatte, für die Gewährleistung der Nuklearsicherheit zu sorgen und somit als Überwachungsorgan der Betreiber:innen zu fungieren, wurde sie vor allem hinsichtlich technischer Aspekte von der JNES unterstützt. Hauptsächlich war JNES daher, mit rund 75 vollzeitbeschäftigten Inspektor:innen, welche zuvor zumeist für Hersteller von kerntechnischen Anlagen tätig gewesen waren, für Inspektionen von Kraftwerksanlagen verantwortlich, wobei der Zuständigkeitsbereich der Organisation auch die Durchführung von Sicherheitseinschätzungen umfasste (vgl. Hatamura et al. 2012: 68f).

Neben den besagten Aufgaben, die die Teilorganisationen des METI übernahmen, hatte jedoch auch das MEXT regulatorische Funktionen speziell im Bereich des Strahlenschutzes. Zusätzlich zur Überwachung des Strahlenschutzes in kerntechnischen Anlagen und dem Erheben von Messwerten stand jedoch auch die japanische Atomenergieorganisation JAEA unter der Überwachung von METI (vgl. IAEA 2015: 64; vgl. OECD/NEA 2017: 18), welche neben Forschungsarbeiten zur Weiterentwicklung von Reaktoren sowie zum Umgang mit radioaktiven Abfällen (vgl. OECD/NEA 2017: 18), in kerntechnischen Notfällen die Entwicklung der Strahlenbelastung im betroffenen Gebiet überwachen und sowohl die Behörden als auch die Zivilbevölkerung über Schutzmaßnahmen beraten sollte.

Während die oben genannten Institutionen für die Einhaltung der Sicherheitsstandards in kerntechnischen Anlagen zuständig waren, war die NSC (Nuclear Safety Commission) zusammen mit dem Repräsentantenhaus (National Diet) jenes Organ, das für das Schaffen von gesetzlichen Rahmenbedingungen sorgte. Dabei gab die NSC meist fachspezifische Empfehlungen zu konkreten Ereignissen wie die Überwachung von Strahlenmesswerten, oder präventiven Maßnahmen gegen Atomunfälle (siehe z.B. Gesetzestext: „Basic Plan for Emergency Preparedness“ 2013) heraus, während das Repräsentantenhaus für das Schaffen von allgemeinen gesetzlichen Rahmenbedingungen verantwortlich war.

Diese Aufgabenverteilung zeigt sich insbesondere im Gesetz für vorbereitende Maßnahmen bei kerntechnischen Notfällen, welches im Jahr 1999 vom Repräsentantenhaus verabschiedet wurde (siehe: Emergency Preparedness Basic Act).

Demnach sind bestimmte Aktionen wie das Informieren der Bürgermeister und Gouverneure der vom Unfall betroffenen Kommunen und Präfekturen verpflichtend durchzuführen, wenn „eine Strahlendosis über dem durch eine Verordnung der Kabinettsbüros festgelegten Höchstwert gemessen wird“ (Act on Special Measures Concerning Nuclear Emergency Preparedness; Artikel 10, Paragraph 1; engl. Version im Anhang). Dasselbe gilt auch für die Ministerien MEXT und METI, welche Details, wie die Gestaltung von Strahlungsmessstationen (Act on Special Measures Concerning Nuclear Emergency Preparedness; Artikel 11, Paragraph 1) oder die Ausstattung des „off-site center[s]“ (Act on Special Measures Concerning Nuclear Emergency Preparedness; Artikel 12, Paragraph 1) deren Vorhandensein durch das Gesetz vorgeschrieben wird, festzulegen haben.

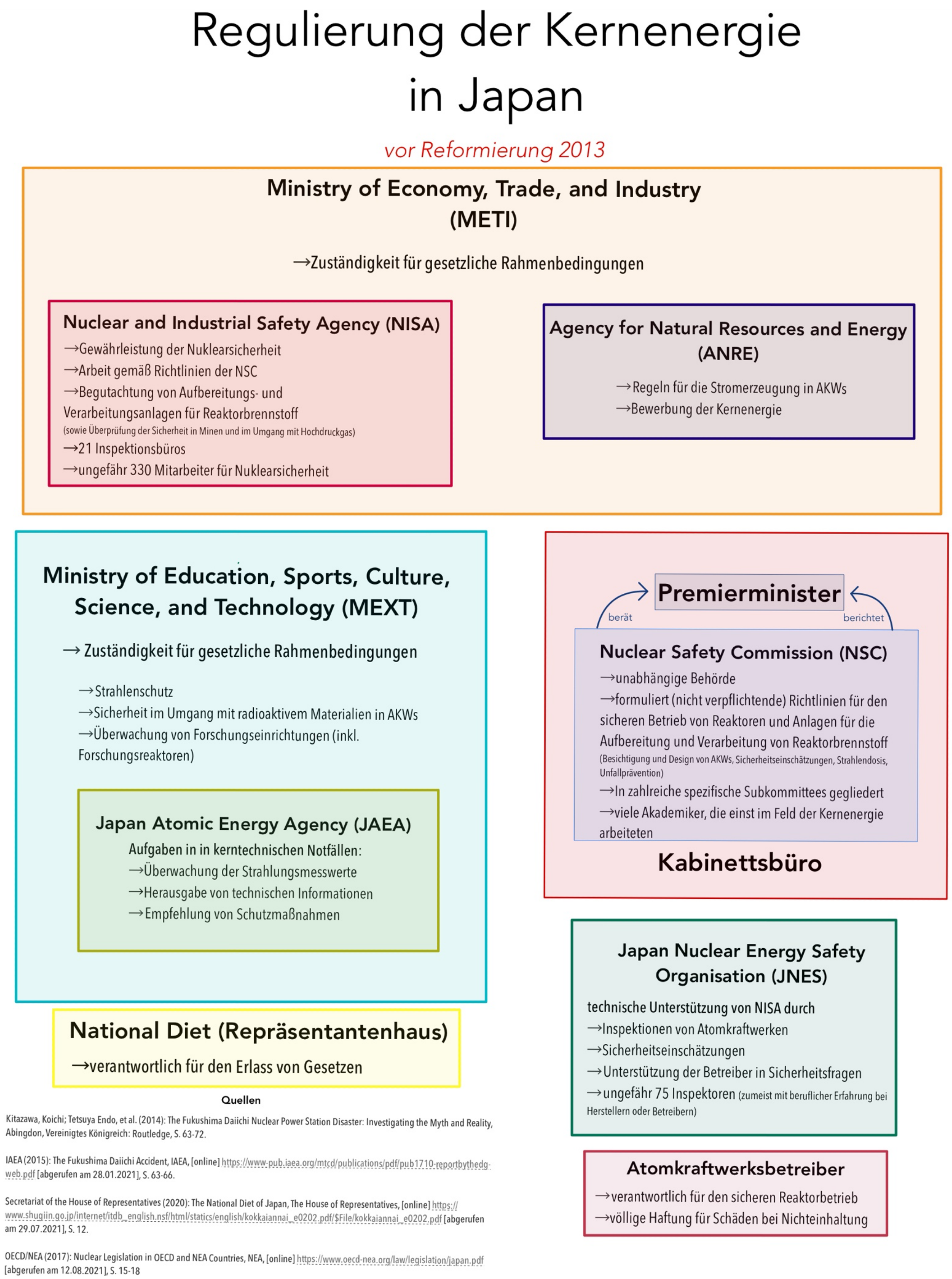

*Abbildung 2 - Vergrößerte Darstellung der Aufgabenverteilung der Regulationsorgane*

# 4. Krisenmanagementstruktur

## 4.1. Vorgesehene Krisenmanagementstruktur

Für die Koordination des Krisenmanagements in kerntechnischen Notfällen waren ursprünglich 3 Haupteinrichtungen vorgesehen, welche jeweils einen bestimmten Zuständigkeitsbereich hatten. Allgemeiner aber kann allgemeiner aber über die Krisenmanagementstruktur gesagt werden, dass man in Japan, im Gegensatz zu anderen kernenergieaffinen Ländern wie den USA eher eine „Top-Down-Strategie" (vgl. National Research Council 2014: 197; vgl. Hasegawa 2013: 28, 43) verfolgte, in welcher die höchste Instanz die vom Premierminister geführten „Nuclear Emergency Response Headquarters" (NERHQ) war (vgl. National Research Council 2014: 207; vgl. TEPCO 2012: 66). Grundlegende Entscheidungen zur Stabilisierung der Lage im Kraftwerk wie beispielsweise das Einspeisen von Meerwasser in die Reaktoren (vgl. IAEA 2015: 39; vgl. Kitazawa et al. 2014: 23f) sowie zum Schutz der betroffenen Bevölkerung durch Evakuierungsbefehle sollten daher von diesen getroffen werden (vgl. IAEA 2015: 79), obgleich dies nicht immer der Fall war.

Zusätzlich zu den NERHQ wurde, allerdings erst 4 Tage nach Beginn der Unglücksserie, im Hauptgebäude des Energieversorgers TEPCO am frühen Morgen des 15. März (um 5:27 Uhr) eine gemeinsame Zentrale zur Krisenbewältigung eingerichtet (vgl. Kitazawa et al. 2014: 30). Der Ort der Entscheidungsfindung verschob sich somit vom Büro des Premierministers in die weniger als 1,5 Kilometer entfernt gelegene Unternehmenszentrale (vgl. IAEA 2015: 80), von welcher aus mit allen beteiligten Institutionen, wie den unterschiedlichen Ministerien, Feuerwehren, der Polizei, den SDF, Behörden auf lokaler Ebene und dem Personal im Kraftwerk Informationen ausgetauscht und über weitere Maßnahmen beraten wurde (vgl. Kitazawa et al. 2014: 192; vgl. TEPCO 2012: 66).

Zusätzlich zur Krisenmanagementzentrale in Tokio waren die beiden weiteren zentralen Einrichtungen für die Koordination der Krisenbewältigung das „on-site center“, welches sich in einem 2007 hinzugefügten, erdbebensicheren Gebäude am Kraftwerksgelände befand, sowie das „off-site center“, welches dem Gesetz entsprechend (vgl. Act on Special Measures Concerning Nuclear Emergency Preparedness 1999) in der 5 km südwestlich entfernt gelegenen Stadt Okuma eingerichtet wurde (vgl. IAEA 2015: 79; vgl. Hasegawa 2013: 25; vgl. Kitazawa et al.: 122; vgl. TEPCO 2013: 69). Einerseits war das „on-site center“ unter Masao Yoshida die Zentrale für die Leitung der Wiederherstellungsarbeiten im Kraftwerk, von welchem ausgehend über ein Videokonferenzsystem mit den „Joint Emergency Response Headquarters“ und dem „off-site center“ kommuniziert wurde, um beispielsweise über folgenreiche Maßnahmen wie die Einspeisung von Meerwasser in die Reaktoren oder die Evakuierung von Personal nach Fukushima-Daini zu informieren. Andererseits war es jedoch auch die Verantwortung des „on-site centers“, vorläufige Pläne für Evakuierungen zu entwerfen (vgl. The Fukushima Nuclear Accident Independent Investigation Commission 2012: 34)

Das „off-site center“ hingegen sollte als Knotenpunkt für die an der Krisenbewältigung beteiligten Institutionen fungieren und dementsprechend das „Joint Measures Council for Nuclear Disasters“ „Council on Nuclear Emergency Measures“ (vgl. Ishibashi et al. 2012: 34) beherbergen, welches sich aus Vertreter:innen aller Verwaltungsebenen von den betroffenen Kommunen über die Präfektur-Regierung bis zur Nationalregierung zusammensetzte (vgl. TEPCO: 66). Da es für einen reibungslosen Informationsaustausch vorgesehen war und zudem von hier aus über Maßnahmen beraten werden sollte, war es unter anderem mit einem Videokonferenzsystem und dem SPEEDI-System ausgestattet, welches eine Prognose des Fallouts ermöglichte.

## 4.2. Tatsächliche Struktur des Krisenmanagements

Aufgrund der Tatsache, dass die Nuklearkatastrophe von Fukushima mit dem Tohoku-Erdbeben und dem daraus resultierenden Tsunami an die fatalste Naturkatastrophe in der Geschichte des Landes (vgl. Hasegawa 2013: 22) gekoppelt war und sich die Konsequenzen des Stromausfalls im Atomkraftwerk Fukushima Daiichi außerhalb jeglicher geplanter Szenarien befanden, kam es auch im Krisenmanagement zu zahlreichen Problemen und Abweichungen von der oben genannten Struktur. Diesen Problemen lag zumeist zu Grunde, dass, wie von der Literatur zum Krisenmanagement hervorgehoben wird (vgl. Sartory et al. 2013: 111f), entweder Kommunikationssysteme oder gar ein gesamter Standort nicht mehr zur Verfügung standen.

Dasselbe war auch für das „off-site center" in Okuma der Fall, zumal es, trotz seiner Funktion als Treffpunkt der Behörden, Einsatzkräfte und Betreiber:innen und Planungsstätte für Schutzmaßnahmen, lediglich kurzzeitig und sehr eingeschränkt genutzt werden konnte. Obwohl Redundanz in der Ausstattung mit Kommunikationsanlagen sichtlich beachtet wurde, war von den vier möglichen Kommunikationswegen durch den Stromausfall und die Erdbebenschäden nur die Satellitentelefonie nutzbar. Weder das Stromnetz noch das eigene Notstromaggregat stellten nach dem Erdbeben Strom zur Verfügung und am Tag darauf fiel auch das Festnetz aus, wodurch neben den Telefonen auch das Internet und somit das Videokonferenzsystem und Emailprogramme unbrauchbar wurden. Hinzu kam, dass Beamt:innen aufgrund der Rettungsarbeiten wegen des Erdbebens und des Tsunami, bereits anderorts beschäftigt waren und, wie auch Vertreter:innen von TEPCO, Schwierigkeiten in der Anreise hatten (vgl. Kitazawa et al. 2013: 123). Diese Beeinträchtigungen bewirkten, dass das „off-site center" nicht mehr seiner Verantwortung nachkam, die Evakuierungen im Umfeld des Kraftwerkes zu planen, sodass diese Aufgabe an den Krisenstab der NISA weitergegeben wurde. Da dieser Wechsel aber eine Verzögerung nach sich zog, schritt das Büro des Premierministers ein, um die Evakuierungsbefehle selbst herauszugeben. Dieses Einschreiten resultierte jedoch darin, dass es zum einen Defizite in der Zusammenarbeit der Behörden auf nationaler und lokaler Ebene gab und zum anderen weder die betroffenen Kommunen noch die Anwohner:innen ausreichend über Details zur Evakuierung beziehungsweise Gründe aufgeklärt wurden (vgl. Kitazawa et al. 2013: 123), was jedoch genauer im folgenden Kapitel thematisiert wird. Darüber hinaus bewirkte die Nähe des „off-site centers" zum Kraftwerk zusammen mit den sich dort verschlechternden Zuständen und dem Fehlen einer Luftfilterungsanlage, dass dessen Arbeitsfähigkeit völlig verhindert wurde, zumal es aufgrund der hohen Strahlenbelastung am 14. März evakuiert werden musste. Als neuer Standort kam jedoch nur das über 50 km weit entfernte Rathaus von Fukushima in Frage, welches jedoch nicht mit dem Dosis-Prognosesystem SPEEDI oder Messstationen ausgestattet war, da andere potentielle Einrichtungen wie beispielsweise in Minamisoma mit der Bewältigung der Folgen des Tsunami und des Erdbebens ausgelastet waren.

Um daher den Ausfall einer gemeinsamen Informationszentrale zu kompensieren und zudem die Atmosphäre des Misstrauens, die zwischen dem Kraftwerkspersonal, TEPCO und dem Premierminister durch das Fehlen einer direkten Kommunikation entstand, zu beseitigen (vgl. Kitazawa et al. 2013: 188ff), wurde am 15. März zusätzlich zu den NERHQ im Büro des Premierministers mit den JERHQ eine neue gemeinsame Krisenmanagementzentrale eingerichtet. Interviews zufolge, die die „Independent Investigation Commission on the Fukushima Nuclear Accident" mit hochrangigen Beamt:innen durchführte, sei dies der Wendepunkt im Verlauf der Krise gewesen, an welchem die Regierung wieder die Kontrolle über die Lage gewonnen habe (vgl. Kitazawa et al. 2013: 30). Auffällig ist hierbei, dass die Gestaltung der JERHQ als gemeinsame

Informations- und Krisenmanagementzentrale für alle primärbeteiligten Institutionen, welche auch von der Untersuchungskommission als „ziemlich effektiv im Erreichen einer maximalen Informationsweitergabe“ beschrieben werden, annähernd deckungsgleich mit der Empfehlung der Strahlenschutzkommission aus dem Jahr 2007 ist (vgl. SSK 2007: 18f, vgl. Interview II).

Neben der Einberufung des JERHQ bewirkte die Tatsache, dass die Kommunikation durch das „off-site center“ und die Folgen der Naturkatastrophe eingeschränkt war, schon zuvor, dass Premierminister Kan deutlich unmittelbarer in die Arbeiten zur Stabilisierung des Kraftwerks eingriff, wie sein persönlicher Besuch im Kraftwerk verdeutlicht. Trotz seiner Intention, die Arbeiten im Kraftwerk zu unterstützen, habe dies gemäß Funabashi und Kitazawa und der Fukushima Nuclear Accident Independent Investigation Commission zu Komplikationen geführt, da die vorgesehenen Informationsabläufe durchbrochen wurden, Mitarbeiter:innen vereinzelt von ihrer Arbeit abgelenkt wurden und das Einschreiten somit keinen Mehrwert für die Entwicklung der Lage hatte.

Ähnliches gilt auch für die Fälle, in denen der Premierminister per Telefon persönlich Equipment für das Kraftwerkspersonal beschaffen wollte. So entsandte Kan noch am Abend des 11. März 40 Lastwägen mit Generatoren und Pumpen (vgl. Independent Investigation Commission on the Fukushima Nuclear Accident/Bricker 2014: 8) auf unterschiedlichen Wegen in Richtung Kraftwerk, doch obwohl viele von ihnen das Kraftwerk am folgenden Morgen auch erreichten, war keine Pumpe und kein Generator mit den dort verbauten Anschlüssen kompatibel. Dieser und ein weiterer Fall, in welchem Kan mit dem Kraftwerkspersonal telefonierte, um die Abmessungen benötigter Batterien herauszufinden, veranlassten die National Diet of Japan Fukushima Nuclear Accident Independent Investigation Commission (NAIIC), sowie die Untersuchungskommission der Rebuild Japan Initiative Foundation (RJIF) zu dem Urteil, dass das Büro des Premierministers durch die Übernahme von Aufgaben, die TEPCO hätte selbstständig bewältigen können, die Gesamtperspektive im Krisenmanagement verloren (vgl. Kitazawa et al. 2013: 191) und die eigene Verantwortung zum Schutz der Bevölkerung vernachlässigt habe:

> *At all times, the government's priority must be its responsibility for public health and welfare. But because the Kantei's attention was focused on the ongoing problems at the plant— which should have been the responsibility of the operator—the government failed in its responsibility to the public. The Kantei's continued intervention in the plant also set the stage for TEPCO to effectively abdicate responsibility for the situation at the plant (vgl. Ishibashi et al. 2012: 35)*

Darüber hinaus sei dadurch eine Umfeld entstanden, in welcher TEPCO versucht habe, passiv zu agieren und die Verantwortung an die Regierung abzugeben. Den Höhepunkt erreichte dies dabei in einem Telefongespräch gegen 1:30 Uhr (vgl. TEPCO 2012: 104) des 15. März zwischen Banri Kaieda, dem damaligen Minister von METI und Masataka Shimizu, dem damaligen Präsidenten von TEPCO (vgl. Ishibashi et al. 2012: 35, 63), der, nach Kaiedas Auffassung (vgl. Kitazawa et al. 2013: 28; vgl. TEPCO 2012: 104f) wegen einer drohenden Wasserstoffexplosion um eine Evakuierung des *gesamten* Kraftwerkspersonals ersucht habe. Obwohl unklar ist, wie fortgeschritten allfällige Pläne zur Evakuierung des Kraftwerks waren und man im eigenen Bericht zum Unfall behauptete, dass „TEPCO nicht versuchte, alle Mitarbeiter:innen vom Kraftwerksgelände zu evakuieren" (TEPCO 2012: 102; engl. Version im Anhang), so führte die Erzählung Kaiedas zu einer intensiven Reaktion des Premierministers. Dieser forderte, dass das Kraftwerkspersonal zur Stabilisierung des Kraftwerks sein / ihr Leben riskieren müsse und behauptete, dass ein Verlassen der Anlage den sicheren Bankrott für TEPCO bedeuten würde (vgl. Kitazawa et al. 2013: 185).

# 5. Maßnahmen zum Strahlenschutz der Bevölkerung

## 5.1. Abfolge der durchgeführten Schutzmaßnahmen

| Datum (Uhrzeit) Jahr 2011 | Kriterien | Anweisungen | Verantwortliche Institution | Zielgruppe | Zugehörige Zone | Quellen |
|---|---|---|---|---|---|---|
| 11. März (20:50) | 2 km Radius von Fukushima Daiichi | verpflichtende Evakuierung | Regierung der Präfektur Fukushima | | „Restricted Area“ (Hasegawa, IAEA, National Research Council) | (Hasegawa 2013: 24; IAEA 2015: 86; vgl. The Fukushima Nuclear Accident Independent Investigation Commission 2012: 36; National Research Council 2014: 86ff) |
| 11. März (21:23) | 3 km Radius von Fukushima Daiichi | verpflichtende Evakuierung | Regierung | | „Restricted Area“ | (Hasegawa 2013: 24; IAEA 2015: 86; National Research Council 2014: 86ff) |
| 12. März (5:44) | 10 km Radius von Fukushima Daiichi | verpflichtende Evakuierung | Regierung | | „Restricted Area“ | (Hasegawa 2013: 24; IAEA 2015: 86; National Research Council 2014: 86ff) |
| 12. März (7:45) | Bis 10 km Radius von Fukushima Daini | verpflichtende Evakuierung für alle Personen im Umkreis von 3 km, Empfehlung zum Aufenthalt in Innenräumen im Gebiert zwischen 3 & 10 km rund um das AKW | Regierung | | „Restricted Area“ | (Hasegawa 2013: 24; IAEA 2015: 86; National Research Council 2014: 86ff) |
| 12. März (15:36) | Wasserstoffexplosion in Reaktorblock 1 | | | | | (siehe: Kapitel 1) |
| 12. März (17:39) | 10 km Radius von Fukushima Daini | verpflichtende Evakuierung für den Fall einer Wasserstoffexplosion | Regierung | | „Restricted Area“ | (IAEA 2015: 86) [6] |
| 12. März (18:25) | 20 km Radius von Fukushima Daiichi (einschließlich des zuvor evakuierten Gebiets) | verpflichtende Evakuierung | Regierung | 78 000 Personen (IAEA) | „Restricted Area“ | (Hasegawa 2013: 24; IAEA 2015: 86; National Research Council 2014: 86ff) |
| 15. März | 20-30 km Radius von Fukushima Daiichi (aufgehoben am 30. September) | Aufenthalt in Innenräumen & Empfehlung zur selbstständigen Evakuierung | Regierung | | „Evacuation Prepared Area“ | (Hasegawa 2013: 24; IAEA 2015: 87) |
| 16. März | Bewohner:innen, die die Sperrzone („Restricted Area“) verließen | Einnahme von Kaliumiodid in Form von Tabletten oder Pulver | (Regierung) | Vorrat für mehr als 900 000 Personen, tatsächlicher Verbrauch sehr gering, da die Evakuierung bereits größtenteils abgeschlossen war (Hamada et al. 2012) | „Evacuation Prepared Area“ | (IAEA 2015: 47; National Research Council 2012: 211) |
| 17. März | Herausgabe von Richtwerten für kontaminierte Lebensmittel und Wasser auf Basis eines von der NSC vorgeschlagenen Screenings, welches am Vortag begann (NRC). Das Ministerium für Arbeit, Gesundheit und Wohlstand (MHLW) ordnete an, dass die Äquivalentdosis ausgehend von kontaminierten Lebensmitteln nicht 5 mSv/Jahr übersteigen dürfe (Hamada und Ogino) (ausgenommen Iod-131 wegen der kurzen Halbwertszeit von ca. 8 Tagen [IAEA]) | | | | | (vgl. Kitazawa et al. 2014: 212, zit. nach: Hamada und Ogino 2012; IAEA 2015: 89; National Research Council 2014: 212) |
| 25. März | 20-30 km Radius (aufgehoben am 30. September) | Empfehlung zur selbstständigen Evakuierung | Regierung | 60 000 Personen | „Evacuation Prepared Area“ | (National Research Council 2014: 211) |
| 22. April | Gebiete mit Äquivalentdosis >20 mSv/Jahr | Verpflichtende Evakuierung binnen einem Monat | Regierung | | „Deliberate Evacuation Area“ | (National Research Council 2014: 211) |
| 16. Juni | Vereinzelte Orte mit Äquivalentdosis >20 mSv/Jahr | Empfehlung zur selbstständigen Evakuierung | Regierung | | „Deliberate Evacuation Area“ | (National Research Council 2014: 211) |

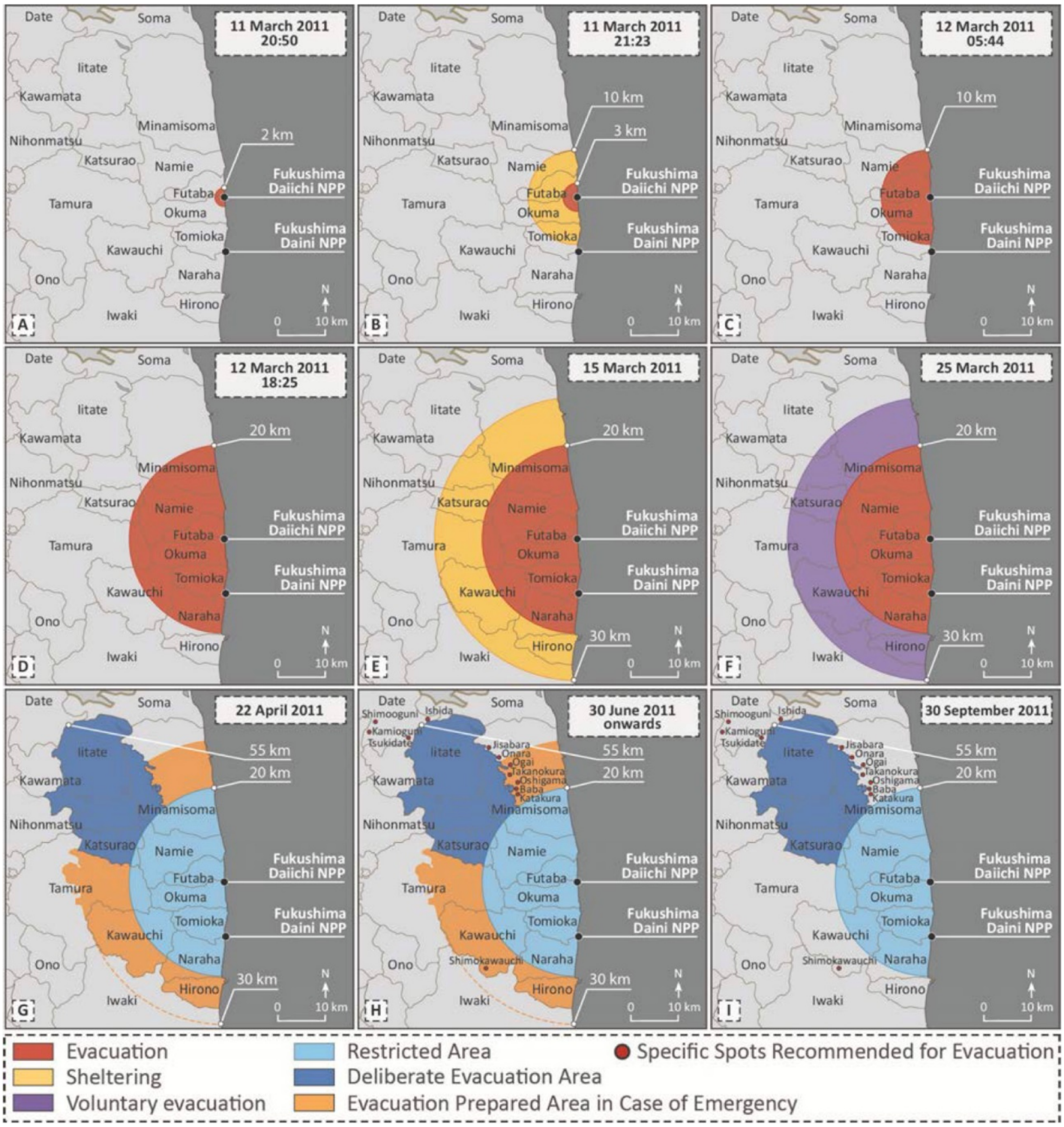

*Abbildung 3 – Veränderung des Evakuierungsgebiets im Laufe des Unfalls*

Wie in der oben angeführten Tabelle ersichtlich, begann die Regierung der Präfektur Fukushima durch eine Evakuierung des Gebietes im Radius von 2 km um 20:50 Uhr (Hasegawa 2013: 24; IAEA 2015: 86; vgl. The Fukushima Nuclear Accident Independent Investigation Commission 2012: 36; National Research Council 2014: 86ff) zeitnah mit dem Setzen von Schutzmaßnahmen, nachdem zu diesem Zeitpunkt noch keine signifikanten Freisetzungen von Radioisotopen in die Atmosphäre erfolgt waren und erst weniger als 2 Stunden zuvor eine kerntechnische Notsituation vom Premierminister ausgerufen worden war (vgl. IAEA 2015: 34f), was eine

Bedingung für das Zusammentreffen der in Kapitel 5 erwähnten Kriseneinrichtungen und folglich auch eine für die Durchführung von Schutzmaßnahmen war. Korrespondierend zur gemessenen Strahlung im Umfeld des Kraftwerkes wurden die Evakuierungsgebiete, welche sich zunächst radial um das Kraftwerk befanden, sukzessive erweitert und umfassten am Abend des 12. März, 3 Stunden nachdem es in Block 1 zu einer Wasserstoffexplosion gekommen war, das Gebiet im Umkreis von 20 km um das Kraftwerk, welches rund 78 000 Menschen beheimatete (vgl. Hasegawa 2013: 24; IAEA 2015: 86; National Research Council 2014: 86ff). Damit war auch das Einzugsgebiet des Kernkraftwerks Fukushima Daini eingeschlossen, in welchem unmittelbar davor der bestehende Evakuierungsradius als präventive Maßnahme für eine Wasserstoffexplosion von 3 auf 10 km ausgedehnt worden war (vgl. Hasegawa 2013: 24; National Research Council 2014: 86ff) Als am Morgen des 15. März am Eingang des Kraftwerksgeländes mit 12 mSv/h die höchste Dosisleistung (ca. ein Vierfaches der durchschnittlichen Jahresdosis) gemessen wurde, forderte die Regierung alle Anwohner:innen in einem Streifen von 10 km rund um das evakuierte Gebiet, in welchem ein Aufenthalt polizeilich verboten war (vgl. Hasegawa 2013: 23), auf, sich nur in Innenräumen aufzuhalten und empfahl der Bevölkerung in weiterer Folge das Gebiet auf freiwilliger Basis zu evakuieren (vgl. Hasegawa 2013: 24; IAEA 2015: 87).

In der Zeit nach dem 15. März zeichnete sich zudem ein Strategiewechsel in der Festlegung der Evakuierungsgebiete ab, da die Regierung über weitere Evakuierungen nun nicht mehr basierend auf dem Abstand zum Kraftwerk, sondern mittels konkreter Messwerte entschied. Aus dem radialen Sperrgebiet heraus ersteckte sich ab dem 22. April daher ein insgesamt fast 40 km langer Streifen, der von einer Äquivalentdosis von über 20 mSv/h gekennzeichnet war und binnen eines Monats verlassen werden musste. Im Juni 2011 wurden darüber hinaus auch vereinzelte Punkte mit derselben Äquivalentdosis definiert, für welche eine Evakuierung empfohlen wurde.

## 5.2. Problematiken der Evakuierung

Eine zentrale Ursache dafür, dass in der Durchführung der Evakuierungen fast ausschließlich auf radiale Zonen gesetzt wurde, war, dass die Prognosen des nach dem Three Mile Island Unfalls eigens entwickelten Fallout-Prognosesystems SPEEDI, welches anhand meteorologischer, aber auch geographischer Parameter und der Menge der in die Atmosphäre geleiteten radioaktiven Substanzen die Verteilung des Fallouts voraus berechnet, nicht für die Planung der Evakuierungszonen genutzt wurden (vgl. Kitazawa et al. 2013: 120f). Führende Entscheidungsträger einschließlich des Premierministers (vgl. Kitazawa et al. 2013: 121) oder des Ministers des Ministeriums MEXT, welches für die Anwendung des Prognosesystems zuständig war (vgl. Kitazawa et al. 2014: 30) , waren sich in der akuten Phase der Krise, in welcher die meisten Evakuierungsbefehle herausgegeben wurden, tatsächlich nicht über die Existenz eines solchen Prognosesystems bewusst und erfuhren teilweise erst durch die Berichterstattung der Medien darüber (vgl. Kitazawa et al. 2013: 121). Obwohl die Fehlkoordination im Einsatz von SPEEDI zu großer öffentlicher Aufregung führte und vom ehemaligen Leiter der New York Times in Japan, Martin Fackler, als „Blitzableiter der öffentlichen Wut“ (vgl. Cleveland et al. 2021: 118, engl. Version im Anhang) bezeichnet wurde, muss allerdings erwähnt werden, dass die Genauigkeit der Prognosen erheblich eingeschränkt war und daher auch nicht in die Planung späterer Evakuierungsgebiete wie die „Deliberate Evacuation Zone“ eingeflossen ist. Gründe dafür waren, dass einerseits das sogenannte „Emergency Response Support System“ (ERSS), welches Einrichtungen wie dem „off-site center“ Daten über den Kraftwerkszustand zur Verfügung stellt und damit auch SPEEDI mit Messwerten beliefert (vgl. National Research Council 2014: 208), aufgrund von Erdbebenschäden nicht funktionsfähig war (vgl. Kitazawa et al. 2013) und zudem ein wesentlicher Teil der freigesetzten Radioisotope nicht über kontrollierbare Wege wie den Schornstein entwichen, in welchen bei einer intakten Stromversorgung

Messungen durchgeführt hätten werden können (vgl. National Research Council 2014: 217). Dass über die radiologischen Folgen des Unfalls eine enorme und polarisierend wirkende (vgl. Rosen/Claußen 2016: 7f) Unsicherheit herrscht, ist ebenso eine Folge dessen, dass, die Gesamtmenge der freigesetzten Radioisotope, der sogenannte Quellterm, nicht exakt ermittelt werden konnte. Auch wurden konkrete Messdaten, die am 12. und 13. März manuell um das Kraftwerk erhoben wurden, bei der Evakuierung des „off-site centers" zurückgelassen, sodass aufgrund der ablandigen Windsituation erst am 20. März Messwerte für SPEEDI erhoben werden konnten. Mit der Begründung, Panik vermeiden zu wollen, wurden die Modelle, welche eine hohe Strahlenbelastung für die Stadt Iitate rund 50 km nordwestlich vom Kraftwerk prognostizierten, nicht berücksichtigt, jedoch wurde das Evakuierungsgebiet mit der „Deliberate Evacuation Area" einen Monat später bis nach Iitate ausgedehnt, da die jährliche Äquivalentdosis 20 mSv überschritt.

### 5.2.1. Mehrfache Evakuierungen am Beispiel Namie

Ein weiteres Problem der Evakuierung, welches neben der Unsicherheit über die Entwicklung der Lage im Kraftwerk teilweise auch auf den Umgang mit SPEEDI zurückzuführen ist, war, dass auch die Evakuierungsmaßnahmen selbst, die betroffene Bevölkerung erheblich beanspruchten. Ungefähr 70 Prozent der Personen im Umfeld des Kraftwerkes mussten mehr als viermal ihren Standort evakuieren, wobei viele von ihnen nicht über die möglichen Konsequenzen des Unfalls informiert wurden und daher häufig ihren Wohnort ohne Gepäck verließen. Wie das Beispiel der Evakuierung der Kleinstadt Namie zeigt, gab es jedoch auch Fälle, in denen Personen in Gebiete mit einer höheren Strahlenbelastung evakuiert wurden. Nicht nur erfuhr der später an Krebs verstorbene Bürgermeister von Namie, Tamotsu Baba wie auch andere Kommunalvertretungen erstmals aus dem Fernsehen Informationen über die Notlage im nahegelegenen Atomkraftwerk, sondern er erhielt aufgrund gescheiterter Kontaktversuche weder von TEPCO noch von der Regierung Anweisungen für die Durchführung der Evakuierung oder anderer Schutzmaßnahmen wie der Vergabe von Kaliumiodidtabletten, weshalb er die Bewohner:innen eigenständig in das gebirgige Gelände nördlich der Stadt evakuierte. Jenes Gebiet, für welches die nicht veröffentlichte SPEEDI-Prognose bereits eine erhöhte Kontamination vorausgesagt hatte (vgl. Cleveland et al. 2021: 131f).

Neben der Bevölkerung von Namie, von der rund die Hälfte aller Personen in ein Gebiet hoher Strahlenbelastung evakuiert wurde (vgl. The Fukushima Nuclear Accident Independent Investigation Commission 2012: 56), waren es jedoch vor allem pflegebedürftige und hospitalisierte Menschen, die die Evakuierung am meisten beanspruchte. Da sich für einige dieser Menschen bereits leichte Veränderungen in der in der Qualität der Pflege, wie sie durch die Transporte verursacht wurden, kritisch auf den Gesundheitszustand auswirken mussten, war es der Umgang mit dieser vulnerablen Personengruppe, der sich am stärksten auf die Bilanz der Todesfälle, die auf den Unfall zurückzuführen sind, auswirkte. So sind mehr als 60 Sterbefälle auf die Evakuierung zurückzuführen, weshalb mehrfach eine differenzierte Vorgehensweise gemäß dem ALARA-Prinzip („as low as reasonably achieveable), bei unterschiedlichen Personengruppen gefordert wird (vgl. Cleveland 2021: 144; vgl. Nomura et al. 2016: 80ff; vgl. Ohba et al. 2021: 7).

### 5.2.2. Konsequenzen der Evakuierung

Auch wenn die Evakuierung, die weitgehend durch die japanischen Selbstverteidigungsstreitkräfte JSDF, aber auch durch lokale Behörden und die Polizei erfolgte, nur sehr begrenzt dokumentiert wurde, bietet ein Paper der Oxford University Press, dass die Protokolle des Krankenhauses Futaba Kosei analysierte, Einblick in die Art der Durchführung der Evakuierung. Das Spital, dass sich im Abstand von 3,9 Kilometern zum Kraftwerk befand, war zum Zeitpunkt der Evakuierung am 12. März der Aufenthaltsort von 136

Patient:innen und 150 Bediensteten (vgl. Sawano et al. 2021: i122ff) und begann mit der Evakuierung, als gegen 6:40 Uhr zwei Polizisten den Direktor des Krankenhauses dazu aufforderten. Obwohl sie keine eindeutigen Gründe für die Maßnahmen nannten, ging durch die Berichterstattung der Medien hervor, dass es im nahegelegenen Atomkraftwerk zu einem Notfall gekommen war und Anwohner:innen im Umkreis von 10 Kilometern zur Evakuierung aufgefordert wurden. Folglich begann man mithilfe von Bussen um 8:30 Uhr mit der Verlagerung von 90 Bediensteten und 53 gehfähigen Patient:innen mit der Verlagerung in das rund 40 Kilometer nordwestlich gelegene Krankenhaus von Nihonmatsu City. Weitere 40 Patient:innen wurden mithilfe eines Lastwagens der JSDF an einen „unbekannten Ort“ (Sawano et al. 2021: i124) gebracht, wobei vermutet wird, dass es sich um das Altersheim in Namie handelte, das allerdings ebenso innerhalb der Sperrzone lag (vgl. IAEA 2015: 47), was bestätigt, dass viele der betroffenen Personen mehrfach evakuiert werden mussten.

Aufgrund eines Informationsmangels über die Lage der Situation herrschte zuerst Unsicherheit über das Verfahren mit gesundheitlich stärker beeinträchtigten Patienten, weswegen diese vorerst weiter in Futaba betreut wurden. Da der Direktor des Krankenhauses allerdings gegen 14 Uhr einen Anruf von einem Freund im Krisenstab erhielt, der die Ernsthaftigkeit der Situation schilderte, beschloss man auch die verbliebenen Patienten mittels einer Luftbrücke mit Hubschraubern von der nahegelegenen High School mit Hubschraubern nach Nihonmatsu zu evakuieren. Da die Evakuierung der Patient:innen im Futaba-Kosei-Krankenhaus in Teilschritten organisiert war und das Verlagern von schwer erkrankten Patienten erst nach einer Evaluierung der Situation und mit eigens zugeteiltem Personal erfolgte, starben von den 136 Patienten lediglich vier am Tag der Evakuierung (zwei Personen an „Krebs im Endstadium“, eine Person an einem „ernsthaften Gesundheitszustand“ und eine Person an einem „schweren Herzversagen“), weshalb sie als „effektiv“ beschrieben wird. Im Kontrast dazu starben von den 338 Patient:innen im nahegelegenen Futaba-Krankenhaus 39 (11,6%) während der Evakuierung – ein Prozentsatz der von den Todesfällen unter den Patienten des Futaba-Kosei-Krankenhauses erst rund 3 Monate später erreicht wurde. Obwohl die konträren Erfahrungen in den beiden Krankenhäusern die Unsicherheit in der Wahl der Vorgangsweise aufzeigen, wird vermutet, dass das Mortalitätsrisiko durch die Evakuierung signifikant angestiegen sein könnte. (vgl. Sawano et al. 2021: i125).

# 6. Ausblick und Zukunftslektionen

Die Nuklearkatastrophe von Fukushima demonstrierte einmal mehr unter enormer medialer Aufmerksamkeit, dass absolute Sicherheit im Betrieb von Reaktoren ein Mythos ist (Safety-Myth) (vgl. Interview II). Auch wenn die Stromversorgung der Reaktornotkühlung mit zwei Redundanzsystemen ausgestattet war und die Reaktoren zudem über einen integrierten Kühlkreislauf verfügten, gelang es nicht, die Reaktoreinheiten 1, 2 und 3 vor einer Kernschmelze zu bewahren. Dies obwohl die Leistungen des Kraftwerkspersonals in der Literatur einheitlich nicht bemängelt und sogar als „herausragend" im Rahmen der Umstände bezeichnet werden (vgl. Interview I).

Aus technischer Sicht sind die Lektionen, die aus den Ereignissen des Unfalls zu lernen sind, eindeutig und wurden sowohl in Japan als auch international vielfach umgesetzt. Die Schutzmauern vor Atomkraftwerken, deren Höhe zuvor 5,5 Meter betrug (vgl. IAEA 2015: 31), wurden erhöht und es wurde sichergestellt, dass Generatoren und Akkus zur Notstromversorgung höher positioniert wurden. Obwohl diese Maßnahmen die zentrale Ursache der Nuklearkatastrophe beseitigten, zeigte der Ereignisverlauf jedoch auch allgemeinen infrastrukturellen Verbesserungsbedarf auf. Während das „on-site center" erdbebensicher war und das Kraftwerkspersonal mit einer Luftfilteranlage vor Radionukliden schützte, traf dies nicht auf das „off-site center" zu und schränkte dessen Handlungsfähigkeit im Krisenmanagement folglich ein. Ebenso waren erschwerende Faktoren bei der Entscheidung über Schutzmaßnahmen, die Erhebung des Quellterms und die Prognose der weiteren Verteilung der strahlenden Substanzen, obgleich sie sich als relativ akkurat herausstellte. Wäre man in der Evakuierung von besonders vulnerablen Personengruppen weniger proaktiv vorgegangen, so hätten sich vermutlich zahlreiche verfrühte Sterbefälle abwenden lassen. Da die Freisetzung von Radioisotopen großteils über undichte Stellen am

Reaktorsicherheitsbehälter erfolgte, wäre eine Möglichkeit der exakteren Näherung des Quellterms, die Reaktorgebäude mit entsprechendem Messequipment auszustatten. Eine manuelle Messung in unmittelbarer Nähe zum Reaktorkern wurde schließlich durch die Strahlenbelastung im lebensbedrohlichen Bereich und die akute Explosions- und Einsturzgefahr verhindert.

Wie aus den Veröffentlichungen der IAEA und anderen Institutionen wie der US Nuclear Regulatory Commission oder dem Nuclear Energy Institute hervorgeht, sind die technischen Empfehlungen tatsächlich sehr umfassend und detailliert. Da freiwillige Empfehlungen wie das Paper „The Way Forward“ des amerikanischen Nuclear Energy Institute allerdings nur technologische Verbesserungen berücksichtigen, bezeichnet Cleveland et al. sie gar als „technokratisch“, zumal sie nur das Risiko von technischem Versagen reduzierten, nicht aber darauf abzielten, die soziale Kompetenz und die Fähigkeit zur Improvisation zu stärken (vgl. Cleveland et al. 2021: 244).

Was aus den Erfahrungen aus Fukushima allerdings insbesondere für die Verbesserung des Krisenmanagements gelernt werden sollte, ist, dass leitende Positionen mit kompetenten Personen besetzt werden sollten (vgl. Interview I), die Kenntnis und einen Überblick über alle verfügbaren Ressourcen und Mittel haben. Damit soll sichergestellt werden, dass, was beim Fallout-Prognosesystem SPEEDI nicht der Fall war, alternative Vorgangsweisen zumindest in Betracht gezogen werden und nicht von vornherein ausgeschlossen werden.

Offener und ausführlicher sollte im Krisenmanagement aber auch die Kommunikation des Krisenstabes mit anderen beteiligten Autoritätspersonen erfolgen. Die Erfahrungen aus dem Futaba-Kosei-Krankenhaus zeigen auf, dass Entscheidungsträger:innen wie Krankenhausdirektoren:innen eine ausführliche Beschreibung der radiologischen Zustände bedürfen, um damit etwaige Schutzmaßnahmen mit dem ALARA-Prinzip begründen zu können.

Ein weiteres zentrales Problem im Hinblick auf die Krisenkommunikation mit der Allgemeinheit war jedoch auch, dass die Regierung die Kommunikation beinahe ausschließlich in eine Richtung führte und nur beschränkt auf die Fragen der Bürger:innen einging. Dies offenbarte sich insbesondere in der Kommunikation über das soziale Medium Twitter, welches das Kabinett des Premierministers zwar für die Kundgebung von Schutzmaßnahmen und Aktualisierungen nutzte, dabei allerdings in keiner Weise auf Rückfragen einging. Besonders problematisch war dieser einseitige Informationsfluss angesichts scheinbar widersprüchlicher Aussagen, wie dass „ausgenommen bei Kleinkindern, keine physiologischen Folgen beim Trinken von Leitungswasser" (Kitazawa et al. 2014: 156ff) gäbe, obgleich man noch am Vortag, dem 23 März, noch behauptete, dass es (allgemein) sicher sei, Leitungswasser zu trinken – ohne die Ausnahme von Kleinkindern zu machen.. Auch TEPCO folgte in der Kommunikation via Twitter demselben Schema und ging nicht auf öffentliche Kommentare ein. Darüber hinaus übten Nutzer Kritik daran, dass das Unternehmen zwar über Stromausfälle in betroffenen Regionen informierte, aber keine Informationen über die von Fukushima Daiichi ausgehende Strahlung herausgab. Den Vorteil von beidseitiger Verständigung nicht nicht zu nutzen, habe, zusammen mit unpräziser Information und unzureichenden Erklärungen von Strahlenmesswerten, laut Kitazawa et al. das Entstehen von Gerüchten begünstigt. Es wurde daher die Empfehlung ausgesprochen, in der Krisenkommunikation dem Dialog mit der Öffentlichkeit mehr Achtung zu schenken. (vgl. Kitazawa et al. 2014: 193f, vgl. Interview II).

Da der Rückbau des Kraftwerkes und die Dekontaminierung des Umlandes (siehe auch: Umgang mit dem mit Radioisotopen besetzten Kühlwasser), noch viele Jahre in Anspruch nehmen werden, ist für die langfristig sichere

Durchführung der Arbeit auch das Vertrauen der Öffentlichkeit erforderlich, wie beispielsweise die Errichtung eines Endlagers in Gorleben zeigt. Da das Vertrauen in die japanischen Behörden während der Nuklearkatastrophe von Fukushima dauerhaft Schaden genommen hat (vgl. Interview I & II) und ein Vertrauensverlust in eine Autorität während einer Krise folgenschwere Auswirkungen auf die Sicherheit der Bevölkerung haben kann, gilt es dieses durch eine umfassendere Einbindung der Öffentlichkeit wieder aufzubauen (vgl. Richardson et al. 2013: 267ff).

# 7. Literaturverzeichnis

# Anhänge

## 8. Originalversionen übersetzter Zitate

*[…] a TEPCO executive who was representing the company at the Prime Minister's Office asked the Site Superintendent on the telephone to stop the seawater injection into Unit 1. That directive was not followed, and seawater injection was not interrupted (IAEA 2015: 39). [S.12]*

*About two hours later, white smoke (or steam) was observed being released from the Unit 2 reactor building near the fifth floor. A radiation dose rate measurement of nearly 12 mSv/h was recorded at the main gate at 09:00 on 15 March, the highest measurement since the beginning of the accident (IAEA 2015: 44). [S.16f]*

*a massive exercise in ticking boxes in documents (Hatamura et al. 2012: 70) [S.21]*

*[…] a radiation dose above the limit specified by a Cabinet Order has been detected (Act on Special Measures Concerning Nuclear Emergency Preparedness; Artikel 10, Paragraph 1) [S.23]*

*TEPCO did not intend to withdraw all of its staff members from the site by any means as can be seen by the undeniable fact that employees remained at the site to control the situation or returned on their own volition (TEPCO 2012: 102) [S.28]*

*March 12 but given no instructions on where to go. The mayor, Tamotsu Baba, chose the tall mountains to the plant's northwest, which he thought would offer sanctuary from the invisible menace of radiation.*

*He found out later that he had led his townspeople right into the heart of the plume, something he believes he would have known had the central government revealed SPEEDI's maps. He says when he saw them later, he was furious because they clearly showed the plume heading toward the mountainous northwest.*

*"We didn't learn about SPEEDI until July," Baba later told me. "The blood ran into my head. What about our lives?"*

*SPEEDI would eventually become a lightning rod for popular outrage at official disinformation and apparent willingness to sacrifice public health to downplay the*

*disaster. This anger crystallized around the story of Namie, a town just north of the plant, whose residents were ordered to evacuate on (vgl. Cleveland et al. 2021: 118). [S.35]*

# 9. Interviewleitfaden I

- Was the risk of severe accidents neglected in the planning of the Fukushima Daini NPP?
- Do you believe that the risk affiliated to the distance of the power stations is justified?
- Did regulations regarding the proximity of NPPs to inhabited areas change since the accident on a global scale?
- How do you see the behavior of plant director Yoshida?
- What is your view on the fact that neither the design of the NPP, nor the emergency manuals considered the event of a long-term station blackout?
- How do you regard the fact that the „off-site center“ had to evacuate from Okuma Town to the town hall of Fukushima? (equipped with the System for Prediction of Environment Emergency Dose Information (SPEEDI) and fallout-monitoring and forecast systems, lacked an air-filtering system)
- Could the evacuation not have been avoided based on the consequences of the TMI and Chernobyl disasters?

## 9.1. Drills

- How do severe accident drills look like in different countries?
- How often are they performed?
- Were you ever involved in the planning or execution of a nuclear accident drill?
- Did the accident affect the requirements for emergency drills?

## 9.2. Regulatory aspects

- To what extent, do you think, is the regulatory system responsible for the course of events?
- Do you believe that the creation of the NRA that meant the reformation of the regulatory system was necessary?
- Which regulatory system for nuclear energy is the most effective from your viewpoint and why?
- What role did the Nuclear and Industrial Safety Agency (NISA) play in the regulatory system and was the NISA charged with doing actual on-site inspections?
- How independent was the Nuclear Safety Commission (NSC) in Japan?
- How much influence could the prime minister exert on it?
- Would you say that the fact that the plant operator was responsible for informing the municipalities surrounding the power plant, was a deficiency in the law? As stipulated by the Japanese law (article 10, paragraph 1), the plant operator is responsible for informing the surrounding municipalities, which didn't sufficiently occur in the case of Fukushima (only 10 percent on the first day).
- How do plant inspections, performed by the IAEA typically look like?
- Does the tendency lie more towards bureaucratic or practical inspections?
- Do the IAEA inspections differ from the ones performed by local authorities?
- How frequently do NPP risk assessments take place and did their procedure or frequency change after the accident?
- Aside from Japan, how much backfitting did occur in NPPs internationally, based on the causes of the Fukushima accident?

## 9.3. TEPCO

- To what extent is TEPCO, along with its organisational structure and management strategy responsible for the catastrophic course of events?
- To what extent would you say that the responsibility lies with TEPCO? (According to the IAEA, TEPCO bears unlimited liability) → "Its liability was unlimited in amount [...]" (IAEA 2015: 169).
- How much could TEPCO and the authorities really demand from staff and emergency response workers (e.g. SDF) in the case of Fukushima?
- How do different countries approach situations where people may have to risk their lives?
- Which approach do you regard as the most sensible?
- What are the legal approaches in different countries?
- How much can plant operators demand from their employees in a nuclear emergency?
- What are the legislative approaches?
- What should be done if legal limits are exceeded?

## 9.4. IAEA

- How important, would you say, was external advice of international institutions, such as the IAEA or the Nuclear Regulatory Commission of the United States (USNRC) in the acute phase of the crisis?
- What was the IAEA's reaction to the evolving crisis in Fukushima?
- What role did the IAEA play during the crisis?
- How and to what extent was the IAEA prepared for dealing with an accident this magnitude?
- Did plans exist regarding internal reorganisations, the activation of task forces or emergency meetings?
- What did they stipulate?

- Were you personally confronted with different tasks during the Fukushima accident?

## 9.5. Crisis management

- Should crisis management in nuclear emergencies be a matter of experts?
- How should a response committee be designed that provides reliability regardless of the people involved, in your view?
- How should it be determined, which information is to be disclosed?
- Do people have a right to know everything in hindsight? (see: Worst-Case scenario developed by the order of the prime Minister Naoto Kan)
- How should stakeholders therefore communicate aspects that are uncertain even for themselves?
- How should stakeholders approach rumors?

## 9.6. Future lessons

- How many technical and regulatory changes did occur in Japan since the accident ten years ago?
- How do they overlap with recommendations issued by e.g. the IAEA, Rebuild Japan Initiative Foundation Investigation Commission, Nuclear Accident Independent Investigation Commission?
- How much did the effectiveness of the regulatory system change within the last ten years, from your point of view?
- Did the behaviour of power companies change since the accident?
- How much did regulators and operators learn on an international scale?
- Do you believe that significant improvements did occur regarding plant safety?

- Do you think that improvements were made with regard to responding to beyond design-basis events?
- To what extent were emergency operating procedures adopted after the Fukushima accident?
- Does operator training now include scenarios like extended station blackouts and loss of monitoring equipment?
- Are operators taught in how to cope with beyond design-basis events?
- From your view, what are the most important (technological, political, or regulatory) changes that should be made based on the experience gained from the Fukushima nuclear accident?
- Would you say that regulatory authorities have learned from the Fukushima Nuclear Accident locally and internationally?

# 10. Interviewleitfaden II

## 10.1. Fragen zur Krisenkommunikation

- Welche Rolle übernehmen Medien in Krisensituationen hierzulande und welche Unterschiede gibt es zu asiatischen Ländern und insbesondere Japan?
- Was sollte man als Stakeholder tun, wenn in einer Krisensituation das Vertrauen in die eigene Institution schwindet oder gar „verloren geht", wie es auch bei TEPCO der Fall war. (interne Widersprüche, vorgegebener und nicht präziser Wortlaut, mehrfacher Wechsel der Pressesprecher:innen)
- Wie sollen Stakeholder mit Aspekten umgehen, über welche auch in eigenen Kreisen Unsicherheit herrscht?
- Gerade Energieversorgungsunternehmen, haben häufig einen Interessenskonflikt von Sicherheit und Wirtschaftlichkeit. Wie kann man erreichen, dass die Sicherheit der Bevölkerung in der Krisenkommunikation den höchsten Stellenwert hat?
- Unter welchen Kriterien sollte ein großes (Energieversorgungs-) Unternehmen wie TEPCO seine Pressesprecher:innen auswählen und was sollte/n die/se Person/en verkörpern?
- Wie kann man diese Personen bestmöglich auf die Kommunikation und den Umgang mit Medien in Krisensituationen vorbereiten?
- Mit welchen Medien erreicht man in Krisensituationen am besten möglichst viele Menschen?
- Ist es sinnvoll, die Kommunikation demografisch differenziert anzugehen, beispielsweise indem man versucht junge Menschen über soziale Medien zu erreichen?
- Welche Verantwortung trägt der Journalismus in Krisensituationen?

- Während des Atomunfalls von Fukushima kommunizierten die Pressesprecher:innen des Kraftwerksbetreibers sehr zurückhaltend und mieden insbesondere Ausdrücke wie „Kernschmelze“, mit der Begründung Panik in der Bevölkerung vermeiden zu wollen. Wie wahrscheinlich sind Paniken tatsächlich und können Sie durch inadäquate Kommunikation begünstigt werden?
- Sollten soziale Medien und Kurznachrichtendienste in Gefahrensituationen als direkter Informationsweg genutzt werden?
- Wie wirkt sich die Regelmäßigkeit der Veröffentlichung von Informationen auf deren Wahrnehmung auf? (Im „Leitfaden zur Information der Öffentlichkeit“ der deutschen Strahlenschutzkommission wird davon gesprochen, alle 2 Stunden eine Pressekonferenz abzuhalten.)
- Wie einflussreich sind Sekundärquellen wie Zeitungen oder Nachrichtensendungen? Erhöhen sie die Gefahr einer Fehlinterpretation von Informationen?
- Wie sehr hat sich die Medienlandschaft seit 2011 verändert und was kann man daraus für die Krisenkommunikation ableiten?

# 11. Abbildungsverzeichnis

## 12. Anmerkungen zu den Interviews

An dieser Stelle möchte ich meinen tiefsten Dank gegenüber meinen Interviewpartnern Oliver Ali und Daniela Ingruber aussprechen, da sie sich trotz eines arbeitsreichen Jahres und schwierigen Pandemiebedingungen dazu bereiterklärten, all meine Fragen detailreich und treffend zu beantworten.

Zur Durchführung der beiden Interviews kann gesagt werden, dass das erste Interview am 5. Dezember 2021 bei einem persönlichen Treffen mit Oliver Ali in Wien stattfand und ich es mit drei Kameras zeitgleich aufzeichnete. Aufgrund der strikten COVID-19 Maßnahmen zum Zeitpunkt der Durchführung am 2. Februar 2022 mussten wir das zweite Interviewgespräch über das Videokonferenztool Zoom führen.

Die beiliegenden Transkripte sind eine wörtliche Wiedergabe des Gesprochen, jedoch wurden Fülllaute und inhaltsleere Versprecher aus Gründen der Lesbarkeit und angesichts der Länge entfernt.

## 13. Transkript I – Interview mit Oliver Ali

Alright, my name is Oliver Ali, I work for IAEA, which is the Atomic Energy Agency in Vienna and as a matter of fact, is a part of the UN-organization. In other words, we are called the watchdog of the nuclear industry. Our job is to enforce the nuclear rules and assure that it is being done as the treaty was written in the 1950s. I came to know Elias from a very young age and I was very happy to see that he is writing a paper on the disaster of Fukushima and lessons learned from it and how we can go forward to improve in a better way. I had read his paper several times. I think it's one of the most outstanding paper given his age of 15 years. To write such a detailed paper and try to document as much as possible, quoting every documentation to a source, which is absolutely brilliant, when come to look at the topic which is very complex and look at the age of a child. So honored to be taking part of it. I will try to answer his questions to the best I can. And of course we are a technical organization, so I keep it most to the technical side. And we are not here to make any judgment on policy or politics.

**So when it comes to the risk of severe accidents, would you say that there were things that were neglected when the Fukushima power plant was designed?**

You have to understand, these power plants are designed in the 60s and 70s and all in full compliance with the regulatory body. To say something is neglected now will be not correct. It has been looked after for years from many different authorities, so it is has it has been in compliance. Of all the safety that it should have, you know, as your question was: “was it neglected in the design?”

Yes, the expansion risk of severe accidents as was the tsunami and the earthquake.

The tsunami and the earthquake are one of those things that still we are learning how to predict. It is called a natural disaster and there is not yet prediction available. We can guess somewhat and the earliest now we have is 3 to 5 seconds early we can detect. So still we are in a very early stage of understanding the law of nature so no one can predict how high, how bad, how severe the tsunami or the earthquake will be. We only know after it happens.

**So were the regulations at the time of the disaster regarding the height of the tsunami within the realm of expectations?**

True, most of the power plants, if we're talking about Fukushima, so we're talking about Japan, are built around the coast, always built near where the water is, that's one kind of a requirement. You have to have supply plenty of waters. So height of the tsunami, I think 15 meters was the prediction at that time of the height of the tsunami, and that's done. There is a paper available.

One can pull it out and read it. It says as a matter of fact is called the "tsunami paper" written by the one of the authors in the US. Many years ago there were talks about how high the wave could be. The thing is, it's very, very difficult to predict the height of a wave because it depends on how the earth crust moves. Either it goes up or it goes down and, based on the movement, the wave is created. So to look back and say was it done correctly? I still think that it's still today sitting now we cannot say how high the next tsunami will be.

**Regarding the Fukushima accident, one aspect that has caused a lot of concerns, the proximity between Fukushima, Daini and Daiichi. Would you say that it is sensible and justified to build 2 power plants so close to one another?**

Generally, power plants are built around the world. If it's more than one, usually they try to build it closer. There is a reason for it. When you build two or three or four or five units, it's much easier and cost effective to build it right close to each other because you can use lot of redundancy from the same and also transfer of fuel and bringing and taking it out and supplying the grid. It all becomes much easier and I say yes, there is some proximity, but most of the world, if you look at it, if there is a cluster, then they are put together also for safety and to protect it. It's easier for security staff to put one, so it's not very uncommon to see the power plants being together.

**You explained why it is sensible to have them close to each other, but were there things that were reconsidered? Because there are not many examples of power plants that are so close. The most common cases are that you have several units in one power plant but not separated, and so did things like rules on where to construct power plants change after the accident?**

The rules where to construct the power plants, it varies from country to country. And it varies from state to state. When I say state to state, I think in a European term we can say, you have a federal law and then you have a state regulator. The federal government may want it one place, the state regulator may say no. Each of the body tries to make the most safe place to build. So all the sites chosen, not only Fukushima or in Japan but also in Germany and Europe. The site study has been done before to make sure that all the basic requirements are in place or in close proximity, before even they think about building something.

**But how are the regulations, let's say in Japan, regarding proximity to residents and did they change?**

The proximity of a nuclear power plant and we are talking about nuclear power plants, differ from country to country. There are countries within Europe that have a 25 or 50 kilometer zone. There are countries in Europe where there is a 5 kilometre zone. So it depends where it is. Japan, we are talking especially 80% of Japan, is mountainous. Not much flat land and where there is flat land, there is agriculture. There are people living in. So the power plant is built, most of them are on the coast. And it just happens that even though there is a perimeter of five or 10 kilometres to be nothing available, but those lands are usually wasted, so agricultures are done on those lands which means even they are within the five kilometer or 15 kilometers own, there are farmers working on it because otherwise the waste of land. There are places in Europe where it's a 25 kilometre zone. But it's not empty. It's a pine forest, so pine is harvested when needed. But people are not leaving.

**One aspect regarding Fukushima that was pointed out in lots of books and literature was that the emergency manuals and the design of the power plant didn't consider long station blackouts that were longer than a few hours. Would you say that was a deficiency?**

Everything looks good from the hindsight. OK, so how long a power blackout should be? Is there a standard now? How long it should be? So a power plant outage is usually last. Few minutes, OK, maybe half an hour max, but there are diesel generators available in the back of every power plant. OK and they are automatic. So what happens? If you have a blackout of more than the requirements automatically, the generator takes over. So, it's not how long the blackout should be because technically there is no blackout. As soon as your power goes out, their generator kicks in. So you are maintaining all the essential equipment, right? The Fukushima case was completely different. If that's what you're looking. Because it said earthquake and tsunami combined at the same time, yeah.

**Now, regarding the equipment failures in Fukushima, what is your standpoint on them? What is your view on the crisis management and the actions of the plant workers and the people involved in dealing with the accident?**

OK, so. In this question you have many small questions inside. So there is a people management you are asking there, you are asking about the failures of the equipment. So let's start piece by piece because it's very difficult to answer a long question. Failures of equipment happen for two reasons. OK, let's put it this way. If there was an earthquake of lower magnitude which means the may wave will not be higher, right? The way it will be under the protective wall. If something tripped, the generator will kick in. But the situation in Fukushima was quite different. This wave was higher than the protective wall built. So the water that came in not only may trip the plant, but it also went into the back where all the diesel generators are. So if you cannot start the diesel generator, you don't have the power. It means that you have a longer blackout. OK, so that was the reasoning why the equipment failed.

OK. Now your other built-in question was, how did the People act or perform? Every power plant in the world doesn't matter if it is a nuclear, it's a diesel, it's a coal fire. They all go through regular training, which means simulated training. Every country has their own schedule, but minimum is one year, once a year. But most of the countries do 2, 3 and 4 depending on how. But they do simulate all kinds of emergencies including people getting hurt and evacuating. They are regularly trained to deal with disasters. They acted as they were trained. But the disaster was of a very different magnitude.

**So would you say that their actions were mostly improvising because as you said, the magnitude was unexpected and even the emergency manuals didn't have advice for the specific things that occur in Fukushima.**

In any disaster, all of these people who were trained. They train if this the manuals are written based on a condition. Then if this happen, this condition happens. You follow page #5, #35 you pick it up there and you follow it. But also the training, it goes beyond. And nobody can guess what goes beyond because? You cannot predict. OK, we can say diesel generator out, but nobody can predict which valve will be open or not. So they do try to do as much as they can. Either predicted or non-predicted. And you have to remember when things are completely gone wrong, human psychology plays the best and the worst at the same time. There are people in the in time of crisis, they performed outstandingly to do the best with what they have available and there are people who are fully trained and just cannot function just because just doing a simulation is one thing, and being in the real one is completely another thing. OK.

**So you're saying that nevertheless, plant operators are also being trained in dealing with beyond design situations.**

They are trained in. Let's put it that way, there is no training available in the world that you can put every kind of scenario in. And if you put every kind of scenario, then you will be training all the time and you don't have it. You create enough based on what it is that you have training power. You train a pilot that if one engine fails you do this way. If 2 engines fail, you land this way, but if the plane starts falling completely apart, nobody but the pilot does the best of his ability given the crisis of the time to that so going back to Fukushima. To the issue. Did the plant people did their best given what they had in hand? I think according to your paper and according to published things, you can see that they have done a really outstanding job knowing that they are in extreme radiation. And yet try to do their best to save as much as possible. Were they successful? No, but they did as much as they could, given they didn't leave the post and ran. They were in the power plant for a very long time.

**But what do these emergency drills consist of? Are they like mostly technical and explain and learn and the workers what to do in what kind of situation as you explained with the pilot and the emergency manual? Or do they also, teach the people on how to, maybe psychologically and also on an interpersonal level how to interact and act with other people and manage the situation on a personal level?**

Many questions again nested into one. Training is done. They're trained technically. What happens if this goes down? Which switch to turn first? Which to do second, which valve to open? Where to dump? These are technicals. OK, what do you do if you open this valve and the water doesn't come because normally nuclear problems usually is that heat builds up too much and one thing is to control the heat, you lower the heat and you need cooling. The cheapest coolant available is water. Training of a technical nature, and that's not only going into the nuclear but also as simple as an ambulance driver, for example. They are trained for psychology because it is a disaster it does affect your thinking. There is a full set of program in every country and every place you go through rigorous training and if it happens in real life, it affects you. You have a whole group of doctors who deal with this psychology. So when you are in it, you are doing it by what we call drilling your training. You will quit, but when it's over, you basically collapse because now the reality hits you. So yes, there is a whole program that they go through to recover backside of it.

**And that program also already existed in the extent of Fukushima?**

That program exists as part of any training program which deals with disaster because one side of the training is to make sure that medical side of it. The other side of it affects people. It affects you as a plant worker. It affects the people living in the village. So you have so-called psychological effect from both sides. You are suffering, you see the suffer of rain of the other people, so it affects you. So yes, there is it. It goes hand in hand, as if there are packets together.

**When we look especially at Japan, did these emergency procedures and drills change and were there things that were adapted after the Fukushima accident?**

With every accident. And Fukushima is not the first one, and I hope it is the last one, but I cannot say that it will be the last one, because as long as human beings are on the face of the earth, we will do mistakes and there are natural disasters. So yes, your question was "were there enough procedures available", yes. Procedures are always available. Trainings are done regularly. OK, so as far as did they follow it? Yes. What can we learn can they improve? Of course after every training, even in the simulation when you do simulated training. After the training is over, you look back to see what went wrong. And from that point you want to say OK for the next simulation. We should improve this area. Similar thing happened in Japan. Lots of things were learned. Not only for only for the Japanese people for the worldwide as a whole, different country looked at it and learned, so there are a lot of lessons to be learned. And a lot of improvement are already made. In Japan and from learning from that, other countries also improve their scenarios. So yes, we still learn because not everything is known. There is a lot of radiation and it doesn't go away in one day. We still have some, let's say cold radiated water holding in the tank, which we are discussing how to get disposal. So, there's a lot of things that we are still learning from it and hopefully improving the next generation.

**But are there specific changes like regarding the frequency of those drills or the Procedures that are being exercised in these fields?**

I can’t tell you because the Fukushima doesn't exist anymore as a plant. But other plants in that particular country and I'm sure in other countries have increased the frequency of training. They have started looking at the training and say: is this enough? Can it be improved? So there if we look at the positive side of it. If there is any positive side of the Fukushima, is, this is the positive side that you really start looking at yourself and say: How can we make it better? And not only that country in most of the countries to start looking and going back and improving. So yes, frequency has increased in almost every place.

**And in general, how often are these emergency procedures performed internationally or, for example in Japan?**

There is no fixed number. It is expected that at least once a year to be done. But it's done several times over. The simulated one is done to assure that every possible scenario they can think of it is covered. Of course there are things that you can only simulate and assume. You literally cannot create a disaster to do it so they try to get as close as possible, as realistic as possible. But reality is completely different. Right? You assume there's a heat and there are people running and doing things chaotic, but when there's really a fire and heat, human being behave completely differently. So yes, they are trained to as realistic as possible.

**And does the IAEA provide guidelines regarding these procedures, or is it completely up to the individual states to formulate these regulations?**

The training is, training is a very, very large topic. When we talk about training, we are talking about what kind of training. If you are talking about plant engineering there is a whole different area that deals with it. OK manufacturing that comes from, let's put example Westinghouse. They are the one of the companies that build the reactor, but there are many companies and training comes from there. So I do not have a role in that. I have another role. This depends between on manufacturer and the plant operator and the country. So IAEA has a safety side of it that looks that, ah, you have the training procedure. You have the training manual, you have been trained and you have a procedure set to practice. We are not the enforcers of this country. It's an independent country. And they can decide for themselves. So yes, there are rules. There is something, but the IAEA cannot inforce. It's about nuclear safeguard, which means misuse does not happen. We are there to make sure that are you doing? Yeah sure, yes you have power plant. You have a manual. You do simulation. You do train and that's why we are not present to judge andake sure are you following exact training. At least my part of job doesn't involve it.

**When it comes to the crisis management in the accident, the management was structured in two in two like centers, there was the offside center in Okuma, which is the city very close proximity to the power plant. And then there was the on-site center directly at the plant, but because of the earthquake and the radiation going out from from the power plant the offset Center had to evacuate to Fukushima quite quickly. I think it was five days after the accident occurred. Would you say that this could have been prevented it considering the radiation coming out of the plant?**

Again, looking from the hindsight. Every power plant has a, let's say, in their disaster recovery plan. That evacuation is part of it. So when things go wrong, based on how much radiation, wind direction, which way the wind is blowing, that's a very important factor. If the wind is blowing toward this, it's not as dangerous in the opposite direction. If it's blowing toward the population, evacuation is needed. How to evacuate people? You have to have just by announcing "please evacuate" doesn't work. How far people can walk? You can get to fifty kilometers, but you have to have buses available and transportable.
Over time, these are protocols which are intermingled with the power plant and the local municipality. These are all internals. This has no responsibility of any foreign entity, including us. We don't enforce, so we know they have a recovery plan. We know that they have a evacuation plan, how they execute it, it's up to them. They were done like you said five days later. At that time, if they told the local people on the ground that this is the time to evacuate. Little reaction.

**But my question focused on the role of the off-site center in the town hall of Okuma, because, as you said, wind is really important and it was equipped with the system for fallout spread and prediction which is called speedy. And it also had other fallout monitoring systems, so it was quite crucial into managing the emergency response, but since it had to evacuate and the workers there had to move to Fukushima, these functions weren't available.**

So there are two functions. One is to operate the plants from off-site and one is to evacuate people. These are two completely separate things. How to operate the plan from a remote location is not available in many places currently. As a matter of fact, there are a few countries, including Japan, who are after learning from Fukushima. Learning from this too. They have already built some and they are building everywhere. A complete remote control site. You would want to do something right according to the procedure, but that thing may not be available to you. Because that thing has been damaged or lost. So what you do, you improvise, you do the best you can, so I think these people did the best they can. And we can step back and reflect and learn. That's what we are doing now. We are learning how to improve and they are improving.

**One limiting factor in the crisis management was that some aspects of these infrastructural parts weren't resilient enough because they, for example, didn't have an air filtering system which eventually led to the fact that that worker needed to evacuate. So were there changes in retrospect, retrospectively, that have been made in order to improve the resilience of of these crisis management infrastructures?**

Again, reflecting that lots of changes, especially I could say I know around the world, has been done, but lots of changes made in Japan. The creation of safety vaults, the creation of NRA and the separation of the management. If you take all of these. There is a lot of improvement. First of all, you know Japan has ordered all plants to shut down. I was there few months after Fukushima. When Tokyo in the city half of the city had light the other half didn’t. Still, there are only 11 reactors running in Japan. Currently, Japan has 50 plus. So each of them is going through full safety review. And once the authority is satisfied with the new changes and new procedure and if there is something to be changed and improved, the plant is not licensed to operate. This is going on currently. So as I guess this year there were only 11 reactors running and as a matter of fact two years ago it was only three or four, so as they go through the safety review and they are fitted with what's needed as a modernization or improvement. They go through a very brief check of safety and once that process is complete the Licensing Board takes over and authorize them to operate. Of course, the local politics is involved, the local code is involved. Even some places the safety has been authorized, but the court has ordered not to operate and that based on people fear. OK so I will leave the people fear and politics out of it. Improvement, yes, a lot almost. So none of the country I know in the world because it hasn't happened this kind of disaster who would shut down completely. And they were running on diesel for many years with half the power if you went to Tokyo. Even two years after the disaster, you won't have air conditioning in the hotel because there was not enough power. So half of the lifts will not operate, trains were running half of them, escalators were shut down so yes, because power you need electric power to live their modern life. Every house has three televisions, 5 computers, all need power to operate and it comes from somewhere and that somewhere is currently the biggest producer is the nuclear energy. It is being licensed as is being approved with the new safety measures, so all of the plants are going through it. Levels are operating and the rest are still going through it and as they become to the current level of safety and procedure, there would be licensed based on the local authority again. It is being inspected and when it's complete they inspect it again. And when they are happy, they give the license to operate. So the operators are basically depending on these licensing people. So yes, they were doing the job at that time. Of course they have much more improved.

**But do you have an insight into this licensing procedures or these inspections because some like workers and former staff complained that there was lots and lots bureaucracy going on that many inspections were less of a technical site, but more of a of paperwork and basically just doing checklists rather than exactly looking at the equipment, so do you, do you know how these inspections were structured and what was done?**

The reason is we are not deeply involved in it. I know the inspections are carried out because it's required. Now the teachers. Because we are never part of this inspection, we are not required to be part of this. We can of course always go if asked as a guest, as a visitor to join. But no, because these are called national authorities, they have to do their own country. And they do it. Where they done right? But let's put it this way, which government in the world that you do know or do not know does not have any bureaucracy involved in it? Yes, there are bureaucrats in Japan. So there are bureaucrats in Europe and US and everywhere else that goes without saying. This you cannot remove that. Even in NISA you have to have this kind of thing so it exists by the law of nature. Where their instructors trained, I don't know because as we don't know, we assume and we respect that if you have agency of checking something, then you should have the right people doing it. This thing that you wrote is that some people say they were not properly trained. I don't know and I don't know who are saying it. Have they looked into their education? Their experience, right? But this is not our part of it we are not involved in the local instruction.

**And does the IAEA also perform inspections of nuclear power plants?**

Yes, that's what we do. We do safeguards of nuclear material. Our main job is to make sure that the plant is not being misused based on nuclear material. That on safety we have a safety section which does the safety, but actually safety falls within the country itself and the regulatory body of that country.

A**nd what do these inspections consist in that the IAEA does? Are they different from the ones that you mentioned that are performed by national authorities?**

Similar and different. Different in a way that the national authority does has the, let's say much more time and much more manpower to do a thorough, in-depth detail based on their own language. IAEA has a very small staff that does it. They do take apart, they look they do look at it, but most of them are as we said, we are not the enforcers. In those countries they are independent. We are not there to tell them that they are supposed to follow a guideline. We are supposed to make sure they are not misusing that particular plant or facility. IAEA is an agency to make sure it's fulfilling the basic treaty. The basic treaty is about misuse of nuclear material, which means, from using for peaceful purposes and not weaponizing it.

**What can you say about the frequency of nuclear power plant nuclear power plant risk assessments and inspections in Japan and in general?**

Two different things, frequency and inspection is completely different because the frequency of instruction is based on type of facility. Some require monthly, some require once a year, some require something in between. So that's the frequency side and the other.

The risk assessment is continuously done and at least, at least once a year for each individual facility. This is again managed by the local authorities. That's why they are there, so they do disaster assessment. They do research assessment and there is a very set proper channels that they submit these and they get approved or sent back for further improvement. They re-inspect and approve until the final license is issued. So when we say licenses issued for that, it doesn't mean the nuclear part licenses issued for each individual parts. Your plan can be functioning, but if you don't have a license for this then you cannot operate. You have to get because there is so many components, so the licensing is very strict which means it tells you that, OK, I have everything is checked up and I'm OK to operate.

**So regarding the frequency. Are there different types of inspections that need to be performed more frequent frequently, and what would they be?**

The overall safety inspection is done regularly at least once or twice for overall. But safety inspection is done more frequently on area by area basis. Some areas every day, some areas a whole set of safety inspection is done before even starting that particular thing so it all depends what the work is involving. And how frequently? Some of the ones I know is done every time before that work is done to make sure everything is in the right place where it should be. Every indicator is in the right place before we even go to, say, turn on the switch. So it all depends, but they are done regularly area by area, but generally for a whole bit once or twice.

**Are there great variations between the countries regarding how profound these inspections are and how often they are performed?**

It differs from country to country. Expectation is the same to maintain the level because rules are different, laws are different, cost is different. And all plays roll into all of this. So yes, in general you can say, one can say yes here's a safety inspection done, and to make sure it's safe. But it is not exactly the same was done in Europe versus was done in Japan or any other country. It's not because I'm saying it is what may be required to do here may be far more required to do in Japan 'cause I know they are far more stricter, so they add many more layers of safety then other country makes it. While these really don't matter, let's do this is most important list with our effort there so they are done based on the need of the facility and based on what that country is. Some are more detailed, more restrictive, others are a bit less, but nevertheless overall they are being done.

**How much backfitting efforts were made regarding nuclear power plants internationally and aside from Japan?**

Backfitting is always done. “Backfitting” what we were talking about, if I understand it correctly, if you find something is wrong, you replace it to improve. That is done regularly. As part of safety inspection new teams are improved, new type of walls is designed and if it can work better it is done so in this respect in Japan, or, let's say other countries, no around the world. Every country’s economy is different. They invest into this based on their financials available. It is very difficult to pinpoint any if a country is doing it and the other is not because they're different.

**But regarding the Fukushima accident, one of the key problems during the response was that, for example, the generators failed, walls were high enough, or some systems were inoperable. And so regarding the things that happened in Fukushima where there are also many technical changes in order with your power plants outside of Japan.**

If you take a Fukushima accident and you said there was no tsunami, OK, the disaster will be completely different simply because those power outage will not happen which means the generator will be working right to provide the power to all the essential equipment. So when designed at that time 15 feet of walls to retain the water was perfectly enough in their calculation at that time during, we are going back some 30 plus years, that everything will operate. So now, we have raised that part of the wall but not every country is right on the sea, but they learn that the generator should be protected far more. Water, and electricity don't mix well together. So they have taken into account and they have improved, so there is a redundancy or there is an improved place or other fitting so they don't simply collapse because there is too much water in there. And the only reason the water was in there because the water breached the wall. The wave was higher itself then the wall could retain it. So building seawalls or making generators waterproof was made in other countries. So yes you do learn from one's failure. The world looks at it and tries to improve it.

**What role and responsibility does the power company TEPCO have with regard to the events that occurred in Fukushima?**

It's a very political question, I would not like to answer it. In a way, I would give you a general idea about it because to ask this kind it's really very hypothetical question. TEPCO is a power distributer, OK, same like in Vienna, the power is generated somewhere. And there are many distributors here, so you can choose in this country. The other countries they don't have that many choices. Of course, some of them own it. Part of it, the power line, which means they pay the cost of building. Some own, all of it, some own part of it, some more pieces of share. There is an engineering team, a scientist team at technical people also in TEPCO that will happen to be one of the suppliers in in Japan. They were a player there's no doubt about it. They were a big player in it because they were involved in building they own part of it, or all of it. I have no idea at this point. But there actually expertise where used. Ah, hard to say. As I said, I cannot answer either way because I was not there and not in one minute. I couldn't see their way how effect 'cause they had to go through the same channel of approval. In of the countries the power distributor owns part of it or all of the power plants. In Japan, it's. Say there's some owned piece of it, so if you own a piece of it, you have a right to say some of the things. TEPCO was involved in it in decision making in giving advice to the ministry on one side and on the other side to the plant operator and the people who were caught inside. So there were communication flowing back and forth.

**Would you say that the relationship between the government and TEPCO was a source of problems?**

The regulatory system and regulatory work are difficult, quite difficult to say from the point of view of a technical side. In hindsight, technically, it should not be any difficult. All parties have one common thing to share which is safety of the plant. Human beings are human beings. There is friendship involved in every place, not only in Japan. There are things involved, so here belief. So caught the technical side of it and stepped into the political side of it. And all kinds of thing goes on the political side of it. Was there any involvement in that that caused the disaster? I would so no, because disaster happened due to natural reasoning. There was a tsunami, there was a earthquake that thing. After the decision making process disaster already happened, so this is basically disaster recovery at this time. Probably there was some element, but I'm not in a place to pass any judgment on any side 'cause that's not our role to do, but that's very well printed. In many newspapers, many magazines, many interviews is all over the place available. And one could decide after reading the story if it's true or not, or which part it is or not.

**In the 2015 report from the IAEA it is stated that the accident should not be solely regarded caused by natural events because there were other factors also involved so would you agree on that?**

The report is their report and those, those are true. Two parts: a national disaster. The next part is human involved in trying to minimize the disaster. Of course many mistakes were made. But then again neither I nor you, nor anybody want itself in that room calling the making the decision. You have to be right in the center when everything is falling apart. When there is a fire and you can feel the heat and yet you have to make a decision right where they make all correct decision, probably not. They did the best at that time to the best of their training and knowledge. So the IAEA is correct. Yes, nature was part of it. Yes, humans were part of it. OK, so natural we cannot correct it because nature happens whenever it happens. The human part, we can correct it. We can train them more, we can simulate more, we can improve it and that part is going on. Not only in Japan, but improvement order. So that is correct. The report is correct and I agree with both sides of the story of points.

**Many reports, including the ones by the IAEA, as you said, state that the actions of the workers were very effective in minimizing the damage. However, it is also stated that the requirements for the power plants were insufficient. So to what extent would you say is the accident caused by the actions shortly after and what, in your view, what role did preparations play?**

The human intervention from the disaster until the few days later, it's all this reason being made by the human being and training deficiency. Of course, you, no matter how much training you you do and you can all never always say that I am 100% train for everything all the time. Right now, with any disaster there are two things happening. One, the human psychology, the shock you go through the initial shock of that disaster, and that as a human being, you feel it. Did something happen that's not supposed to happen? And the training is that how quickly you recover from this shock, so you can perform to start doing things. So this is the training this you cannot remove because human beings are human being. You will go through that shock and then the training takes over and they say OK under these circumstances you're supposed to do you start looking at it and say what's available to me, what's not functioning and what is functioning and what is functioning, what can I do or minimize the damage as much as possible? Hindsight, you can say yes, we could have done more but it's always on the hindsight. The people in there. No light, no functioning, nothing calculation on the wall by hand comes through down to that point. Everything dark. Trying to go through a room by feeling and by just some flesh. Ah, they were trying to minimize the disaster as much as possible to their best. Uh, could they have done more, of course. If you had more people, yes. If you had more of this, yes, but remember inside it's a very small group of people. The control room is not manned by hundreds of people. It's only you know it's a handful of people who are the main and then the restaurant. So yes, it could have been done much better. OK, but again it's on the hindsight. We learn from it. At that time they could have minimized the damage, yes, but everywhere they turn to something is wrong and they're trying to make the best. Of the worst scenario, and it was the worst scenario.

**How much could TEPCO and the authorities demand from from their workers, because they were only contracted workers and had a private contract. So how much in in this crisis situation could they exert force on their workers on things that they really have to do?**

In short, it is not our place. At least my place to even judge them or tell them what this is. The independent contract that is made between one company of a nation to another company. Right? What is written in their contract? How much to do. See there, they are trained to do certain things. If they're trained and you were saying, oh this the company they should say you should go and do this even you  lose your life, but it says on the contract you have to do it at that time. During the disaster the person said, oh sorry I can't do it. OK, right? So this kind of thing, this is on a very local level that we are not involved in at all. This is how they, what's the contract, how much they can ask the local people to do. That's between the contractor and the people and the law of the area. We are not part of it. They have made a documentary movie and out of most of the people who did stay, most of them volunteers. They were asking: This is a life threatening radiation thing and you want to go and turn that wall off in that place? And who wants to go? And you see some people raised there so yes, there are lots of heroic moment like that. Only I can also see from the hindsight and say that we are not involved in those contracts at all.

**But are these contracts in general very different to usual working contracts because of the possible dangers that that one might have to face or are they mostly similar to regular ones.**

Very difficult, at least for me to give a detail, because I don't know the detail of those contracts. What I know in general about the contracts, I can tell you. Any  person working into a nuclear facility regardless of the work, they are going through very simple, similar brief, they are told this is radiation related. It affects your life, it shortens your life. It hurts you. It's your eyes. It's your body, it’s your DNA. All of them are explained to them so it is not that they are not told, so that's part of hiring. Let's say most of the countries it is part of the thing that when you sign up. It's explained to you and you willingly agree that OK, I will do it. It's the same. I always give you example that the pilot knows: Every time I fly, chances are I may lose my life. But yet the person chooses to fly right? So dangerous, all of them are explained as part of the contract. So I am 100% sure that those things were explained to the people who are part of the contractor. When you work in this kind of industry, these are the things you should know.

**The IAEA gives recommendations regarding dose limits for humans. Or how different radiation doses it could affect health, however radiation limits in countries are different to the suggestions of the IAEA.**

Very good question. This is the international standard. There is an international body which has an international standard of by body organ type. Which means how much radiation is too much for the eyes? How much is too much for the skin? How much? That's one layer, the second layer is the minimum. The country has to follow that international standard. Some countries are far more strict, so if they say for example the 10 is the limit. That country may choose that 8 is the limit because we like our people and we like to keep them safe. We will change them more frequently. We will have a shorter working period so this will help. IAEA has even much more. It is international level monitored internationally. And even in Austria there are international levels. This guideline to follow and then it's set by the locals. So the locals have to follow the minimum standard level or improve. Also most of the places usually have improved upon it. So yes, there is international body that monitors. And this involves radiation experts, medical experts and there are many brochures, charts and books available which can tell you exactly limits by organ type how to reduce it, how you can protect yourself. All of them are available. Open market

**What institution is it that prescribes these limits?**

Yeah, OK. The idea is behind any radiation protection as low as possible. Right? That's simple ALARA as low as possible now. You know this is a radioactive thing that you have to work. See that and you have to work on it. So how do you reduce the dose minutes? So, there are many things to reduce right? You put the shielding. One thing you have a protective shielding on yourself, you protect yourself. That's one way of looking at it or you find another way of distancing because the further you move away, the less it is. Or you use the time. So, these are all calculated. If this thing will, let's say for example, will take one hour to take care of it. But it's a high dose. You you can send very safely one person to work as fast as possible for 15 minutes and the person has to come out. And the next person goes. So, there are ways that you have to work to limit it by properly managing either people, distance or shielding. The idea behind all of this is as low as possible. That is internationally known.

**But you were saying that. There is a maximum limit that all countries need to follow.**

They there is a standard limit that all countries have to follow. And others have taken them and improved upon. Basically, they have lowered their own limit. So, let's see what happens if you get more than this standard. You don't work anymore. So for example, if there is a yearly dose of, let's say, I pick a simple number for injury 5. And the first three months you have achieved by measuring the dose and all this stuff that you have received 5. When you have nine more more months of work, you simply cannot. They will just put you in the office to sit. So, these are the ways of controlling, not overdosing the people and letting me so. This is why people can work safely all their life in a nuclear industry and not have the problem because it was properly managed.

**But is there a sort of treaty that requires the countries to follow these limits?**

There is a standard set. Like many things, there are international standards that you have to file. There's no enforcer out there or policing out there that if you don't do, I put you in the jail or you we find there's not most of the treaty and other things are done by honor. You sign a treaty that you will follow. You are expected to follow it and most do. Because it relates to the safety of their own people, so most do.

I think you were doing great. You have done far more than any of the kids I know in your age group. So, by being in 15 and buying doing this, it's an outstanding job.

Thank you.

I can tell you your teacher, your professor, so yes.

**And during the crisis in Fukushima, how important would you say was externally advised by international institutions such as the IAEA?**

Great question, the international advice was coming from all over the world. It's from everyone. The IAEA had a team of experts already waiting here on this side. We stand by. So, as I know other countries stand by you call upon when you need it. What kind of advice do you need. The knowledge was available to the Government of Japan saying we are here for your support, should you need anything. Not from only, technical advice, but also equipment and other things. We are ready to provide, and that's how you manage because you don't want to go and start interfering in other management. Standby and say, we are here. Just tell us what you need when you need it. We will provide and so it was available.

**And how exactly was the IAEA involved in the days when the accident was the most severe?**

Involved in a way we have people in Japan. We have people here and as I say, it is meant 24 hours. There was daily briefing going on in the IAEA. Including ambassadors of every country or they were invited to come. And IAEA was providing all kinds of information made available that time to all the governments during the situation because many country have lots of questions which way the wind is blowing. What will happen to the water where the fishery is going to go and there was everyday briefing that continues. At least, I don't remember now but at least couple of days of before it became a weekly and then monthly, and then on demand. So the IAEA was doing its role and providing information and getting information at the same time from the Japan government to us. So yes, it was both very information was flowing and also earlier rules to provide information to larger community around the world, including Austria.

**And were there any plans or procedures within the IAEA on what to do? What reorganizations to do in case there is a severe accident going on?**

Quite difficult question in the way what you ask what I said to you in the first question I answered you is that is a very standard procedure. In a disaster in the country where it happens because there's a set of procedures already set to take care of it. You let the local people who know the place the best resolve and do you stand by and when you are called upon you go and IAEA went after many weeks after their OK and other countries were all involved. There were lots of footage coming from one of the country where it was easier for them. Their drones were available to fly over and give them more light information. So yes. Not only IAEA but there were other countries involved in offering. So the government of Japan at that time knew if they need something it's available and they can always call upon.

**Were you personally involved with doing other tasks than usual during the accident phase in Fukushima?**

No, I was involved in our assignments that we have to do regularly. There is an emergency team that's made and they're involved and we provided any information depending what's needed to this team. The team can draw information based on what they need from each area of the IAEA. So, not me directly involved into the crisis time but I was there ater the crisis. I had been to Fukushima several times after that.

**Is this emergency team always available?**

24 hours. Always available.

**And they are also prepared for different types of emergencies?**

We don't know what kind of emergency comes from where. When it does come, the team draws upon what's needed. They have a pool of list of everything available to them if this happened then we call upon this kind of people service.

**Is it specific for nuclear power plant accidents or do they also? Are they also prepared for other types of accidents?**

Oh mostly nuclear related because that's our job.

**But is it mostly for our power plants or also for other nuclear facilities?**

All nuclear facilities, which we call all nuclear cycles which basically means from mining from the ground to almost the end of the cycle, reprocessing and burial so the full cycle we are involved in in this area, yes.

**And out of how many people does this team consist?**

It's very small and very big depending on duty. 24 hours is a small team. You know it's a shift work, then it is small team and when I say it's a very large team they have the backing of the full IAEA plus they can draw people out of request other countries to help. So it all depends on the nature, you can't keep 200 people sitting in the room waiting for a disaster to happen. You have few people who monitor it and they follow up on when it's needed. A very central aspect in the disaster of Fukushima is crisis management. **Would you say that crisis management in emergencies should be a matter of experts?**

It depends of course. Crisis management is not only used in nuclear industry, but almost everywhere. Every company, everything related to disaster recovery, uses crisis management depending on the type of work or type of nature of business. You have to have a certain group of people who know the business to help you in the management by default. Yes, you would need people who are expert or who are educated in that area because who do you ask to help you for management of that crisis? Somebody who knows how to manage.

**How should a crisis management team or a emergency response committee be designed that is reliable and independent?**

The management team is designed to have independent opinion and expert opinion and not get swayed by one or the other group. So, when people are selected to be part of this team. They are selected because they have a certain skill of expertise or several skill that's needed. So you do a so-called non biased selection, which means you try to get the best people in the place. That's the idea. Does it happen all the time in real life? Probably not. But the intention is there to have it. Remember: we are human beings. We are not perfect so we do tend to be slightly biased one way or the other way, but that's not the intention. It is always to put the best people in the place who can provide the best invoice. And such is done.

**Nevertheless, in the case of Japan, the Prime Minister always was sort of the highest person of in the command line. Do you think that it is sensible to have these emergency situations managed by the leading politicians?**

Every country designed in their own way. What I think has not much meaning to it because Austria will decide his way. So will Japan. So will the rest. I think it should be in the group of people who know the subject matter. So it really depends: country by country. In generic terms, you would want the best people. If the president happened to be the best expert on the subject matter, why not? So it's very difficult to say because it involves independent countries and their way of living and how they function. We cannot force them to do this way or that way, because that's the only right way.

**Especially in nuclear emergencies, it is the most important to secure people from the dangers of radiation. So how should people involved in crisis management determine on what information during the phase of crisis should be disclosed and what information should be kept secret? Or what information is absolutely important in the in the situation of danger and what is might even be counterproductive?**

Yes, it depends on country, by country again and also the knowledge of the people living around the facility. How well they are educated because there is a negative side or information. People do not know how to react and what to do so you trust on your team on the people of experts who know at what time with information to give and how to move people out if needed. Because if you provide too much information at the wrong time when it's not needed, you will create another crisis of chaos where there is no control and you lose everything. So how much is needed and where it's needed depends on the culture of the people. Educational people and the country so very different. How a culture of Japan will react? Austria or the US are very different. The rules which apply there may not work at all here.

**However, especially in the field of fire protection, it is known that you should only provide people with instructions on what to do in an emergency like leaving things behind, getting outside as as fast as possible, so direct instructions and no additional information or explanation. Would you say that this principle also applies to radiation protection?**

This principle applies to many other things. You cannot instruct only people and expect people to do so. For example, very simple thing: I tell you a comparative is there. So there is already because tsunami is a common thing in Japan. People are trained from right from the kindergarden upward. So when this siren goes on. Which way to walk? Where to find safety? It's all marked on the road so people know how to follow and how to get to a safer place, right? They have volunteers available in every village and people that who are there to direct people to go this way. So these are all internal thing how you train the local people. Just giving information. It's not enough. If you are giving information you have to have people in place to direct them to that place. You have to have the traffic signal even working like green all the way from this side and red all the way from coming. All the simple things add up to the management because people by nature would stop when the serial lights and one start even for a few seconds, created that model. So it's not only how you give information, but do you have people in place who will see that it follows through. So training is needed both side and and information provided needed. What, when and how much? What was it? Also the case that people gave instructions on how to evacuate and what to do then or. Was this situation to all overwhelming? The there is a very set procedure of tsunami which is drilled pretty much regularly in Japan. Because this is a common thing happens there so people are trained for that they move. The part of the nuclear disaster was the added one so as far as people can follow, they did, but the disaster was such that the roads, they are supposed to take to escape were gone, the bridges were gone, there was water everywhere. It added to the complexity of it. It's not that many countries don't have this because they don't have tsunami or typhoon that all the time. People in Texas and other places know when Tornado came what to do. They are trained for that. That was helpful to this country because lot of life got saved because they were trained for tsunami.

**It was very unique about the disaster in Fukushima, as you explained that it was three events at once. An earthquake, a tsunami, and the nuclear power plant blackout. Would you say that now there are procedures on what on what to do if nuclear accidents occur simultaneously to nuclear disasters?**

They are procedures available and the only thing none of us know is the kind of disaster or how much you can have a total collapse and none of the procedure works because the infrastructure is not there to carry it out even. So yes, they are there. They have been improved. I'm sure there will be people who will be trained more now. Now, everything is based on a certain limit. When you exceed that limit and when the infrastructure collapses, then it's a different story, and that's what happened. In conclusion, 3 accident at one time. You were told to escape on this road to that place, and if the road doesn't exist anymore, it's all water.

**And now one part of crisis management is giving people the information that they need in the acute phase of an emergency. But once the danger is over and people are again in a safe place, would you say that people have to right to know all the information surrounding the accident and events, because regarding Fukushima there were many things that weren't disclosed and for example there was a worst-case scenario written by the government which is kept secret, and of course it is not sensible to publish these types of in the face of danger because they might have a negative effect. However, do people have the right to find out these things afterwards?**

The right of the people is always there to ask for information and goes and should be made available to people when needed during the disaster phase. It is kind of unimportant you, what is your first response? Safe people life. Everything is secondary until you cross the barrier and you made them to the safe place. What is the next one? Make sure that people have food and water. What is the next one? They have a place to sleep. The time to provide that added information, what you want that should be made available depends on the country and depends on the people and the law of the country, when it should be made available so people can read, understand and have all information, yes. So time needed for disaster recovery and their information yes.

**A thing that is inherent with nuclear problems is that it is very hard to estimate how a situation will develop and what might happen. So, when coordinating an emergency response. Many uncertainties might be there and might require different actions. Should the stakeholders and people involved also address uncertainties to the public?**

Uncertain, by definition means uncertainties which mean you can only prepare for so much if you knew. Everything there is, is not an uncertainty anymore. It's a known fact. This is problem with not only nuclear but all other major accidents. There is a calculated prediction that this happens. When that happens, this could have these consequences. And then there is a beyond. And when that beyond happens you have to depend on your people on your expert who is on the ground to call on what they need and how they want to recover. You cannot document everything in one place. It is impossible to think about all scenarios you can do, and nature has its own way of dealing with it. When it comes, it goes completely differently. So as seen when we always improve, we learn so. Yes, it should be documented as much as possible and stakeholders should prepare for it. For the known ones. The unknown is unknown.

**So the crisis management was structured in two centers and one was the offside center that was located in Okuma and it had the function of monitoring the spread of the radiation and had a crucial role in planning evacuations because there everything was monitored. But because of the spread of radiation from the power plant that was very close, it had to be evacuated, which meant that these functions weren't available anymore. Would you say that this is that this was a big deficiency in the management of the evacuations?**

I would not say it for any country that there were deficiencies? What I would say too, because each country does based on what's required by their local law and authority. So coming to this question was there any deficiency you have to think two different things? No one can predict how bad the disaster will be. How high the radiation will be, so everything is based on a certain number. No one can predict the wind direction of that time. So all of these are assumptions. And especially when you combine this: One is a nuclear disaster, the second one is a tsunami disaster. We are completely forgetting in this interview the second disaster. Tsunami has its own effects in general with people in Japan because they are trained how to evacuate, where to go, what to do. And the disaster of that tsunami is available in every television footage you can watch. How much water came? How many villages were lost and so on. They were dealing with two major crises at the same time. One is the actual tsunami crisis. And one is the nuclear power plant going from functioning to non-functioning to a disastrous state. They had this off-site center where they thought they would be able to handle it. Yes, but at certain point human life becomes more important than anything. When you know the radiation is going high and you have to evacuate the people, otherwise their life is in danger you regulate people first. And you say next, wherever we have the safety we can manage from there. So that's exactly what happened when it came to a certain point. They have to make a decision.

Human life is more important. If you keep them longer, they will have a long term effect.

**But when we reflect the Three Mile Island accident we also had fallouts that were even greater than the one in Fukushima so these aspects of radiation in the proximity of the plant could have been, you know, anticipated.**

You got Three Mile Island and Chernobyl. OK, there is some common factor that something went wrong. Higher radiation, but a completely different disaster based on different circumstances. So as far as we learn a lot from Chernobyl, I happened to be part of Chernobyl-team and was in Chernobyl so I have seen both side of it. I've seen Fukushima accident. The evacuation there was different, but also we did not knew there was a accident in general. Several days later the Russian government did not inform. It was the Finnish who picked up on their monitor that there is something wrong with the radiation is high, and they are the first one to alarm the Europe. So yes, what we know from that side. Completely different. There was an evacuation. There was no tsunami there. So completely different evacuation and yes, there were local authorities and the buses were available to take the people out. So there was a different set of scenarios here. Your whole infrastructure has collapsed. The road was underwater, everything was, power line was out so a completely different set of scenarios. So there is a common factor, but that here even you want to evacuate people, but you cannot because the bridge is gone, the road is gone, the water is there, but there is no power available in the village anymore so they have to deal with whatever they had at the time to the best. Yes, reflecting back, we can learn a lot and we are learning a lot and I think that's the good side of it that. If it happens, I hope it doesn't happen then we are well prepared. If it happens, will be completely different.

**But the thing is, the off-site center even considered blackouts and had redundancy for that so it was operable even though there was an energy blackout for days. But the crucial point for the evacuation was only the radiation. Even though the systems were all operable, the people couldn't stay there because of the radiation.**

Yes. It is a silent enemy. You cannot see it. You cannot feel it. You can be standing in a very high radiation area without knowing. So at some point someone had to decide that this dose limit is going out of the normal and we have to do something about it. You have to also know human behavior in the time of crisis. We can always look back and say guys there was a road right next to you. Why didn't you take it? Crisis is completely different. People think differently. You are trained to think correctly and follow, but if everything is collapsing around you, especially in this case, it was not just only a nuclear disaster. It was combined with the tsunami and all the damages that came with that. So at that time, if they decided that it is time to move, right or wrong, at this time. I cannot judge if they made the right decision or the wrong decision, because it is not fair to judge it afterward because you were not in the crisis. If you are in the crisis, your behavior will change.

**Now, aside from the technical aspects in the disaster one key pillar of nuclear energy in Japan is the regulatory system, and there were many changes that have been done since the accident, but how far would you say is the regulatory system responsible for the course of events that occurred?**

I would say at this time again we cannot go back and recreate so the regular system currently slightly changed and is different and modified and modernized and improved. And that's all coming out of the lesson learned. To go back and say that the regulatory system of the previous one was not existing or that's the reason it caused the disaster. No, that would be wrong to say for any place or any country as I have no alliance with any of them. The regulatory system was there. The training was there, simulation was there. If the current regulatory system would be in place and we go back in time and say it will be different? I don't know because I have no idea what kind of next tsunami there would be now. The current safety standard is a 25 meters high seawall from 15 to 25 meters high. That's a long one for any wave to breach. But who is there to say that the next wave will be less than 25 and it will not be more? We don't know. That's a human calculation. Estimation that based on this kind of 9.0 under the sea kind of earthquake would not create more than 25 meters or so, the current every seawall is being has been raised and being raised now. So, to go back and say that the 15 meter wall was not enough. No, at that time the regulatory body thought that's more than enough for given all the scenarios. So we learn by our mistakes and we learn how we can improve so given that scenario. The previous regulatory body did what they thought was the best. And the plan worked for more than 40 years, safely, that's the positive aspect on their side. The new one is definitely far more rigorous, far more improved, far more modernized and I hope it will perform much better. If there is anything, I hope there is none.

**After the accident, the Nuclear Regulatory Authority was created in Japan as a sort of umbrella organization that had more rights and was independent and should take over all the functions of previous authorities like the Nuclear Safety Commission or the NISA (Nuclear and Industrial Safety Agency). Would you say that the changes made are positive and what kind of changes are the most notable?**

The one that you see after the disaster. How much improvement has been done? Creation of this nuclear resolution? An independent one and trying to put many of the same element under one umbrella, so the flow of communication is much more clear and direct looking at that to now. It's a big improvement. It is a structure that comparatively, if you compare this one or the other one where the managements were interlaced and different people have different authority, yes. It is comparatively far better and improved and time will tell how. One thing to remember: You can write the best book on the best procedure of a given point in time. But time ticks away. Would that procedure with that management with that training, the same training will be good in the next five years because things have moved, things that will improve, so you have to retrain, reprogram, rewrite. All of the things is are continuous improvement and that is needed everywhere because you cannot just write one safety procedure and say follow it because it's good for this year. Maybe the next five years, but definitely not the entire time because things have improved and moved on. So you must keep continuously doing it, so in comparison, yes, a lot of improvement.

**And when we look at the structure of the regulatory system prior to the accident one major institution was NISA. The Nuclear and Industrial Safety Agency. What was its specific responsibility? For example, did they actually perform site expectations, site inspections?**

Yes, they do it regularly and most almost every country has it. It's not that they have to do it because the operator of the plant cannot do anything unless they have it. Authority or we call it licensing from the licensing board and this body is called by a different name in different countries. So here it’s “nuclear inspection safety authority”, if you want even as simple as this. If you want to do a change, a simple change. The operator has to apply to this body: A licensing firm. Basically it asks: I want to do this. They send a team of people to check it. Is it first of all needed? Second of all is it done the right way. And then it's approved.

**After the accident of Fukushima, how many legislative phases changed in Japan? For example, regarding the responsibilities of operators and of the government, because one very controversial aspect is who should be responsible in informing the people about the accident. Before the**

**accident it used to be the way that actually the plant operator was the one who should, by the law, inform the municipalities.**

Lots of changes. Lots of legal laws change to form this NRA and Safety Commission and rules and laws so again in internal rules and internal legal laws we do not get involved. Each province has their own. So for example. One I can tell you about one thing: It’s the province’s rules. That anything happens in that particular facility. For example, somebody gets a hand broken a simple access must be reported to this local authority and to the news media, so anything that happens in this place is reported to the local authority. So that's the record. So each prefecture or each province is different based on their needs, but in general, yes. Lots of changes. And lots of improvement. On the hierarchy again, the local authority is different, simply by nature. If you want to build a power plant. Let's say in the 22nd District, the central government agrees and I would say yes: It's OK, go ahead and build it over there. But the magistrate of this area will say: If you built it, you have to fulfil this, and even then, the people of this area may say: No. You want it, we don't want. So there are many layers, so lots of local changes are based on local need. Some people, some areas are very highly populated. It's a very different opinion. Some areas are very deserted so less opinions may be easy to get things passed through. But overall, lots of changes and in a good way.

**Could you give an example of a legislative Change that was very important after the accident?**

Not in particular. I can give you a general view of it, but not in particular because we are not involved in the legislative side of it. That's the Parliament and the local people who decide on the outcome that is several new agencies. Several more direct, safety rules and regulation is that we see from being outside that lots of improvements are being done. The creation of a new agency that looks over everything like a big umbrella. Line of communication and how they should move, so these are so-called external views from where we are. Of course these are passed and done through the local assembly and congressmen and their local things. Now, we are completely not part of it at all. It's not our place that's the choice of the people who live there and decide how they want it.

**What interactions are there between the IAEA and local nuclear safety authorities, especially in Japan?**

We realistically do not deal with locals because we only deal with states. So every country, including Japan, has an agency office that is directly responsible to talk to us, so we call them the state authorities and we are here. So we have a communication line with them. Then they are responsible to follow through the locals, so we are not involved in local aspects at all as far as I know in our work.

**What can you say about the frequency of nuclear power plant nuclear power plant risk assessments and inspections in Japan and in general?**

Two different things, frequency and inspection is completely different because the frequency of instruction is based on type of facility, some requires monthly. Sample requires once here. Some requires something in between. So that's the frequency side. And the other one was so It was the question. The risk assessment is continuously done and at least. At least. Once a year for each individual facility. This is again managed by the local authorities. That's why they are there, so they do disaster assessment. They do research assessment and there is a very set proper channels that they submit these and they get approved or sent back. For further improvement and reinspect and recently and approved until the final license. OK, so when we say licenses issued for that, it doesn't mean the nuclear part licenses issued for each individual parts. Your plan can be functioning, but if you don't have a license for this, but you cannot, you have to get because there is so many components, so the licensing is very strict which means. It tells you that, OK, I have everything is checked up and I am OK to operate.

**So regarding the frequency are there different types of inspections that need to be performed more frequent frequently, and what? Would they be?**

OK so I am assuming based on your question you were talking about safety inspection. You're not talking about inspection. The state line inspection is completely different. Then the safety, the overall safety inspection is done. Regularly at least once or twice for overall. But safety inspection is done more frequently on area by area basis. Some area every day, some area a whole set of safety inspection is done before even starting that particular thing. So it all depends. What the work is involved? And how frequently some of the one I know is done every time before that work is done to make sure everything is in the right place where it should be. Every indicator is in the right place before we even go to, say, turn on the switch. So it all depends, but they are done regularly. On an. Area by area, but for generally for a whole bit once or twice.

**But when it comes to things like for example in Fukushima, it was for a long time uncertain if there was a core meltdown or if one is likely to happen. So should communicators address the possibility of a meltdown in front of the public or is it counterproductive to do?**

How educated are your people? Because saying something very technical to a layman: What does it mean? Not much. Right? So if your population is trained, they understand it when you say it. In many countries a nuclear term is taken as a bad thing, including this one. We don't train our people. We don't tell them what this is. So the more they know, the better it. People should be able to understand, so either you make the information simple, so you don't tell them it's a meltdown. Or you tell them something that they understand, otherwise it will be meaningless.

**So when it comes to crisis communication, it is known that the upcoming of rumors should be avoided as much as possible, however often times it is inevitable. So how should, especially in a nuclear emergency, stakeholders approach rumors and uncertainties within the public?**

Rumors cannot be controlled. Rumors are by definition rumors. People talk about things without knowing much information and that's why rumors are there. How do you control the situation? Either you can provide too much information, but then you create another problem. So these are part of the game. There will be rumors. Human lives are needed to be saved first. At that point, a rumor doesn't matter. If that is the basic thing, you do that. The next step: Give them medical treatment. Give them food. Keep them watered. Rumors will be there. When we look at the Fukushima accident in hindsight plant operators and regulators learn internationally from the accident. This is a case study that will be studied for years to come. We are still studying Chernobyl and we are learning a lot from it and there are long-term effects. There are short terms. There are medium terms. A lot is has been learned, a lot will be learned and this is case history will be studied by many young generations like yours, who will come and use this an example and learn. It provides immense information for the future. How to prevent it how to do a better job? How to manage the crisis? What can be learned from it? And it’s not just for now.

**And what safety improvement or change that was implemented would you say is the most notable?**

There is not a single piece. There has been lots of improvement. Lots of enhancements. Lots of rigorous training. I wouldn’t pinpoint one, because all small parts comes together to work as a one big mechanism. Overall, there has been a lot to be learned. And it is still being studied. There are many things are still being studied all around the world. Lots of universities are involved in case studies and all kinds of research is done. So yes, not only TEPCO but many, many companies around the world have learned a lot of lessons from it. As far as other people's behavior goes, if they have changed, I have no idea. I hope it has changed for better and much more because I do not know if internally something has changed. How much would you say that Fukushima has influenced the public interest or public view on nuclear energy, and what could this mean for the future of nuclear energy? It has influenced public as a lot all around the world, both in a good and a bad way. Bad really means people don't want it. That’s one side of it. The second side of it is that we need power. We have to find alternative solutions and until that alternative solution is found this is the source we have. You can use wind and all the available, but still the consumption is too high in the world. It has helped people to understand both sides of it, and I think it will continue to help people understand for a long time because, as I said, the case studies are still going on. Lots of researchers still going on so. It has helped both sides of it.

**Thank you very much.**

You're welcome, you're welcome. Thank you for the longest interview of my life, but I can tell to your people and your audience and your teacher that it is one of the best paper written from a 15-years-old on a very complicated topic.

# 14. Transkript II – Daniela Ingruber

**Könntest du dich eventuell kurz vorstellen, damit meine Betreuerin auch weiß, mit wem ich spreche?**

Okay, ich heiße Daniela Ingruber, ich arbeite an der Donau Universität Krems im Research Lab for Democracy and Society and Transition und hab früher im Parlament gearbeitet. Als parlamentarische Mitarbeiterin war ich an einer Universität von der UNO der "UPeace" in Costa Rica und arbeite an verschiedenen Universitäten im In- und Ausland und gleichzeitig aber auch für Filmfestivals und berate auch manchmal in Bezug auf Politik.
Dankeschön.

**Die erste Frage, die ich stellen wollte, ist, weil es in meiner Arbeit auch um Kommunikation in kerntechnischen Notsituationen geht wollte ich fragen, wie man sozusagen am besten möglichst viele Menschen erreicht in Situationen, in denen eine akute Gefahr ansteht und wie man sozusagen, worauf man bei der Kommunikation achten muss, um der Sicherheit die höchste Priorität zu geben.**

Also das Erste ist natürlich einmal das Technische, aber ich glaube, das hast du schon besprochen. Inzwischen in Japan wird, wird ganz, ganz viel übers Handy gemacht. Da gibt es eigene Programme, die in speziell in Ländern wo es zu Vulkanausbrüchen oder Erdbeben kommt, gibt es solche Apps auch, wo man rechtzeitig informiert wird. Das ist ja mal das eine und damit erreicht man glaube ich inzwischen am meisten Menschen, weil sich immer mehr Menschen auch daran gewöhnen, so erreicht zu werden. Das andere ist, dass gerade in einer Krisensituation einerseits natürlich wichtig ist zu beruhigen, aber andererseits auch möglichst die Wahrheit zu sagen. Man sieht das auch bei der Covid-19 Pandemie ganz gut, dass diejenigen, die die Wahrheit gesagt haben, auch wenn es nicht immer schön war, sich am leichtesten getan haben und auch jetzt nachträglich am glaubwürdigsten dastehen.

Bei Japan hat man damals gesehen, dass es einfach nicht erlaubt war, die Regierung zu kritisieren und regierungsnahe Unternehmen zu kritisieren. Und die Journalisten, die es getan haben, sind sehr schnell verurteilt worden und haben Ihre Jobs verloren.

Jetzt nachträglich, dass es aufgearbeitet wird und eigentlich jene Journalisten besser dastehen, die die Wahrheit auch versucht haben, herauszufinden.

Das ist natürlich auch kulturell so eine Sache. Wo kann man die Wahrheit sagen und wo nicht? Und das ist einmal das Eine.

Eine Sache mancher Gesellschaften ist auch eine Hierarchie in der Demokratie, auch etwas Kulturspezifisches, dass man bei uns leichter mit der Wahrheit rausrücken kann als in einer anderen.

In asiatischen Ländern besonders,  allerdings bei uns nicht so leicht wie in den USA, wo man es viel leichter aussprechen kann, einfach, weil die Menschen das so gewohnt sind.

In einer Krisensituation ist es aber ganz wichtig, dass man bei der Wahrheit bleibt, weil dann tut man sich auch leichter, weil eine Lüge braucht dann die nächste Lüge und irgendwann kommt man aus dem raus. Jetzt ist es natürlich oft so, bei der Dynamik einer Notsituation, dass man auch oft nicht genau weiß, wie sich eine Situation entwickelt und dass sehr viele Unsicherheiten gibt.

**Wie geht man mit denen am besten um, mit wenn man mit Leuten kommuniziert, und was soll man sozusagen, welche Aspekte soll man, wenn es jetzt wirklich darum geht, Menschen in Sicherheit zu bringen, soll man dann auch wirklich alle Unsicherheiten ansprechen im Sinne der Wahrheit oder ist es dann sinnvoll, zuerst direkte Anweisungen zu geben und dann erst im Nachhinein sozusagen über diese Dinge zu sprechen?**

Es ist, da gibt es natürlich auch unterschiedliche Stufen. Bleiben wir einfach bei einem Atomunfall. Ja, wenn wir wissen, dass der Atomunfall so schlimm ist, dass die meisten Menschen sterben werden, dann werden wir es nicht sagen, also wir kennen das auch aus Hollywood Filmen.

Wenn alle zugleich ins Auto steigen und die Stadt verlassen, dann gibt es einen Mega-Stau. Das ist natürlich nicht hilfreich, das heißt da sagt man natürlich nicht unbedingt den Grad der Gefahr, aber man kann die Leute natürlich auch nicht im Glauben lassen, dass das alles in Ordnung sei. Das heißt, da gibt es einfach verschiedene Abstufungen. Jetzt habe ich leider vergessen, dass ich noch den ersten Teil der Frage da wollte ich noch was dazu sagen.

Die Frage war sozusagen, wie genau man über die Situation berichten muss und ob man auch auf Unsicherheiten eingehen soll und die eigenen Unsicherheiten sagen soll.

Da gibt es auch in der Wissenschaft oder auch eben der Kommunikationstheorie sowohl auch in als auch in der Praxis zwei unterschiedliche Wege. Die einen sagen, man sollte nicht einfach zugeben, dass man es nicht weiß, sondern sollte immer so tun, als wäre man Herr der Situation, weil man sonst eine gewisse Schwäche zeigt und von der Bevölkerung nicht mehr ernst genommen wird.

Ich glaube, dass gerade die Pandemie genau das Gegenteil gezeigt hat, dass es wesentlich besser ist zu sagen "wir wissen es einfach nicht genau, aber wie bemühen uns". Damit ist zum Beispiel der ehemalige Gesundheitsminister Anschober wahnsinnig gut gefahren, was er eben gesagt hat, war "wir wissen es nicht ganz genau, aber ich tue alles, was ich kann, und meine Leute tun alles, was ich kann", das geht eine gewisse Zeit lang, wenn man das jetzt besonders lang spielt, funktioniert es auch nicht mehr, denn irgendwann muss aus der Inkompetenz Kompetenz werden, sonst braucht man Berater und da sind wir auch dann gleich beim nächsten Thema.

**Wen lässt man sprechen? Also nicht nur wen lassen die Medien zu Wort kommen, sondern auch als Regierung oder als verantwortliches**

**Unternehmen lasse ich den Präsidenten sprechen, lasse ich den Pressesprecher sprechen?**

Das sind zwei unterschiedliche Dinge. Ja, weil der Präsident der Firma oder der der höchste Manager wird dann anderes Gewicht haben als eine Pressesprecherin. Oder hole ich mir überhaupt einen sogenannten unabhängigen Experten, der im besten Fall wirklich ein unabhängiger ist, oder eine Expertin, die natürlich auch wieder ganz anders sprechen können?

Und die Bevölkerung neigt eigentlich dazu, Expertinnen in solchen Situationen fast lieber zu haben. Während sich die Politiker und Politikerinnen dann eher für das emotionale und für das moralische zuständig sind. Also, insofern ist es ganz gut, mehrere Personen sprechen zu lassen und die unterschiedliche Rollen durchführen zu lassen.

Und ein Schlagwort sozusagen im Krisenmanagement in Japan war die Panik. Ist es wirklich möglich, dass eine Art panische Reaktion der Bevölkerung überhaupt auftreten kann, und ist es ein eine berechtigte Sorge, in der Kommunikation mit den Personen zu sagen, dass man eben sich zurückhält mit den Informationen, um Panik zu vermeiden? Oder ist es unrealistisch, dass das eine eine Panik in der Bevölkerung auftritt.

Das ist eine sehr gute Frage, weil die auch wieder zweischneidig ist, einerseits ja, Panik kann passieren, man braucht nur überlegen, wenn man kennt, das vielleicht doch selber, wenn es irgendwo einmal ein Erdbeben gibt oder ein Feuer ausbricht, oder? Ja, ich war gerade kürzlich in einer Situation, wo jemand, einfach vom Stuhl gefallen ist, gestolpert ist. Alle waren in Panik und richtig reagiert haben einfach nur 2 Personen, obwohl mehrere im Raum waren.

Das ist jetzt im Vergleich zu einem Atomunfall eine ganz kleine Panik, aber die haben einfach nicht gewusst "was tun". Man sieht es bei Autounfällen, also bei ganz kleinen Situationen, bis hinzu wirklichen Katastrophen. Menschen neigen zunächst einmal zu Panik und insofern muss man natürlich beruhigen.

Andererseits kann zu wenig Informationen oder sehr gefilterte und offensichtlich falsche Informationen noch zu größerer Panik führen und das kann vor allem auch zu dazu führen, dass man sich eher an Verschwörungslegenden hält. Die hat es rund um den Atomunfall gegeben und gibt es bei jeder Katastrophe immer auch und die sind natürlich dann noch viel gefährlicher.

**Das heißt, wie geht man dann im Endeffekt am besten vor, um derartige panische Reaktionen zu vermeiden?**

Ich glaube, das ist in Wirklichkeit wahnsinnig schwierig ist, weil man selbst ja auch menschlich ist, ja, und vielleicht sogar selber in Panik ist, sonst selber viel zu wenig Informationen hat und dann muss man sich auch noch um die anderen kümmern. Aber im Grunde genommen ist das Ganze ähnlich wie in einer Kriegssituation und da gibt es auch für das Bundesheer, nicht nur das Bundesheer, sondern für jede Armee der Welt gibt es so eigene Handbücher, was man tun soll und im besten Fall holt man das nicht dann raus in der Situation, sondern kennt sie auswendig und so gibt es natürlich auch bei Atomunfällen ganz genaue Abläufe und die man sich halten kann, so dass man da auf etwas zurückgreifen kann, was schon vorbereitet ist. Ich würde auch sagen, würde ich zum Beispiel einer solchen Firma oder eine Regierung, die mit Atomenergie zu tun hat, hätte ich immer schon eine ganze Box mit verschiedenen Antworten parat, die ich dann rausziehen kann und dich dann nur noch anpassen muss, damit ich dann, wenn ich selbst in Panik bin, gut reagieren kann. Das Wichtigste ist dabei dann sicher behutsam vorzugehen, also das klingt jetzt ganz, ganz banal, aber behutsam vorsichtig, auch vorsichtig formulieren, aufpassen, dass man nicht selber panische Worte verwendet.

Also man kann den Leuten ja auch die Wahrheit sagen, aber mit Worten, die sie ertragen, oder ich kann sie anschreien und sagen “Ihr werdet alle sterben”, wie es in den Filmen immer vorkommt. Wenig hilfreich, aber ich kann das Gleiche auch sagen, sodass die Menschen nicht ausflippen. Das hat ganz viel mit Wortwahl zu tun.

**Das heißt, du hast gesagt, es kommt vor allem auf die korrekte Vorbereitung an und die Art und Weise, wie man dann letztlich die Botschaft übermitteln möchte. Worauf muss man dann eben achten, wenn man mit der Bevölkerung spricht? Was ist sozusagen wichtig zu beachten, um wirklich sicherzustellen, dass alles so verstanden wird, wie es eigentlich vorgesehen war und dass auch keine Missverständnisse sozusagen auftreten?**

Sehr gescheite Frage und du hast du die Antwort im Grunde genommen schon selber gegeben? Das heißt ganz einfache Sprache, es hilft also nichts, wenn ich mit irgendwelchen wissenschaftlichen oder fremden Worten daherkomme, sondern ganz, ganz einfache Sprache. Kurze Sätze, ganz genaue Anweisungen nicht irgendwie Nebensätze oder Wenn-Sätze oder sowas, sondern ganz ganz konkrete. Das ist die Situation.

So zum Beispiel dieses Motiv, das nennt sich Kiss, kurze Informationen und die, die einfach so kondensiert, dass sie auch jeder wirklich verstehen kann, sodass das wird ihnen jeder Krisen Situation auf der ganzen Welt macht und eigentlich die Leute, die dann die Meldungen bringen sollten, müssten theoretisch darauf vorbereitet sein. Und die müssen auch all das schon längst vorbereitet haben.

**Welche Rolle kommt jetzt dem Journalismus zu, wenn man sich Krisensituationen wie jene in Fukushima anschaut, welche Rolle, sprich welche Aufgaben hat oder sollte der Journalismus haben und wie wirkt sich sozusagen die Tätigkeit von Journalisten dann auf die Situation aus?**

Ich habe das damals total interessant erlebt, ich war nämlich nicht in Europa, sondern in Costa Rica, in dieser Universität von den Vereinten Nationen und ich hatte Studierende aus Japan, die natürlich in Panik waren weil, wenn man weit weg ist und die Eltern sind dort, wo die ganze Familie und Freunde sind alle dort und man kann nichts tun, ist es besonders schlimm. Was sie aber gesagt haben ist, sie haben in Costa Rica und dann auch in den internationalen Medien viel mehr reelle Informationen bekommen als die Journalisten in Japan es verbreitet haben und das sieht man auch immer wieder, je näher man dran ist, desto schwieriger ist es, darüber zu berichten und, dass es aus der Distanz meistens einfacher ist. Und da sind wir wieder bei diesen Dingen, über die wir am Anfang gesprochen haben, aber die Rolle der Journalisten ist natürlich prinzipiell immer aufzuklären und zu schauen, dass auch von anderen richtig aufgeklärt wird, also Fragen zu stellen, immer wieder nachzufragen, lästig zu sein. Gleichzeitig aber auch wichtige Meldungen wie was zu tun wäre das so an die Bevölkerung weiterzugeben. Das ist so eine zweigeteilte Rolle. Das eine ist so ein bisschen, wie politische Bildung zu machen oder Bildung generell zu machen und zu schauen und das andere ist natürlich zu fragen: “warum ist das passiert? Wer ist zuständig, wer ist verantwortlich und so weiter?” Das ist eine Sache, die sicher nie ein einziger Journalist machen kann, und in jedem Medium müssen die Rollen ein bisschen aufgeteilt sein. Ich habe in meiner in meiner Recherche eben auch gesehen, dass das einerseits natürlich, wie du auch schon erwähnt hast, es sehr wenig kritische Berichterstattung aus Japan gegeben hat am Management der der Regierung.

**Während eben die ausländischen Medien oftmals als erstes neue Dinge herausgefunden haben und die Situation kritisch beleuchtet haben. Worauf kann man das zurückführen? Wie unterscheidet sich die die Medienlandschaft in Japan im Vergleich zu der international oder in Europa zum Beispiel?**

In Japan wird sehr viel auch beachtet, was man wem sagen darf. Das ist im asiatischen Raum in einigen Ländern so, dass wie zum Beispiel in Thailand könnte noch immer die Todesstrafe darauf stehen, wenn man den König beleidigt, ja, obwohl man den aktuellen König sowohl beleidigen sollte, weil er, weil er illegale Dinge macht, ja, aber man darf einfach nicht, und das ist gefährlich. Und in Japan ist es nicht mehr so streng, aber es ist noch immer so gegen gewisse Personen und gegen die Regierung sagt man nichts, sondern sagt auch so wenig wie möglich gegen ältere Leute. Das ist das eine, dieses kulturelle, das andere ist aber auch, wenn man selber betroffen ist, tut man sich viel schwerer, über etwas zu schreiben, wenn ich als Journalist, ich kann viel leichter neugierig sein, wenn sich das ganze woanders abspielt. Ich kann viel leichter kritische Fragen stellen, als wenn ich selber betroffen bin und das hat schon auch damit zu tun. Die Medienlandschaft ist in Japan jetzt nicht schlechter als woanders. Was besonders spannend war damals, war dieser zivile Journalismus, der sich da entwickelt hat, wo einfach die Leute selber sich aufgemacht haben, hingefahren sind oder wo sie waren Videos gepostet haben. Wie verbreitet also dieser Grass-Root Journalismus, hat er früher einmal geheißen jetzt, heutzutage ist sowieso jeder sein eigener Regisseur, ist das hat sich ja auch massiv verändert, aber damals war das ganz, ganz beachtlich und die haben natürlich, da gibt es auch diesen Ehrenkodex nicht gegenüber.

Oder die, den Journalisten eigentlich haben? Dadurch konnten wir auch viel freier berichten, das heißt vieles, was in Japan aufgedeckt wurde, wurde zunächst einmal von privaten Leuten und nicht von Journalisten aufgedeckt.

**Ein wichtiger Punkt im Krisenmanagement ist auch Vertrauen und Autorität und ist natürlich sehr wichtig für jene Organisationen oder Personen, die, die sozusagen Entscheidungsträger sind. Wie sollte man oder wie hängt jetzt sozusagen das Vertrauen oder die Autorität einer Person damit zusammen? Welchen Einfluss haben Medien und Journalisten auf das Vertrauen in die in die Autoritäten und Entscheidungsträger und haben Journalisten auch sozusagen eine Art Verantwortung in einer solchen Situation, in der, in der es darum geht Menschen aus der Gefahr zu bringen und Menschen zu schützen?**

Du hast wirklich gute Fragen Elias. Sie haben dann natürlich ganz wesentliche Aufgaben, sie können mit ihrer Berichterstattung dafür sorgen, dass Leute jemandem vertrauen oder nicht vertrauen, die Art, wie jemand dargestellt wird, wie Hintergrund, Berichte über die Person kommen, wie jemand zum Beispiel fotografiert wird ja, wie also welche Fotos über jemanden gebracht werden kann ganz ganz viel in der Bevölkerung auslösen, indem wie jemandem vertraut wird oder nicht? Nicht umsonst hat, kleiner Exkurs, Sebastian Kurz immer darauf bestanden, dass sein persönlicher Fotograf die Fotos macht, weil der sie aus einer bestimmten Perspektive aus einem Licht Blickwinkel gemacht hat, die er normaler Journalist aus professioneller Sicht einfach nichts getroffen hätte oder gewählt hätte. Und so ist es natürlich auch bei den Personen, die in einer Krisensituation sprechen, welche Berichte da kommen werden ganz viel darüber sagen, wie die Leute vertrauen oder nicht vielleicht also es kann in beide Richtungen gehen. Medien können da Vertrauen aufbauen und sie können aber auch dafür sorgen, dass man nicht vertraut. Soziale Medien spielen da inzwischen natürlich auch eine ganz, ganz große Rolle. Da braucht man nicht nur über normalen Journalismus zu reden, aber Vertrauen in der Bevölkerung ist etwas vom Wichtigsten und gleichzeitig das teuerste gut, das es gibt, weil es wahnsinnig schwierig zu bekommen ist. Das sieht man auch in den letzten Jahren, wie weltweit überall die Vertrauensindizes runtergehen, je länger eine Krise dauert. Das trifft natürlich auch für so eine Katastrophe wie einen Unfall und Fukushima zu, dass, je länger das dauert, desto geringer ist das Vertrauen. Also Fukushima hat eben gezeigt, dass das Vertrauen in die japanische Regierung und generell die Behörden drastisch gesunken ist. Das war längerfristig also das, das ist noch immer nicht ausgestanden.

**Ist es möglich, sozusagen wenn es zu einem Vertrauensverlust kommt, in einer Krisensituation dann noch immer handlungsfähig zu sein und auch sinnvolle Maßnahmen umzusetzen oder ist das sozusagen eine Situation, in der, in der man als Institution oder als Person, in die das Vertrauen verloren wurde, oder ist man dann sozusagen chancenlos und muss, dass das Steuer dann sozusagen anderen Menschen übergeben?**

Als Person ist man sehr schnell einmal chancenlos. Als Institution kann man genau vom Loswerden einer Person durchaus profitieren und dann wird man quasi die Schuld auf das Opfer.

Im Englischen das „scapegoat“, das dann irgendwie schuldig wäre, obwohl es in Wirklichkeit gar nichts dafür kann und vielleicht eh alles richtig gemacht hat, aber irgendjemand wird ausgewechselt und damit glaubt man, dass das Vertrauen steigt.

Das kann man ein paar Mal machen, wenn man kann es nicht zu oft machen, wenn diese Personen ununterbrochen ausgetauscht werden, und das passiert ihnen in Krisensituationen häufig, dass, dass sich eine Person gar nicht richtig einarbeiten kann, ist sie schon wieder weg und kommt schon die nächste Person, nur weil es mal schlechte Medienberichte gegeben hat? Das hat aber natürlich auch keinen Sinn, das heißt, man muss eine Person auch Chancen geben, und da sind wir wieder unter dem Punkt je besser man vorbereitet ist, desto eher wird man richtig reagieren und auch das, das halbwegs über die Bühne bringen, selbst in einer Krisensituation. Auch wenn es schwer ist oder man schickt eben absichtlich, das machen auch manche und in Japan war das auch ein bisschen sichtbar, man schickt absichtlich schwächer aussehende Personen vor, damit die zuerst einmal unter Anführungszeichen bestraft werden können und dann einfach weggeworfen und entlassen in dem Sinne werden, damit dann eine andere Person, die wichtiger für das Unternehmen ist, als Retter auftreten kann.

**Da du das erwähnt hast, könntest du da vielleicht etwas genauer darauf eingehen, auf welche Personen du dich beziehst oder auf welche Ereignisse konkret?**

Also in Japan, die Unternehmerfirma des Atomkraftwerkes, die haben ihre Pressesprecher innerhalb kürzester Zeit ausgetauscht. Ja, und die haben einfach immer wieder jemanden vorgeschickt und die politisch, irgendwie Verantwortlichen haben sich zunächst einmal zurückgehalten. Sie haben nicht sofort etwas gesagt, das hat da haben Sie zuerst einmal Minister verlassen. Die Minister kann man austauschen. Der Regierungschef kann aber dann noch bleiben. Solche Dinge ja. Das wird bei wird weltweit so immer wieder so gemacht, dass die Person, die eigentlich gemacht hat, dann so ein bisschen im Hintergrund bleibt, die anderen ein bisschen vorher reden soll, die vorne sind austauschbar, der der oder die im Hintergrund weniger. Gibt es auch im in Kriegssituationen ist auch ganz, ganz spannend jetzt mit der Ukraine-Krise zu sehen? Wer spricht, wer wird für was vorgeschickt und so? Also das sind Krisen erstaunlicherweise austauschbar.

Es war in Japan, da du das eben angesprochen hast, eben auch so, dass als sich die Situation im Kraftwerk verschlechtert hat. Tatsächlich ist auch der Premierministerin das Kraftwerk geflogen ist und uns sozusagen versucht hat, direkt in das Geschehen einzugreifen. Teilweise hat er auch selbst Feuerwehrautos weggeschickt oder mit den Mitarbeitern dort gesprochen, war sehr scharf kritisiert wurde, weil ihm sozusagen eine Art Micromanagement vorgeworfen wird und er sozusagen die Gesamtsituation vernachlässigt hat. Ist diese Kritik berechtigt? Also sollte man sich als Präsident aus den aus den Details raushalten und versuchen, die Situationen globaler zu managen, oder ist es durchaus verständlich, dass manchmal Personen direkt in das Geschehen eingreifen?

Also man kann natürlich nichts irgendwo in seinem Palast, er hat keinen Palast, aber sagen wir mal so, man kann nicht dort bleiben, wo man selber in Sicherheit ist, sondern man muss quasi an die Front man sieht das bei jeder Überschwemmung, da muss immer jemand also Regierungschefs, egal ob jetzt auf bundes- oder nationaler oder regionaler Ebene, die müssen hinkommen und die müssen etwas sagen und die müssen der Bevölkerung reden.

Da kommt man natürlich nicht herum, aber es gibt da auch wirklich viele dumme Aktionen hat er nicht sogar auch noch Wasser dort getrunken?

**Soweit ich weiß, ist mir das nicht untergekommen, aber das kann ich jetzt nicht ausschließen.**

Ich habe irgendwie in Erinnerung, dass er das Wasser dort getrunken hat und gesagt hat, wenn ich es trinke, dann ist das Wasser gut.

Ich habe zumindest gelesen, dass er ganz normal Nahrung konsumiert hat, ohne sich die Hände zu waschen und eigentlich ohne Bedenken sozusagen dort Lebensmittel konsumiert hat.

Okay gut, genau sowas ist natürlich wahnsinnig dumm.

Weil man dann einfach nur leugnet und weil das nicht hilfreich ist für die Leute, die nicht wie er dann wieder wegkönnen, ja also für die, die dort leben. Das ist eine Katastrophe. Das heißt, man muss schauen, dass man emotional für die Leute da ist, aber dieser Vorwurf des Micromanagement, den du genannt hast, ist natürlich vollkommen richtig, den machen in der Situation ganz, ganz viele, weil das Bild mit dem Feuerwehrauto oder mit der Schaufel bei der Überschwemmung oder sonst etwas, das sind lauter Bilder, die sich gut verkaufen lassen.

Das sehen die Presseleute häufig, dass aber gleichzeitig ganz, ganz dumme Bilder sind. Das heißt, es wäre gescheiter gewesen, er wäre in einem richtigen Schutzanzug dort gewesen und hätte Schutzanzüge hätte dafür gesorgt, dass die anderen auch Schutzanzüge bekommen und hätte, wäre dabei zu sehen gewesen, wie er tatsächlich etwas tut, was für die Situation hilfreich ist, anstatt Bilder für die Bevölkerung und für den nächsten Wahlkampf zu sammeln und solche Bilder in Katastrophen sind leider immer wieder auch sowas. Ja also Trump hat das mit seiner Frau gemacht, dass sie auch bei einem Hurricane hingegangen sind, und sie ist dann im Stöckelschuh und in irgendeiner komischen Jacke hergekommen, wo dann nur noch über die Jacke geschrieben worden ist. Ich weiß nicht mehr, welcher Spruch draufgestanden ist, aber irgendwann wirklich dummer Spruch und es wird dann damit einfach abgelenkt, ja und da kann man als Politiker extrem viel falsch machen. Wenn man sich falsch verhält, wenn man, der Laschet, glaube ich in Deutschland hat bei der Überschwemmung gelacht. Ja, natürlich kann man in eine Situation kommen, in der man auch lacht, auch jetzt auch dort nur darf ich mich dort nicht filmen lassen oder fotografieren lassen, wenn ich gerade über etwas lache. Das heißt, da muss man extrem aufpassen, und das gehört zu den schwierigsten Situationen überhaupt für Verantwortliche. Ob das jetzt der der Chef einer großen Firma ist, die Verantwortliche sind.

Aber ein bisschen sind wir dann wieder bei dem, was man ganz am Anfang gesagt habe, einfach authentisch sein und ehrlich sein die Menschen schätzen, dass immer mehr und grad in einer Krise braucht man genau das.

**Zu den Medien wollte ich auch noch Fragen erhöht eine Wiedergabe, der Ereignisse, oder der Situation durch andere Personen oder Journalisten. Wie wirkt sich das auf das Verständnis der Situation aus, oder ist es wahrscheinlicher, dass Missverständnisse entstehen, wenn Informationen durch andere Personen durch Mundpropaganda an Menschen gebracht werden, erhöht das sozusagen die die Wahrscheinlichkeit, dass Informationen falsch verstanden werden?**

Also die Stille Post wird es immer geben und in der Stillen Post wissen wir alle verändert sich Information. Gerade Menschen in Panik oder nehmen wir mal die Panik weg, einfach in Sorge werden auch Informationen weniger konkret aufnehmen und hören nur das, was sie hören wollen. Das heißt, es hängt wirklich davon ab, wie berichtet wird. Und da haben es, dann wird wahrscheinlich jeder zu seinem Medium des Vertrauens greifen. Man kann sich schon vorstellen, dass auch bei uns in der Kronen Zeitung etwas anderes steht als in der Presse oder im Standard. Wichtig wäre, dass die Menschen sich aus verschiedenen Quellen Infos holen, gerade in einer Krisensituation, aber wiederum gerade in der Krisensituation hat man manchmal gar nicht die Zeit, sie zu konsumieren. Das heißt, es erweist sich eigentlich immer wieder das Radio in solchen Situationen als hilfreich, aber funktioniert in Krisenzeiten nach wie vor auch deshalb, weil es mehr als ein Smartphone, das unendlich viel Energie verbraucht und dauernd aufgeladen werden muss, kann man halt beim Radio einfach noch mit Batterien betreiben.

**Eine Frage, die ich mir auch aufgeschrieben hab, war die Regelmäßigkeit in der Veröffentlichung von Informationen, sprich wie oft die der Krisenstab beispielsweise Aktualisierungen gibt, über das Radio oder über die Medien oder über Pressekonferenzen, da wird teilweise davon gesprochen, in der akuten Krisensituation alle 2 Stunden eine Pressekonferenz zu machen. Wie sollte man das beurteilen? Und wie wirkt sich sozusagen die Häufigkeit mit der Information wiederholt oder herausgegeben wird auf die Wahrnehmung dieser Informationen aus?**

Also man kennt das ja selber, wenn irgendwo auf der Welt etwas Interessantes passiert, man schaut ununterbrochen nach und drückt ununterbrochen auf „Refresh", weil man hofft, dass schon wieder etwas Neues da ist. Diese Liveticker, die sich da jetzt überrall eingebürgert haben, die funktionieren auch genau mit diesem System. Eine Regierung oder ein Unternehmen, wie das da war, muss natürlich auch relativ schnell sein, also ich würde auch sagen etwas zwischen ein und 3 Stunden wäre schon gut. Ich würde eher von eineinhalb als zwei Stunden sprechen, in der höchsten kritischen Phase. Je länger eine Krise dauert, desto länger können diese Abstände natürlich auch bleiben. Aber es muss auch nicht jedes Mal eine Pressekonferenz sein.

Aber Meldungen rauslassen muss man am Anfang schon häufig und Wiederholung macht da nicht wirklich etwas, so lange immer wieder etwas Neues dabei ist, beziehungsweise Wiederholung von den wichtigsten Dingen, also wie man sich zu verhalten hat, das sollte sowieso häufig in den verschiedensten Formulierungen stattfinden. Ja und apropos, verschiedene Formulierungen: Gleichzeitig müssen alle in der Regierung oder in seinem Unternehmen dann auch gleich sprechen, ja, also, andere Formulierungen heißt nicht, dass jeder was anderes erzählen soll, das heißt, die Message muss die gleiche sein und der Premierminister muss das gleiche wie seine Minister sagen und nicht unterschiedliche Dinge.

**Welche Rolle spielen dahingehend Schlagwörter also in Fukushima? Ein Wort, das einerseits mit sehr viel sehr viel Aufmerksamkeit verbunden war, „Kernschmelze“, andererseits wurde dieses Wort möglichst vermieden und auch versucht, das euphemistisch zu umschreiben also welche Rolle spielen solche Wörter die welche Rolle spielen Schlagwörter und wie sollte man mit ihnen umgehen?**

"Kernschmelze” ist hier ein brutales Wort, weil da weiß auch derjenige der oder die nicht viel über Atomkraft weiß, dass Kernschmelze so ziemlich das Schlimmste ist, ja also das das kennt man einfach zu sehr, das heißt, dieses Wort würde ich auch so lange wie möglich vermeiden, wenn ich ehrlich bin. Gleichzeitig, wenn ich die Kernschmelze vermeiden kann, dann würde ich das Wort sehr wohl in den Mund ja also wir haben jetzt die Kernschmelze verhindert. Das ist ein gutes Wort. Die Kernschmelze passiert in einer halben Stunde. Wenn sie in einer halben Stunde passiert, dann und man das erahnt, dann ist wahrscheinlich schon gescheiter Mann sagt aber Wörter, die vielleicht nicht wirklich sowas werte, die nicht sofort notwendig sind, weil man auch noch nicht weiß, ob es zu einer Kernschmelze und bei dem zu bleiben kommen wir. Ich würde es zunächst einmal vermeiden, um nicht die Panik weiter zu schüren. Das heißt Bevölkerung zu berühren, aber ihr zu sagen, alle Vorsichtsmaßnahmen zu treffen, und in Japan weiß die Bevölkerung, was man tut. Also schon seit dem Zweiten Weltkrieg gibt es da Trainings, was meiner Situation tut um die Leute dazu zu bringen, dass sie es dann tatsächlich auch tun, wäre schon wichtig, aber solche Worte würde ich möglichst, also diese ganz großen Schlagworte würde ich vermeiden. Sie gehen nämlich auch niemals weg, diese großen Worte. Damals hat George Bush das Wort “War on Terror” so ganz nebenbei gesagt und CNN hat es sofort übernommen und das Wort ist bis heute nicht mal weggegangen. Ein ganz ein grauenhaftes Wort und für sich.

**In einer Krisensituation oder in der Krisenkommunikation geht es darum, dass man möglichst viele Menschen erreicht. Ist es dabei sinnvoll, dass man differenziert vorgeht? Das heißt, dass man unterschiedliche Altersgruppen oder unterschiedliche Personengruppen auf unterschiedlichen Kanälen und möglicherweise auch mit unterschiedlichen Botschaften erreicht. Oder sollte man in der Kommunikation möglichst einheitlich sein?**

Die Botschaft muss dieselbe sein, aber sie muss anders formuliert sein, weil einem
Universitätsprofessor, der sich an die letzten 30 Jahre mit der Thematik beschäftigt hat, brauche ich nicht das Gleiche zu erklären wie jemanden, der dort direkt vor Ort wohnt und jetzt gerade rausgeht, um seine Schafe einzusammeln, also da muss man unterschiedlich reden. Auch mit Jugendlichen würde ich anders reden. Mit den Menschen, die weiter weg sind, wird man vielleicht anders reden und so und deshalb ist es ja auch wichtig und das wird auch so gemacht in Krisensituationen, dass man auf unterschiedlichen Ebenen kommuniziert und natürlich auch mit unterschiedlichen Medien.

**Du hast ganz am Anfang angesprochen Apps, die, mit denen man also in Japan ist, ja sind ja Tsunamis eine Bedrohung und es gibt auch tatsächlich von der japanischen Regierung eine App, mit der über Naturkatastrophen wie Erdbeben und Tsunamis Auskunft gegeben wird. Allerdings stelle ich mir die Frage, wie effektiv derartige Apps sind. Zum Beispiel auch die Rotes-Kreuz-App in Österreich gibt es ja auch, weil diese Apps eben sehr, sehr wenige Downloads haben, nicht nur in Österreich, sondern auch in Japan ist es so, dass eigentlich nur sehr wenige Menschen tatsächlich über solche Apps verfügen.**

Also es wird in verschiedenen Ländern schon verwendet, inzwischen ist es weniger über Apps, sondern oft einfach über einen ganz banalen WhatsApp-Kanal, wo minütlich Informationen weitergegeben werden, und so erreicht man die Leute niederschwelliger als über eine App, die man sich erst runterladen muss, allerdings in einer tatsächlichen Krisensituation. Es kann so eine App ganz, ganz hilfreich sein und dies ist dann aber auch was anderes als diese App, die man in
Deutschland, Italien oder Österreich. Diese Rotkreuz App, wie wir sie in Spanien zum Beispiel haben, hat ja super funktioniert, aber da ist sie auch mit anderen Infos gefüttert worden ja, und das hängt davon ab, wie man sie von Anfang an füttert. Es gibt auf den Philippinen und so gibt es auch solche Apps, weil man eben öfters mit Katastrophen rechnen muss und die Leute verwenden sie tatsächlich. Es wird inzwischen auch für die verschiedensten Krisensituationen sowas und wenn ich selber vor etwas Angst habe, dann werde ich die auch benutzen. Insofern können die schon gut sein, aber die müssen halt dann wirklich auch mit guten Infos gefüllt werden, guten Warnungen. Ich glaub damals bei dem Unfall hat es weniger über die App funktioniert als über sonst Social-Media-Kanäle.

Bei den Social Media Kanälen war Twitter von zentraler Bedeutung. Bei der App ist es ja meistens so, dass Informationen sozusagen nur in eine Richtung herausgegeben werden, also, dass die Menschen nur Informationen bekommen, aber nicht Rückfragen stellen können, oder Dinge nachfragen können, die vielleicht nicht vollständig verstanden worden sind. Wie wichtig ist der Dialog in einer Krisensituation? Ist es wichtig, auf Fragen einzugehen oder ist es eher wichtig, einmal nur in eine Richtung Informationen fließen zu?

Ich würde sagen, das muss auch wieder geteilt werden. Irgendjemand muss Fragen beantworten, Menschen haben Fragen und jeder will die gleiche Frage aber für sich selber beantwortet wissen also das ist natürlich ganz, ganz wichtig, aber als zum Beispiel Premierminister muss ich nicht jede Frage beantworten, sondern da habe ich mein Team dafür, dass das macht das Gleiche, bei einem Unternehmen. Ich muss zunächst einmal auch im Voraus gehen, damit die Leute überhaupt wissen, was sie fragen können und manche Frage wird sich dann vielleicht auch erübrigen? Aber es muss beides parallel laufen unbedingt. Deshalb braucht es auch immer ein Krisen Team und niemals eine Person, die alles allein macht.

Wie ich herausgefunden habe, war es allerdings so in Japan, dass das Büro des Premierministers zwar einen eigenen Twitter Account hatte, der auch sehr viele Follower hatte und sehr viele, also die Nachrichten wurden sehr zahlreich gelesen. Allerdings war es so, dass in den ersten Tagen der Krise tatsächlich gar keine Antworten herausgegeben worden sind und danach auch erst sehr, sehr spärlich geantwortet wurde. Was hauptsächlich darauf zurückgeführt wird, dass zu wenig Personal da gewesen ist, um auf die Fragen einzugehen. Heißt das, man möglichst viele Ressourcen haben soll, um auf die Fragen einzugehen und dahingehend sozusagen sicherstellen, dass das Fragen auch wirklich. Ich glaub, man muss einfach vorbereitet sein und es gab diese ganz große Arroganz: „bei uns passiert eh nichts", wie man das immer wieder einmal hat und mit dieser Arroganz bereitet man sich dann auch nicht vor und dann hat man in der Krise nichts und auch zu wenig Personal. Zunächst einmal, wenn ich alles schon im Vorfeld mitgedacht hab, aber das wird meistens nicht passieren, weil dann eingespart wird und man eben meint „wird schon nix sein". Wird viel zu selten Geld investiert, dann hat man eine Krise und weiß nicht recht, wie man damit umgehen soll.

**Hat sich das inzwischen geändert? Also die Nutzungszahlen von sozialen Medien sind natürlich seit 2011 deutlich angestiegen und Twitter ist ja jetzt in vielen Fällen das Sprachrohr von Politikern ist. Ist es heute so, dass die Ressourcen für soziale Medien und für Kommunikation darüber heute besser sind als früher?**

Ja, ja, gibt natürlich wird von Anfang an mehr in das hineingesteckt aber, ich würde bezweifeln, dass man in der Krisensituation wieder dann trotzdem besser ist als damals weil man sich meistens noch immer mit anderen Themen beschäftigt, wenn es irgendwie geht. Ich denke, dass die Pandemie da auch ein ganz gutes Beispiel dafür war, dass welche Infos dann geteilt worden sind und dass eigentlich, die die Bevölkerung selber und gute Journalisten dann meistens für die guten Informationen sorgen und gar nicht unbedingt immer die, die eigentlich verantwortlich wären.

Am sehr kontraproduktiv in in Krisensituationen sind natürlich Gerüchte und Verschwörungstheorien. Wie sollte man als Organisation oder als Krisenstab mit solchen Gerüchten umgehen? Wenn es konkrete Gerüchte gibt, sollte man die dann auch wirklich ansprechen oder sollte man einen anderen Zugang zu dieser Situation haben? Das hängt wirklich vom Gerücht ab, wenn es einfach so ein kleiner Shitstorm ist, wo man denkt, der wird bald weg sein, dann würde ich nicht drüber eingehen (vgl. Fiederer/Ternès 2016: 4), auch wenn es unter die Gürtellinie geht. Ganz persönlich würde ich es auch nicht tun, aber wenn es etwas ist, was ich mit Fakten aufklären kann, dann würde ich das tun. Das ist hilfreich, aber was immer das Hilfreichste ist, und das ist leider hilfreicher, als die als die Fakten, ist, die Leute bei der Emotionen. Wenn ich über die Emotionen Zugang zu den Leuten hab, dann kann ich ihnen auch die unbequeme Wahrheit vermitteln. Gerade diese Verschwörungslegenden arbeiten wir sehr viel mit Emotionen und holen die Leute bei ihren Ängsten ab und wenn auf diese Ängste nicht eingegangen wird, dann haben die eine viel größere. Wenn ich diese Ängste aber von vornherein ernst nehme und auch genau zuhöre, was die Leute brauchen und eben Fragen beantworte, bevor noch Gerüchte überhaupt aufkommen können, dann ist das hilfreich also ich glaube, man muss einfach schneller sein. Die Gerüchte entstehen ja meistens im Vakuum von Informationen.

**Und welche Rolle spielen dahingehend, die sozialen Medien?**

Ja, keine so tolle, wie wir wissen. Ich glaube auch nicht, dass man das so ganz, ganz schnell unter Kontrolle kriegt, aber ich werde immer dafür selber auch die sozialen Medien zu nutzen und die nicht einfach nur den jenen zu überlassen die 15 Minuten Berühmtheit wollen und deswegen irgendetwas erzählen, was ein Märchen ist oder eine Verschwörungslegende oder so. Also weil Twitter zum Beispiel Dritte benutzen, wie Politiker und das benutzen vielfach Journalisten, Intellektuelle und so weiter. Aber der Großteil der Bevölkerung ist eher auf anderen Medien, und die muss ich bespielen und gerade in einer Krisen Situation auch schauen, dass dort meiner Information also so lange nicht hoffentlich die Wahrheit, zack, auch dorthin kommt.

**Weil du angesprochen hast das, dass man mehr oder besser vertreten sein sollte in sozialen Medien. Dort ist es ja so, dass das sozusagen Informationen oder Berichte natürlich durch Algorithmen oder durch Empfehlungen leichter an viele Menschen kommen als andere und häufig sind es auch jene Informationen, die die emotional eben gestaltet sind oder sehr pompös sind. Sollten Politiker, Krisenstäbe oder Institutionen, die eben am Krisenmanagement beteiligt sind, denselben Zugang zu diesen Medien haben wie wie Personen, oder sollte es da eine Unterscheidung geben?**

Also prinzipiell ist eine Krise gibt es wäre das Wichtigste. Einmal versuchen, die Krise zu bewältigen und den Menschen Informationen zu geben, die sie brauchen, aber auch nicht selber ins Märchen erzählen, kommen oder Geschichten drucken, wie man so in Wien war sagen würde, das heißt, man muss nicht alles zugleich bespielen. Man wird auch nicht die Kapazitäten haben, ich glaube vernünftiger und näher an der Bevölkerung meine Antworten und meine Empfehlungen findest du eher werde ich die Chance haben, dass sie mir auch zuhören und dann wären sie vielleicht auch weniger auf Kanäle ausweichen, die ich dann nicht bespielen kann, weil sie es einfach zeitlich nicht ausgeht, budgetär nicht ausgeht, weil es auch sinnlos ist also ich brauch nicht in ein in einem Wettkampf mit jemanden gehen, der sowieso überhaupt nichts glaubt von dem was ich sage das heißt ich sollte jenen Medien nützen, wo ich glaube, dass ich den Großteil der Bevölkerung erreiche und auf sinnvollere, weiche Weiße erreiche. Alles andere kann man sich dann vielleicht später kümmern, aber ich werde niemals 100% der Bevölkerung erreichen und mit meinen Wahrheiten durchkommen. Aber die Frage ist doch immer, in welcher, an welchem Punkt einer Krise ist man und in der massivsten Krise? Von einem Atomunfall sind Verschwörungslegenden zunächst einmal ganz egal, ich werde wahrscheinlich andere Prioritäten haben und auch haben sollen.

Abgesehen von den Fehlern, die beim bei der Krisen Kommunikation und bei der beim
Krisenmanagement in Japan passiert sind, wie sehr hat sich die Wiese hat sich sozusagen die Medienlandschaft und die Situation mit den mit sozialen Medien verändert, also im Vergleich zu heute und sollte man heute anders vorgehen, als es beispielsweise 2011 sinnvoll gewesen wäre?

Also Medien haben sich einfach entwickelt, vor allem die digitalen Medien, dann wiederum die sozialen Medien und auch die sogenannten Mainstream-Medien spielen ja viel mehr mit diesen Kanälen beziehungsweise überhaupt mit neueren Technologien. Das Werkzeug und die Tools sind da und auch das Wissen für den Umgang mit den Tools, also insofern ist doch vieles ganz anders. Dadurch verändert sich natürlich Journalismus sowieso, also ich glaube, dass man heute in Japan hätte man nochmals die gleiche Situation, würde man heute vielleicht anders agieren, hätte man die Situation von damals erst heute könnte sein, dass man wieder die gleichen Fehler machen wird ist man lernt ja leider meistens erst aus der Katastrophe und nicht ihm im Vorfeld, weil man sich mit anderen Dingen beschäftigt.
Aber ich glaube, was einfach bleibt oder was vielleicht sogar stärker geworden ist, dass Menschen vielmehr zweifeln an Informationen, die von oben kommen, unter Anführungszeichen und da hat man es heutzutage sicher viel, viel schwerer. Gleichzeitig geht alles viel schneller und man muss schauen, dass man schneller ist. Als Premierminister habe ich heute viel weniger Zeit zu reagieren als im Jahr 2011 hatte.

**Als Außenstehender hat man natürlich wenig Einblick in die internen Abläufe von dem Unternehmen TEPCO damals oder von dem Krisenstab von der von der Regierung. Ist es üblich, dass mit der Öffentlichkeitsarbeit und mit der Kommunikation mit der Bevölkerung spezielle Leute beauftragt werden, die eigentlich mit dem tatsächlichen Krisenmanagement nichts zu tun haben, also Menschen, die eben die eben sozusagen eine Ausbildung in Kommunikation oder Öffentlichkeitsarbeit haben und sozusagen nicht überhaupt nicht in das tatsächliche Krisenmanagement und die tatsächliche Krisenbewältigung involviert sind?**

Ja, sieht man immer wieder, dass Pressesprecher und Pressesprecherinnen gut darin sind, das was sein soll und das Bild, das man malen möchte, zu vermitteln, das machen sie super professionell, aber die wenigsten von ihnen können Krisenkommunikation, weil das Ganze ein anderes Feld ist. Und insofern sind dann meistens die falschen Leute an diesen Positionen in einer tatsächlichen Krise. Man bräuchte eigentlich ganz, ganz andere Leute, die auch psychologisch arbeiten, die also nicht umsonst hat es in einigen Ländern Leute vom Militär gegeben, die auch in der Pandemie eine wichtige Rolle gespielt haben. Im Krisenmanagement hat Österreich erst ganz spät reagiert und jemanden genommen, und ich will da jetzt gar nichts dazu sagen. Aber in anderen Ländern ist das viel früher genommen worden, weil natürlich das Militär Krisenkommunikation theoretisch besser können müsste, aber auch nicht immer. Es gibt natürlich auch Politiker und Politikerinnen, die quasi so ein Naturtalent sind. Die Merkel war, ganz egal, was man politisch von ihr hält, sie war eine exzellente Krisenkommunikatorin, das hat sie einfach gekonnt. Die hatte richtigen Moment genau den richtigen Tonfall, das richtige Gesicht, die richtige Geste. Und vor allem immer die richtigen Worte gefunden und das können ganz wenige Politiker und Politikerinnen.

Wenn man jetzt auf die Krisenkommunikation schaut, wie Politiker kommen und gehen natürlich und es gibt natürlich schon Leute im oder Experten, die die ihr ganzes Berufsleben in einem Feld wieder der Kernenergie bleiben und Leute in im Militär, beispielsweise die Krisenstäbe leiten oder die häufiger in Krisensituationen sind. Wie gewährleistet man sozusagen, dass das Krisenmanagement auch funktioniert, wenn man andere Leute hat, oder eben an eine komplett andere politische Lage oder Situationen?
Das ist eigentlich eine Frage, die ich unmöglich beantworten kann, weil jede Krisensituation dann doch wieder ein bisschen anders ist und was auch viel davon abhängt, ob es eine Krise ist von 3 Tagen oder eine von mehreren Jahren oder auch wen ich zu Beginn zur Verfügung habe. Aber was man aus solchen Krisen, wo es schief gegangen ist, sehr gut lernen kann, ist, dass es eben zu wenig ist zu wissen, wie ich jemanden hübsch, erfolgreich und gut aussehen lassen kann oder wie ich mein Unternehmen hübsch und erfolgreich aussehen lassen kann, dass das es einfach mehr dazu braucht und, dass ich jemanden brauche, der oder die wirklich mit der Bevölkerung auch kommunizieren kann, dass das könnte man lernen also auch, dass das Zuhören wichtig ist und dass Diskussionen oder Kommunikation in beide Richtungen gehen muss und das können ganz viele Pressesprecher überhaupt nicht, können auch viele Manager nicht. Es gibt dabei in jedem Unternehmen und sicher auch in jeder Regierung Personen, die das sehr wohl können. Ja, aber so ein Krisenstab, den kann ich dir jetzt leider nicht genau sagen, weil das einfach ganz unterschiedlich ist, je nach Situation. Aber ich, ich hätte in meinem Team immer Leute, die ganz unterschiedliche Fähigkeiten haben.

**In einer Krise wird es häufig zu einem Interessenskonflikt kommen. Und in Japan war es natürlich so, dass, das Unternehmen das Energieversorgungsunternehmen eine große Rolle gespielt hat im**

**Krisenmanagement und natürlich andere Interessen verfolgt als Verkäufer sozusagen von Atomstrom als, die Regierung. Wie sollte man in einer Krise mit einem solchen Interessenskonflikt, den es vermutlich immer geben wird, weil unterschiedlicher Institutionen in einer Krise beteiligt sind, wie sollte man damit bestmöglich umgehen?**

Es wird in jeder Situation immer Leute und Institutionen geben, die andere Interessen haben. Also natürlich hat ein Unternehmen andere Interessen als eine Regierung, das lässt sich auch nicht hundertprozentig zusammenzubringen, aber auch in einer Krise wird man danach unterschiedlich kommunizieren, also die, die Verantwortung wird immer auch eine politische sein. In Japan war das natürlich besonders dadurch, dass die das Unternehmen der TEPCO nicht allzu fern von der Politik. war. Ja, da, gab es sehr, sehr enge Bänder. Das ist häufig so bei großen und mächtigen Firmen, und insofern wird immer die Politik eine wichtige Rolle spielen in der Kommunikation. Wenn es zu einem Unfall kommt, will von dem Unternehmen eine Entschuldigung und so weiter, aber bei einer großen, nationalen Krise, da brauch ich natürlich die Regierung, die mir sagt, was zu tun ist und die mich auch beruhigt. Und da muss auch, weil du sagst Interessen, dann muss die Regierung auch das Interesse haben, die Bevölkerung zu schützen, das so steht auch in fast jeder Verfassung so drin, dass das eines der Hauptinteressen des Staates.

Es ist natürlich schwierig, jetzt Vergleiche zwischen unterschiedlichen Ländern anzustellen, aber wie wichtig war für die Entwicklung der Situation in Japan sozusagen die politische Landschaft und die Medienlandschaft und die Kultur in Japan, dass ich die Dinger so ergeben haben und wie würde der Umgang mit einer solchen Krisensituation in anderen Ländern aussehen.

Ich bin keine Japanologin und traue mich da jetzt auch nicht wirklich Stellung zu nehmen. Ich weiß nur, dass viel mit den japanischen Traditionen mit den Hierarchien diskutieren, mit dem wie man, wie man spricht, das, was eben anfangen unseres Gesprächs auch kurz diskutiert hatten, dass das eine Rolle spielt. Das heißt aber nicht, dass es woanders besser gelaufen wäre. Vielleicht bei uns oder in Frankreich, oder wo auch immer etwas anderes schief gegangen wäre, man hätte auch vielleicht zu langsam reagiert zu wenig an die Öffentlichkeit gegeben. Man hat sich ja damals gesehen mit Tschernobyl, dass da ganz viel vertuscht worden ist. Vielleicht hätte man bei uns inzwischen aus der Krise von damals gelernt und würde heute mehr an die Medien geben und mehr Wissen rauslassen mehr Informationen veröffentlichen. Ich hoffe es, aber man weiß es nie ganz genau. Dass so etwas ganz ohne nachträgliche Kritik ablaufen kann, wird nie der Fall sein, weil allein die Tatsache, dass es passieren kann, ist ein Skandal.

**Im Falle von der Nuklearkatastrophe von Fukushima war es so, dass natürlich nur Japan akut von dem Unfall betroffen war. Es ist natürlich nicht auszuschließen, dass sich eine solche Situation in anderen Ländern ereignet. Also in im Falle von Tschernobyl war es natürlich so, dass auch umliegende Länder betroffen waren. Aber es ist natürlich auch möglich, dass das Vorfälle in Grenzregionen passieren, oder, dass auch unterschiedliche Provinzen oder Bundesländer von Katastrophen betroffen sind. Wie sollte man vorgehen, dass das Krisenmanagement und die Krisen Kommunikation einheitlich abläuft? Sollten sich alle Beteiligten treffen und zusammen mit der Öffentlichkeit kommunizieren?**
Es gibt, es wird nie stattfinden, das wäre ein Traum, aber da hat dann jede Regierung ihre eigenen Dinge zu tun und ihre eigene Agenda und deshalb wird das nie stattfinden. Aber, dass man sich zusammenschließt, immer wieder abstimmt und so, das ist natürlich schon wichtig. Ich meine, es war ja auch nicht so, dass nur Japan betroffen, weil es hat ja auch diese Diskussionen geben was passiert mit dem Meerwasser und so weiter und so fort, was macht der Wind wieder mal und so weiter. So weit ist Japan nicht von anderen Ländern entfernt. Österreich hat es halt zum Glück für die Österreicher und Österreicherinnen nicht so sehr getroffen. Nähere Länder zu Japan waren schon ziemlich besorgt. Ich kann mich noch gut erinnern, dass da ganz viel auch sofort eingegriffen wurde, also in den Informationsaustausch. Auch ganz viele Ideen waren sofort da, was könnte alles verseucht sein und da war in allen Ländern die Bevölkerung sehr aktiv. Die Grünen haben damals in ganz Europa bei den nächsten Wahlen dazugewonnen, und es ist zum Teil auch zu mir ist gesagt worden, es sei darauf zurückzuführen. Also insofern ist so etwas immer auch eine globale Katastrophe.

Dass wir heute dieses Gespräch führen, ist ja ein eigenartiger Zufall, weil ja heute die EU,
Nuklearenergie, klimafreundlich dargestellt hat und wir da jetzt auch sehen, wie unterschiedlich das in den einzelnen Ländern behandelt wird. In Österreich darf man es sonst nicht einmal aussprechen, ja in anderen Ländern wie Frankreich sagt man das gerne und ist froh darüber. Also da hat jedes Land die eigenen Traditionen.

**Würdest du sagen, dass das Krisenmanagement im Falle von Atomunfällen besonders ist, oder besonders schwierig ist, weil Kernenergie doch ein sehr polarisierendes Thema ist und Menschen dazu einen häufig sehr fixierten und emotionalen Zugang haben? Wie wichtig ist, bei der Kommunikation mit der Bevölkerung deren Verständnis und Bildung über Kernenergie zu berücksichtigen?**

Kernenergie ist so ein Thema, wo jeder meint, etwas darüber zu wissen und man weiß im
Durchschnitt wahrscheinlich relativ wenig. Es ist aber auch deshalb ein emotionales Thema, weil so etwas selbstgemacht ist. Das ist nicht der Vulkan, der ausbricht und dadurch einen Tsunami quasi auslöst, ist ganz etwas anderes. Jemandem die Schuld geben und es wird auch jemand schuld sein, in welcher Weise auch immer, dadurch ist das noch viel emotionaler. Weil man, wenn man zum Opfer von anderen Menschen wird, wenn es die Natur macht, also eine sogenannte Naturkatastrophe, dann schaut das anders aus.

**Ich, was ich noch fragen wollte ist, welche Lektionen kann man lernen oder was kann man aus dem Unfall ableiten für die Zukunft und im Hinblick auf**

**das Krisenmanagement? Was kann man sozusagen präventiv für andere Länder lernen, aus dem Krisenmanagement der TEPCO und von der japanischen Regierung?**

Im Grunde genommen ist ja das genau das, was du versuchst, herauszufinden mit deiner Arbeit, also werden wir dann alle deine Arbeit lesen und das hoffentlich danach dann wissen und deshalb finde ich es so toll, dass du das das machst, weil das ein total wichtiges Thema ist. Da wird jetzt wahrscheinlich jeder was anderes sagen, weil jeder natürlich immer die Dinge aus seiner eigenen Branche sieht. Also man sieht natürlich einerseits, wie sehr wir meinen, dass, dass wir alles unter Kontrolle haben. Also man könnte lernen, dass wir die Dinge nie wirklich unter Kontrolle haben und dass man das auch vorbereitet sein muss und dass es, wenn man zur Krisenkommunikation geht, dann zeigt es einfach deutlich, dass die Bevölkerung ernst genommen werden will und das zeigt sich bei jeder einzelnen Krise: Menschen wollen nicht für dumm verkauft werden. Sie wollen Informationen, sie wollen nützliche Informationen, sie wollen sie schnell und sie wollen nicht beruhigt werden, weil das beruhigen so, dieses „Hab keine Panik und alles wird gut“, das ist nicht glaubwürdig und deshalb ist es noch immer am besten, möglichst so wie wir es am Anfang gesagt haben, bei Fakten zu bleiben und die Leute zwar vor zu großen Worten zu schützen, aber nicht vor der Wahrheit. Eigentlich, wir Österreicher haben mit der Ingeborg Bachmann, die diesen schönen Satz gesagt hat, dass „den Menschen die Wahrheit zumutbar sei“, dass das trifft, sich immer wieder gerade in der Krisenkommunikation.

**Dankeschön für die Beantwortung meiner Fragen!**